\documentclass[epsf,prb,twocolumn,showpacs,longbibliography,nofootinbib, superscriptaddress]{revtex4-1}
\usepackage{amsmath,amssymb,amsthm,amscd,latexsym,epsfig,bm,bbm}
\usepackage{float}
\usepackage{graphicx}
\usepackage{color}
\usepackage{framed}
\usepackage[colorlinks,bookmarks=false,citecolor=blue,linkcolor=red,urlcolor=blue]{hyperref}
\usepackage{verbatim}
\usepackage[normalem]{ulem}

\usepackage[all]{xy}

\usepackage{tikz}
\usepackage{tikz-cd}
\usetikzlibrary{arrows}
\usetikzlibrary{intersections}
\usetikzlibrary{shapes.geometric}
\usetikzlibrary{decorations.pathmorphing, patterns,shapes}
\usetikzlibrary{decorations.markings}

\makeatletter
\newcommand\notsotiny{\@setfontsize\notsotiny{6.5}{7.6}}
\makeatother
\newcommand{\overbar}[1]{\mkern 3mu\overline{\mkern-3mu#1\mkern-3mu}\mkern 3mu}

\newcommand{\R}{\mathbb{R}}
\newcommand{\Z}{\mathbb{Z}}
\newcommand{\GH}{\mathfrak{GH}}

\def\Z{\mathbb{Z}}
\newcommand{\GP}{P}
\newcommand{\GPinv}{E}
\newcommand{\tGPinv}{\widetilde{E}}

\begin{document}

\title{Flow of (higher) Berry curvature and bulk-boundary correspondence in parametrized quantum systems}
\author{Xueda Wen}
\affiliation{Department of Physics, University of Colorado, Boulder, CO 80309, USA}
\affiliation{Center for Theory of Quantum Matter, University of Colorado, Boulder, CO 80309, USA}
\affiliation{Department of Physics, Harvard University, Cambridge MA 02138}
\author{Marvin Qi}
\affiliation{Department of Physics, University of Colorado, Boulder, CO 80309, USA}
\affiliation{Center for Theory of Quantum Matter, University of Colorado, Boulder, CO 80309, USA}
\author{Agn\`{e}s Beaudry}
\affiliation{Department of Mathematics, University of Colorado, Boulder, CO 80309, USA}
\author{Juan Moreno}
\affiliation{Department of Mathematics, University of Colorado, Boulder, CO 80309, USA}
\author{Markus J. Pflaum}
\affiliation{Department of Mathematics, University of Colorado, Boulder, CO 80309, USA}
\affiliation{Center for Theory of Quantum Matter, University of Colorado, Boulder, CO 80309, USA}
\author{Daniel Spiegel}
\affiliation{Department of Mathematics, University of Colorado, Boulder, CO 80309, USA}
\affiliation{Department of Physics, University of Colorado, Boulder, CO 80309, USA}
\affiliation{Center for Theory of Quantum Matter, University of Colorado, Boulder, CO 80309, USA}
\author{Ashvin Vishwanath}
\affiliation{Department of Physics, Harvard University, Cambridge MA 02138}
\author{Michael Hermele}
\affiliation{Department of Physics, University of Colorado, Boulder, CO 80309, USA}
\affiliation{Center for Theory of Quantum Matter, University of Colorado, Boulder, CO 80309, USA}

\date{\today}

\begin{abstract}
This paper is concerned with the physics of parametrized gapped quantum many-body systems, which can be viewed as a generalization of conventional topological phases of matter. In such systems, rather than considering a single Hamiltonian, one considers a family of Hamiltonians that depend continuously on some parameters. After discussing the notion of phases of parametrized systems, we formulate a bulk-boundary correspondence for an important bulk quantity, the Kapustin-Spodyneiko higher Berry curvature, first in one spatial dimension and then in arbitrary dimension. This clarifies the physical interpretation of the higher Berry curvature, which in one spatial dimension is a flow of (ordinary) Berry curvature. In $d$ dimensions, the higher Berry curvature is a flow of $(d-1)$-dimensional higher Berry curvature.  Based on this, we discuss one-dimensional systems that pump Chern number to/from spatial boundaries, resulting in anomalous boundary modes featuring isolated Weyl points. In higher dimensions, there are pumps of the analogous quantized invariants obtained by integrating the higher Berry curvature. We also discuss the consequences for parametrized systems of Kitaev's proposal that invertible phases are classified by a generalized cohomology theory, and emphasize the role of the suspension isomorphism in generating new examples of parametrized systems from known invertible phases.
Finally, we present a pair of general quantum pumping constructions, based on physical pictures introduced by Kitaev, which take as input a $d$-dimensional parametrized system, and produce new $(d+1)$-dimensional parametrized systems. These constructions are useful for generating examples, and we conjecture that one of the constructions realizes the suspension isomorphism in a generalized cohomology theory of invertible phases.
\end{abstract}

\maketitle

\tableofcontents

\section{Introduction}
\label{sec:intro}

\subsection{Parametrized quantum systems}
Gapped phases of quantum matter are a source of surprisingly rich physical phenomena and beautiful mathematical structure. In particular, many gapped phases are topological, in the sense that their universal low-energy properties are described by a topological quantum field theory (TQFT). The understanding of topological phases has undergone tremendous advances over the past several years.\cite{Hasan2010,Hasan2011,Qi2011,turner2013band,Senthil2015,Witten2016,Wen2017}

A recent prominent theme is the discovery and study of generalizations of topological phases of matter. One generalization comprises topological phases in periodically driven, many-body-localized quantum systems, sometimes referred to as ``Floquet topological phases.\cite{harper20Floquet}'' Another generalization are the fracton phases, in which excitations of restricted mobility are associated with the lack of a low-energy TQFT description.\cite{nandkishore19fractons,pretko20fracton}

This paper is concerned with a different kind of generalization of topological phases. Namely, we study parametrized families of gapped systems, and the universality classes (phases) of such families. The basic idea is first to choose a space $X$ that we will refer to as the parameter space. In general, $X$ can be any topological space, but we usually have in mind nicely behaved compact spaces such as the $n$-dimensional sphere $S^n$. Next, for each point $x \in X$ of parameter space, we specify a gapped finite-range Hamiltonian $H(x)$ defined on a $d$-dimensional spatial lattice, and the Hamiltonian is assumed to vary continuously with $x$. Formally, this information can be packaged into a continuous map $H : X \to \GH$, where $\GH$ is some space of finite-range gapped Hamiltonians.

We refer to such a parametrized family, with parameter space $X$, as a (gapped) \emph{system over} $X$. The spatial dimension $d$, whether the system is bosonic or fermionic, and the symmetry that is imposed (if any), are all fixed attributes of the system that do not vary as a function of $X$ (even if $X$ has multiple connected components). It is important to emphasize that, in our terminology, the parameter space $X$ is distinct from (and far smaller than) the space $\GH$ of gapped Hamiltonians, or an even larger space of all finite-range Hamiltonians. These latter spaces are also sometimes referred to as parameter space, but to avoid confusion we will not do so in this paper. Intuitively, it is helpful to think of $X$ as a space of parameters that can be tuned by a hypothetical experimentalist who can vary some number of ``knobs'' that control certain terms appearing in the system's Hamiltonian, such as the strength of a magnetic field.

When we study ordinary gapped quantum systems, we are usually interested in phases of such systems, and universal properties that are the same throughout a phase. The same is true for systems over $X$, where we have a related notion of a \emph{phase over} $X$. Roughly speaking, two systems over $X$ are in the same phase if they can be continuously deformed into one another without closing the gap anywhere in parameter space. A more detailed and precise definition of phases over $X$ is given in Sec.~\ref{sec:parametrized}.

Before proceeding, it is helpful to anchor our discussion by mentioning some familiar examples. First of all, in the case where $X = \mathrm{pt}$ is the topological space containing a single point, a gapped system over $X$ is nothing but an ordinary gapped quantum system, which can be referred to as a \emph{system over a point}. Moreover, phases over $X = \mathrm{pt}$ (\emph{phases over a point}) are the same as ordinary phases.

A more interesting example is a single spin-1/2 in a Zeeman magnetic field of fixed magnitude, where the field direction is parametrized by $X = S^2$ viewed as a space of unit vectors.\cite{berry1984Berry,SimonBerryPhase} This system is characterized by a non-trivial Berry curvature and associated non-zero Chern number. Familiar examples are not limited to zero dimensions; the Thouless charge pump is a classic one-dimensional example with a conserved ${\rm U}(1)$ charge.\cite{Thouless1983} The parameter space is a circle ($X = S^1$),
and as the parameter is adiabatically cycled around the circle, a quantized amount of charge is pumped across the system. These two familiar examples are reviewed in more detail in Sec.~\ref{sec:review}.

A number of recent works have begun to explore the physics of parametrized families beyond a handful of classic examples,\cite{TeoKane2010,Kitaev2019differential,Cordova2020A,Cordova2020B,Kapustin_2020,kapustin2020pump,Hsin_2020,Hermele2021,shiozaki2021adiabatic} and these recent developments are also reviewed in Sec.~\ref{sec:review}. Among these works, Kapustin and Spodyneiko's higher Berry curvature and the associated quantized invariant -- which we refer to as the KS invariant -- will play a particularly important role in this paper.\cite{Kapustin_2020}

There are a number of reasons that parametrized families of gapped systems are of physical interest. Quite generally, if we view $X$ as a subspace of a system's phase diagram lying entirely within a gapped phase, the corresponding phase over $X$ captures universal global properties of the phase diagram. A related idea is that we often think of ordinary gapped phases as connected components of a space $\GH$ of gapped Hamiltonians. Non-trivial phases over $X$ can be viewed, at least roughly, as probing more subtle features of the topology of $\GH$. A more specific perspective, which applies when $X = S^1$ (or, more generally, $X = S^1 \times Y$), is to observe that systems over $X$ are in correspondence with adiabatic pumping cycles where the circle-valued parameter is cycled slowly as a function of time, as in the Thouless charge pump. We may thus expect to find quantized pumping phenomena associated with non-trivial phases over $X$.

A more abstract reason to be interested in parametrized families comes from Kitaev's proposal that gapped invertible phases are classified by a generalized cohomology theory,\cite{kitaevSimonsCenter1,kitaevSimonsCenter2,kitaevIPAM} which builds on earlier work on the $K$-theory classification of gapped free-fermion phases.\cite{kitaev2009periodic}
Given a non-negative integer $d$ and any topological space $X$ (not necessarily with the physical interpretation of parameter space), a generalized cohomology theory assigns an abelian group $E^d(X)$. Taking $X$ to be a single point, in Kitaev's proposal $E^d(\mathrm{pt})$ is the group of ordinary $d$-dimensional invertible phases. More generally, $E^d(X)$ is the group of $d$-dimensional invertible phases over $X$. Establishing Kitaev's proposal remains an open problem, and we discuss its status in Sec.~\ref{sec:classification}.

Understanding whether or not gapped invertible phases are classified by a generalized cohomology theory, and if so which cohomology theory, is important for illuminating the relationship between gapped phases and quantum field theory. Invertible topological quantum field theories (TQFTs) have been shown to be classified by a generalized cohomology theory.\cite{freed2021reflection}
It is generally believed that invertible gapped phases are well-described by invertible TQFTs, and one way to establish this would be to show the corresponding cohomology theories are the same.

\subsection{Overview of the paper}

This paper consists of two intertwined threads that we now describe; this discussion also serves as an outline of the remainder of the paper. One of these threads, in Sections~\ref{sec:1dmodel-bbc}, \ref{sec:chernpump} and~\ref{Sec:HigherChernPump}, is the development of a bulk-boundary correspondence for Kapustin and Spodyneiko's higher Berry curvature, which is a closed $(d+2)$-form $\Omega^{(d+2)}$ on parameter space $X$, where $X$ is a differentiable manifold and $d$ is the spatial dimension. This bulk-boundary correspondence clarifies the physical interpretation of the higher Berry curvature and the associated quantized KS invariant, and relates it to anomalous boundary properties of non-trivial phases over $X$. Sections~\ref{sec:1dmodel-bbc} and~\ref{sec:chernpump} discuss systems in one spatial dimension ($1d$), while Section~\ref{Sec:HigherChernPump} discusses higher dimensions.

In a semi-infinite one-dimensional system, we show that $\Omega^{(3)}$ can be understood as a flow of 2-form Berry curvature to/from the spatial boundary of the system (Sec.~\ref{sec:1dmodel-bbc}). This is illustrated in a solvable $1d$ spin chain over $S^3$ denoted as $H_{1d}$ (Sec.~\ref{Sec:1+1d_lattice}). Spatial boundaries of this model are anomalous, with a single gapless Weyl point over $S^3$, which is impossible for a strictly $0d$ system over $S^3$. The boundary phenomena are associated with a bulk non-zero quantized KS number of $2\pi$, showing that $H_{1d}$ is in a non-trivial phase over $S^3$. The KS number is computed in Sec.~\ref{sec:hbc-1d} following a review of Kapustin and Spodyneiko's results. The bulk-boundary correspondence, which gives $\Omega^{(3)}$ the interpretation of a flow of Berry curvature, is developed in Sec.~\ref{Sec:Semi-infinite}, and is further illustrated in $H_{1d}$ by a ``clutching construction'' in Sec.~\ref{Sec:Clutching}. Finally, Sec.~\ref{Sec:Inverse} presents an inverse system for $H_{1d}$, in the sense that $H_{1d}$ stacked with its inverse is in the trivial phase over $S^3$.

As a consequence of the bulk-boundary correspondence, we should expect that it is possible to construct a $1d$ system that can be understood as a Chern number pump. In Sec.~\ref{sec:chernpump}, we construct such a system over $S^2 \times S^1$, which is closely related to the system $H_{1d}$ over $S^3$. Upon cycling the periodic parameter taking values in $S^1$, a quantized Chern number (over $S^2$) is pumped to/from the spatial boundary.

More generally, in $d$-dimensions, $\Omega^{(d+2)}$ has the interpretation of a flow of $(d+1)$-form higher Berry curvature to/from a $(d-1)$-dimensional spatial boundary (Sec.~\ref{Sec:HigherChernPump}). After reviewing higher Berry curvature in systems of dimension two and higher (Sec.~\ref{Sec:KS_higherD}), we introduce a solvable $2d$ model over $S^4$ denoted $H_{2d}$ (Sec.~\ref{sec:2dmodel}). Just like $H_{1d}$ has a boundary termination with a single gapless point that is a source of Berry curvature (a Weyl point), similarly a boundary of $H_{2d}$ has a gapless point that sources the $1d$ higher Berry curvature $\Omega^{(3)}$. We develop the bulk-boundary correspondence for the higher Berry curvature $\Omega^{(4)}$ in two dimensions, and use this to compute the KS number of $H_{2d}$, showing it takes the non-trivial quantized value $2\pi$ (Sec.~\ref{Sec:2d_KS_bbc}). $H_{2d}$ can be obtained from $H_{1d}$ by the suspension construction (see below) and it is not a coincidence that these two systems have the same KS number. Indeed, in Sec.~\ref{sec:ndKSnumber} we show that upon applying the suspension construction to a $d$-dimensional system over $X$ (where $X$ is a $(d+2)$-dimensional oriented closed differentiable manifold), the resulting $(d+1)$-dimensional system has the same KS number as the input $d$-dimensional system. It follows that there is a sequence of $d$-dimensional systems over $S^{d+2}$, all with KS number $2\pi$, of which the first three members are (in $d=0$) a single spin-1/2 in a Zeeman magnetic field, $H_{1d}$ and $H_{2d}$. The $d$th system in this sequence is obtained from the $d=0$ system over $S^2$ by applying the suspension construction $d$ times. Finally in Sec.~\ref{subSec:KSpump} we discuss $(d+1)$-dimensional systems over $X \times S^1$ that pump $d$-dimensional KS number to/from the spatial boundary.

The second thread discusses general results and expectations for gapped parametrized systems, and introduces a pair of general quantum pumping constructions, based on physical pictures introduced by Kitaev,\cite{kitaevSimonsCenter1,kitaevSimonsCenter2,kitaevIPAM} that take as input a $d$-dimensional parametrized system and produce a $(d+1)$-dimensional system. The key property of the output system is a pumping or flow of the input $d$-dimensional system to/from a spatial boundary. These constructions can be used to generate many examples of parametrized systems.

Section~\ref{sec:parametrized} discusses parametrized quantum systems in general, and in particular gives a detailed account of the notion of ``phases over $X$,'' \emph{i.e.} phases of parametrized systems over the parameter space $X$. We also discuss the notion of invertible systems and phases over $X$, discuss their expected anomalous boundary properties, and formulate a conjectured bulk-boundary correspondence for invertible systems. We formalize the idea that a system's Hamiltonian near a spatial boundary can depend on additional parameters that do not affect the bulk Hamiltonian.

In Sec.~\ref{sec:classification}, we discuss the expectation that phases of invertible parametrized systems are classified by a generalized cohomology theory.\cite{kitaevSimonsCenter1,kitaevSimonsCenter2,kitaevIPAM} First in Sec.~\ref{sec:gc1}, assuming this proposal is true, we discuss its consequences. Our focus is on using generalized cohomology classification to guide the construction of new examples, emphasizing the important role of the suspension isomorphism from this perspective. Then in Sections~\ref{sec:gc2} and~\ref{sec:gc3}, we review the generalized cohomology proposal and sketch a potential strategy to construct a generalized cohomology theory of parametrized invertible phases.

The heart of the second thread is Section~\ref{Sec:GeneralConstruct}, where we introduce the general quantum pumping constructions and study their boundary physics. One of these constructions, which is the \emph{suspension construction} referred to above, takes as input  a $d$-dimensional invertible system $H_d$ over $X$, and produces a $(d+1)$-dimensional invertible system $SH_d$ over the (unreduced) suspension $SX$. The suspension $SX$ is a quotient space formed from the product $X \times [-1,1]$ by identifying each of $X \times \{ 1 \}$ and $X \times \{-1 \}$ to single points. If $X = S^n$, then $SX \cong S^{n+1}$. Making slightly different choices, we instead obtain a $(d+1)$-dimensional system $PH_d$ over $X \times S^1$. While the explicit examples studied in this paper are bosonic systems without symmetry, we emphasize that these quantum pumping constructions are not limited to this setting, and can be applied to invertible phases with any desired attributes. We conjecture the suspension construction gives a concrete realization of the \emph{suspension isomorphism} in a generalized cohomology theory classifying gapped invertible phases. 

After introducing the quantum pumping constructions in Sec.~\ref{subsec:SP-constructions}, we discuss their boundary physics in Sec.~\ref{subsec:general-bdy-physics}. In particular we show that $PH_d$ can be viewed as a pump of the $d$-dimensional phase invariant of $H_d$ over $X$, from a spatial boundary into the bulk, upon cycling the periodic parameter in $S^1$. Similarly, $SH_d$ can be understood in terms of flow of the $d$-dimensional phase invariant from the boundary to the bulk; the essence of this phenomenon is the same as in $PH_d$ but is not pumping in a strict sense, due to the absence of a suitable periodic parameter. Finally, Sec.~\ref{subsec:universality} provides a roadmap toward potentially establishing that the suspension construction realizes the suspension isomorphism in a generalized cohomology theory of invertible phases.

The paper concludes with a discussion in Sec.~\ref{sec:discussion}. Some technical details -- on computation of the KS number, properties of the solvable models, and the locality of the higher Berry curvature -- are contained in appendices.

\section{Chern number, the Thouless charge pump, and generalizations}
\label{sec:review}

Here we review two well-known examples of parametrized quantum systems, as well as some recent work on parametrized quantum systems and related topics. Our first example is a $0d$ spin-$\frac{1}{2}$ quantum system over $S^2$.\cite{berry1984Berry,SimonBerryPhase} We parametrize points of $S^2$ by $w = (w_1,w_2,w_3)$ with $w_1^2 + w_2^2 + w_3^2 = 1$. The Hamiltonian is
\begin{equation}
\label{spinhalf}
        H(w) =  w_1 \sigma^1 + w_2 \sigma^2 + w_3 \sigma^3 \text{,}
\end{equation}
where $\sigma^{1,2,3}$ are the usual Pauli operators.
The eigenvalues are $E_\pm = \pm 1$ for all $w \in S^2$, so the Hamiltonian is gapped and defines a $0d$ system over $S^2$.  

It is well known that there is no way to consistently define the phase of the ground state globally over $S^2$; the obstruction to doing so is captured by the integral of the Berry curvature over $S^2$. The Berry curvature is a $2$-form $\Omega^{(2)}$ on $S^2$ (see \eqref{F2} for a general expression), and for this system is given by
\begin{equation}
\Omega^{(2)} = \frac{\sin \theta}{2} d\theta \wedge d\phi \text{,}
\end{equation}
where $\theta$ and $\phi$ are the usual spherical polar coordinates. 

The Chern number
\begin{equation}
C = \int_{S^2} \Omega^{(2)} = 2\pi
\end{equation}
measures the obstruction not only to defining the phase of the ground state globally over $S^2$, but also -- at the same time -- to deforming the parametrized family $H(w)$ to a trivial (\emph{i.e.} constant) family over $S^2$ without closing the gap for some $w \in S^2$. In more detail, suppose we have a homotopy $H(w,t)$ taking values in Hermitian $2 \times 2$ matrices, where $t \in [0,1]$ and $H(w,0) = H(w)$. If $H(w,1)$ is constant (\emph{i.e.} does not depend on $w$), the associated Chern number would be zero, which is only possible if $H(w,t)$ is gapless for some value of $(w,t) \in S^2 \times [0,1]$. 

In more mathematical language, the ground states form a line bundle over $S^2$ which does not admit a nowhere vanishing global section, due to the nonvanishing first Chern class which generates $H^2 (S^2, \Z)$. It is impossible to deform $H(w)$ to a family which is constant over $S^2$, because the ground states of a constant family of Hamiltonians form a trivial line bundle. 

We note that defining the Chern number requires us to specify an orientation on $S^2$; this is required for the integral $\int_{S^2} \Omega^{(2)}$ to make sense. We choose the orientation given by the locally-defined 2-form $d\theta \wedge d\phi$. This is the same as what we refer to as the outward-normal orientation on $S^2$, which is given by viewing $S^2 \subset \R^3$, and defining a basis $(b_1, b_2)$ of the tangent space $T_x S^2$ at a unit vector $x \in \R^3$ to be positively oriented if $(x,b_1,b_2)$ has the same orientation as the standard orientation on $\R^3$.

Our second familiar example is the Thouless charge pump.\cite{Thouless1983}
In our terminology, this a $1d$ quantum system over $S^1$ with a ${\rm U}(1)$ symmetry. A parametrized Hamiltonian for the Thouless charge pump is  
\begin{multline}
        H(w_1,w_2) = -\frac{w_1}{2} \sum_{i \in \Z} (-1)^i c^{\dagger}_i c_i 
        + \sum_{i \in 2\mathbb{Z}} g^+(w_2)  c^{\dagger}_i c_{i+1} \\
        + \sum_{i \in 2\mathbb{Z}+1} g^-(w_2) c^{\dagger}_i c_{i+1}+ \text{H.c.}
\end{multline}
where $w = (w_1, w_2)$ with $w_1^2 + w_2^2 = 1$ parametrizes points of $S^1$ and 
where the functions $g^\pm(w_2)$ are chosen as follows:
\begin{equation}
        g^+(w_2) = \begin{cases} w_2, & w_2 \geq 0 \\ 0, & w_2 \leq 0 \end{cases}
\end{equation}
\begin{equation}
        g^-(w_2) = \begin{cases} 0, & w_2 \geq 0 \\ -w_2, & w_2 \leq 0 \end{cases}.
\end{equation}
The $c_i$ and $c^{\dagger}_i$ are fermion creation and annihilation operators indexed by lattice sites $i \in \mathbb{Z}$, 
satisfying the anticommutation relations $\{c_j,c_k\}=\{c_j^{\dag},c_k^{\dag}\}=0$, and
$\{c_j,c_k^{\dag}\}=\delta_{jk}$. The Hamiltonian clearly has a ${\rm U}(1)$ charge conservation symmetry at all points in parameter space. This free-fermion system is a two-band insulator with flat bands for all $(w_1,w_2)$, and is therefore gapped at a filling of one fermion per lattice site.

This system over $S^1$ is nontrivial because it exhibits quantized charge pumping. Interpreting the $S^1$ as a time parameter to be cycled adiabatically, one unit of charge will be pumped through any point of the lattice over a single cycle. It was argued early on that this phenomenon is robust to interactions and to disorder;\cite{Niu_1984}
any deformation of the system over $S^1$ that preserves the gap and the ${\rm U}(1)$ symmetry will pump the same quantized amount of charge.

One simple way to understand the robustness of the quantized charge pumping is to consider a semi-infinite system, \emph{i.e.} a system with a spatial boundary. For instance, one can define a system with boundary by retaining all lattice sites with $i \leq N$, taking $N$ even, and dropping all terms in the Hamiltonian that couple to sites with $i >N$ while keeping other terms unchanged. One finds that the energy spectrum is gapless at the single point of parameter space $(0,1) \in S^1$ and gapped otherwise. Upon tuning $w \in S^1$ across the gapless point, the ${\rm U}(1)$ charge of the ground state jumps by unity. As a result, there is no way to consistently define the ground state charge as a function of $w \in S^1$; this would be impossible for a strictly $0d$ system over $S^1$, and the boundary is anomalous in this sense. In the semi-infinite system, such a charge jump is allowed, because it is compensated by a quantized pumping of charge from the bulk to the boundary (or vice versa). Finally, while modifying the Hamiltonian near the boundary can introduce additional gapless points where the charge jumps, this cannot change the total net quantized jump in charge as $w$ is cycled around $S^1$.

We now briefly mention some recent works that pertain to generalizations of the two familiar examples reviewed above, especially in higher-dimensional interacting systems. In free fermion systems, higher-dimensional analogs of the Thouless charge pump were considered by Teo and Kane.\cite{TeoKane2010} Among other results, they classified parametrized families of $d$-dimensional free fermion systems, in particular over $D$-dimensional spheres $S^D$.

Building on a proposal of Kitaev,\cite{Kitaev2019differential} Kapustin and Spodyneiko generalized the Berry curvature (a 2-form for $d=0$ systems) to families of gapped many-body systems in $d$ spatial dimensions.\cite{Kapustin_2020}
The higher Berry curvature is constructed as a closed $(d+2)$-form on the parameter
space. The cohomology class of the higher Berry curvature defines a topological invariant which we refer to as the KS invariant. Moreover, when the parameter space $X$ is an oriented $(d+2)$-manifold, the higher Berry curvature can be integrated over $X$, resulting in the KS number, which was argued to be quantized for invertible systems.

Reference~\onlinecite{kapustin2020pump} studied generalizations of the Thouless charge pump to $d$-dimensional systems with ${\rm U}(1)$ symmetry over $S^d$.
Reference~\onlinecite{Hsin_2020} studied parametrized families of quantum field theories and their associated topological terms, focusing in particular on many-body diabolical points in phase diagrams that generalize Weyl points in $0d$, and on boundary phenomena.  
See also Ref.~\onlinecite{Thorngren_Else_2016} for a classification of topological phases with crystalline symmetry using parametrized families of TQFTs.

Recently, the generalization of quantum pumping to parametrized families of systems with discrete symmetries and various types of parameter spaces $X$ was reported in Ref.~\onlinecite{Hermele2021}.
Later in Ref.~\onlinecite{shiozaki2021adiabatic},
the quantum pumping in invertible quantum spin systems with discrete symmetries was studied. Focusing on $X=S^1$, it was shown that a lower dimensional symmetry protected 
 topological phase is pumped to the boundary after a nontrivial cycle of quantum pumping. In addition, it was shown that a symmetry protected topological phase in one dimension lower can be trapped at a spatial texture in the parametrized system.

From the perspective of quantum field theory, so-called anomalies in the parameter space
were recently studied in parametrized families of quantum field theories.\cite{Cordova2020A,Cordova2020B,TTT1703,Zohar1705,Zohar1706,Yuji1710,Yuji1803} In a theory with global symmetries, a 't Hooft anomaly can be diagnosed by coupling to background gauge fields and finding that the resulting partition function is not gauge invariant. In parametrized families of quantum field theories,  by viewing the scalar coupling constants as background fields, the
notion of 't Hooft anomalies can be extended to include these fields.  Quantum field theories with such anomalies arise as effective field theories for spatial boundaries of the gapped invertible parametrized quantum systems that are the focus of this paper.

\section{Parametrized systems and phases}
\label{sec:parametrized}

Here we give a more detailed account of what we mean in this paper by systems and phases over $X$. We also discuss the notions of trivial and invertible systems and phases over $X$. We do not aim to be mathematically precise or to give a complete discussion of all the issues that arise in attempting to define phases. Instead, our intent is to give enough detail for a theoretical physicist reader to understand what we mean by systems and phases over $X$. Indeed, some degree of detail is necessary given that these notions are not standard in condensed matter physics.

First we recall the definition of a $d$-dimensional gapped quantum system over $X$. To specify such a system, we take the $d$-dimensional lattice $\Z^d$, and place some degrees of freedom at each lattice site, as usual for quantum lattice systems in $d$-dimensions. (If $d=0$ the lattice is a single point.) These degrees of freedom can be either bosonic or fermionic. We may choose to impose some symmetry; to avoid subtleties, we only consider unitary or anti-unitary internal symmetries (\emph{i.e.} symmetries that do not permute lattice points) throughout. We then imagine that we have a space $\GH$ of gapped, finite-range Hamiltonians respecting the symmetry (if any). We also sometimes write $\GH_d$ when we want to emphasize the spatial dimension. Having made these choices, we fix a topological space $X$ (the parameter space), and a gapped system over $X$ is a continuous map $H : X \to \GH$. More generally, a (not necessarily gapped) system over $X$ is a continuous map $H : X \to \mathfrak{H}$, where $\mathfrak{H}$ is a space of all finite-range Hamiltonians.

It is convenient to denote a system over $X$ by $H$, where $H$ is the Hamiltonian map $H: X \to \GH$. However, it is important to keep in mind that this is a shorthand notation. More precisely we should say that a system over $X$ is specified by the following data:
\begin{enumerate}
    \item The spatial dimension $d$.
    \item The local degrees of freedom attached to each lattice site.
    \item The symmetry imposed, if any.
    \item The parameter space $X$.
    \item The continuous map $H : X \to \GH$. (Note that we suppose the first three pieces of data give us the space $\GH$.) 
\end{enumerate}

A system $H$ over $X$ is said to be trivial if the map $H$ is constant as a function of $x \in X$, and if the Hamiltonian $H(x)$ is that of a collection of decoupled zero-dimensional systems, one for each lattice site. The ground state is thus a product state (for bosonic systems) or an atomic insulator (for fermions). In defining trivial systems, it might be tempting only to require that $H(x)$ is a collection of decoupled $d=0$ systems for each $x \in X$, but \emph{not} to require that $H(x)$ is constant. However, this choice is unnatural as it would cause problems in $d=0$, where it would lump the example of a spin-1/2 in a Zeeman field into the trivial phase (as defined below), even though this system is characterized by a non-zero Chern number.

There is an important binary operation defined on systems over $X$ of the same spatial dimension referred to as stacking, where two decoupled systems $H_1$ and $H_2$ are placed ``on top of one another.'' For instance, in the case of bosonic systems, the Hilbert space at each lattice site is a tensor product of the two systems being stacked. The Hamiltonian is $H_{{\rm stack}}(x) = H_1(x) \otimes \mathbbm{1} + \mathbbm{1} \otimes H_2(x)$. We often denote the stacked system by writing $H_1 \ominus H_2$. Note that stacking two trivial systems produces another trivial system.

Now we can describe gapped phases over $X$ as equivalence classes of gapped systems over $X$. First of all, we do not consider \emph{all} gapped systems over $X$ when discussing phases. Instead we fix the spatial dimension $d$, whether fermionic degrees of freedom are allowed, and the symmetry imposed (if any). We also fix the parameter space $X$. Phases are equivalence classes defined on the set of systems with the preceding fixed characteristics. The equivalence relation is then generated by three operations -- deformation, stacking and isomorphism -- which we describe in turn.

First, two systems $H_1$ and $H_2$ over $X$ are in the same phase if their Hamiltonian functions can be deformed continuously into one another while keeping the gap open and preserving the symmetry. More precisely, we have a homotopy between the maps $H_1 : X \to \GH$ and $H_2 : X \to \GH$. That is, we have a continuous map $H : X \times [0,1] \to \GH$, with values written $H(x,t)$, so that $H(x,0) = H_1(x)$ and $H(x,1) = H_2(x)$. (Note that by demanding $H(x,t) \in \GH$, the Hamiltonian is automatically gapped and symmetry-invariant for all $(x,t) \in X \times [0,1]$.) When $X = \mathrm{pt}$, this reduces to the usual notion that two gapped systems whose Hamiltonians are joined by a path are in the same phase, as long as the path is through gapped, symmetry-respecting Hamiltonians.

Second, two systems $H_1$ and $H_2$ over $X$ are considered to be in the same phase when they become the same upon stacking with trivial systems $T_1$ and $T_2$. That is, more precisely, $H_1$ and $H_2$ are in the same phase if there exist trivial systems $T_1$ and $T_2$ such that $H_1 \ominus T_1 = H_2 \ominus T_2$. The motivation for this equivalence operation is the same for parametrized systems as for ordinary systems (\emph{i.e.} systems over a point). Lattice models of quantum systems are always idealizations that ignore some effectively inert degrees of freedom, like the atomic core levels in a solid. Whether or not we ignore some such degrees of freedom is a choice we make in our theoretical description, so it should not affect which phase a given system is in. This equivalence operation ensures that phases are not sensitive to such choices.

Finally, and perhaps least familiar, we should allow for a notion of isomorphism between systems, and consider isomorphic systems to be in the same phase. Isomorphism should capture the rough idea that two different systems as defined above may in fact be different mathematical descriptions of ``the same'' physical system. For example, we may have two systems that are mapped into one another by a unitary transformation (constant over $X$), where the unitary is a product of single-site unitary operators. Such a unitary is nothing but a local basis change, and two systems thus related are clearly different descriptions of the same physical system. Another example is that two systems related by a translation of the lattice should be considered isomorphic, since the only difference between such systems is the arbitrary choice of origin. Beyond these two examples, additional kinds of isomorphism may ultimately be needed to give a mathematically precise and physically sensible definition of phases (whether ordinary phases or phases of parametrized systems). However, this issue is not important for our present purposes and will thus be left for the future.

We can compare phases over two parameter spaces $X$ and $Y$ if we are given a continuous map $f : X \to Y$. Then given a system $H$ over $Y$, we can form the pullback system $f^* H$ over $X$, whose Hamiltonian map is simply $H \circ f$.  The pullback gives a function $f^* : \GP^d(Y) \to \GP^d(X)$, where $\GP^d(X)$ and $\GP^d(Y)$ are the sets of $d$-dimensional gapped phases over the corresponding spaces. It is straightforward to show that if $f : X \to Y$ is a homotopy equivalence, then $f^* : \GP^d(Y) \to \GP^d(X)$ is a bijection. It is an important point that, in this sense, only the homotopy type of $X$ plays a role in studying phases over $X$.

The trivial phase over $X$ is defined to be the unique phase containing any trivial system over $X$. This is well-defined because any two trivial systems $T_1$ and $T_2$ are obviously in the same phase, because $T_1 \ominus T_2$ and $T_2 \ominus T_1$ are isomorphic (trivial) systems. Note that our terminology distinguishes between the trivial phase and trivial systems, and, in particular, a typical system in the trivial phase will not be a trivial system.

An important subset of phases are the invertible phases, which form an abelian group under the stacking operation. A system $H$ over $X$ is invertible (or, synonymously, in an invertible phase) if there exists another system $\overbar{H}$ over $X$ such that $H \ominus \overbar{H}$ is in the trivial phase. In this paper, we denote the group of $d$-dimensional invertible phases over $X$ by $\GPinv^d(X)$. It is easy to show that if $H$ is invertible, then, for each $x \in X$, $H(x)$ is an invertible system over a point. While the converse statement is physically reasonable, it is not obviously true. 

Physically, invertible systems over a point are characterized by a lack of fractional bulk excitations, and by spatial boundaries with a range of interesting properties. Roughly speaking, the boundary of a non-trivial $d$-dimensional invertible phase (over a point) is anomalous in the sense that it cannot occur on its own as a $(d-1)$-dimensional quantum system. Another sense in which the boundary physics is anomalous is that a QFT description of the boundary is characterized by a non-trivial quantum anomaly.

We expect a similar situation for invertible systems over $X$, but before stating our expectations, some general remarks about systems with spatial boundaries are needed. Suppose we have a $d$-dimensional system over $X$ (not necessarily invertible), and we truncate the lattice to expose a $(d-1)$-dimensional spatial boundary. Sufficiently far away from the boundary, the local terms in the Hamiltonian $H(x)$ are the same as before the boundary was introduced. However the Hamiltonian near the boundary is not naturally given in terms of $H(x)$. Moreover, the Hamiltonian near the boundary can depend on additional parameters that do not affect the bulk Hamiltonian. This can be formalized by introducing ``bulk'' and ``boundary'' parameter spaces $X_{{\rm bulk}}$ and $X_{{\rm bdy}}$, with a surjective map $\pi : X_{{\rm bdy}} \to X_{{\rm bulk}}$. The system with boundary is a system (not necessarily gapped) over $X_{{\rm bdy}}$. However, away from the boundary, the local terms in the Hamiltonian only depend on $x \in X_{{\rm bdy}}$ through $\pi(x) \in X_{{\rm bulk}}$. The simplest possibility is of course $X_{{\rm bulk}} = X_{{\rm bdy}}$ with $\pi$ the identity map, but other situations will arise in the discussion of later sections.

Now suppose we have a non-trivial $d$-dimensional invertible system over $X$, in the presence of a spatial boundary, in the simple case $X = X_{{\rm bulk}} = X_{{\rm bdy}}$. We expect that the $(d-1)$-dimensional spatial boundary is anomalous if viewed as a $(d-1)$-dimensional system over $X$, in the sense that some property of the boundary is impossible for a strictly $(d-1)$-dimensional system over $X$.  There is an important difference from systems over a point, which is clearest when $X$ is connected and when $H(x)$ is in the trivial phase (over a point) for each $x \in X$. In such cases, which include the concrete examples studied in this paper, the anomalous nature of the boundary will only be apparent globally, \emph{i.e.} by examining the boundary physics as $x$ varies over all of $X$. In contrast, the local boundary physics in a contractible neighborhood of some $x_0 \in X$ is not anomalous and can be realized in a strictly $(d-1)$-dimensional system. In different language, we expect that a QFT for the boundary will be characterized by a non-trivial anomaly in the space of coupling constants.\cite{Cordova2020A,Cordova2020B}

Another way to phrase our expectations is in terms of a bulk-boundary correspondence, namely:
\begin{framed}
\textbf{Bulk-boundary correspondence:} Two invertible systems $H_1$ and $H_2$ over $X$ are in the same phase if and only if there exists a trivially gapped spatial interface between them.
\end{framed}
By a spatial interface, we mean a third system over $X$ whose Hamiltonian is locally that of $H_1$ on one side of a co-dimension one boundary region, and locally that of $H_2$ on the other side of the boundary. By trivially gapped, we mean that the interface, in addition to being gapped, lacks fractional excitations and spontaneous symmetry-breaking for all points $x \in X$. We obtain a special case of the bulk-boundary correspondence by thinking of vacuum as a system in the trivial phase, then the bulk-boundary correspondence implies that an invertible system over $X$ is in the trivial phase if and only if it admits a trivially gapped boundary (\emph{i.e.} trivially gapped interface to vacuum).

\section{Flow of Berry curvature and bulk-boundary correspondence in $1d$ parametrized systems}
\label{sec:1dmodel-bbc}

In this section, we construct a parameterized family of 1$d$ lattice models over $X = S^3$ (Sec.~\ref{Sec:1+1d_lattice}), and study its bulk and boundary properties. The spatial boundary of our model has a single gapless Weyl point over $S^3$, which is impossible for a strictly zero-dimensional system over $S^3$ and is thus anomalous. In Sec.~\ref{sec:hbc-1d}, after reviewing Kapustin and Spodyneiko's work on the higher Berry curvature and the associated quantized KS invariant, we compute these quantities for our solvable model over $S^3$. We then develop a bulk-boundary correspondence relating the bulk higher Berry curvature to the flow of Berry curvature to/from a spatial boundary (Sec.~\ref{Sec:Semi-infinite}). In Sec.~\ref{Sec:Clutching} we describe a clutching construction that provides another way of understanding the relationship between the 1$d$ KS invariant and the Chern number in zero dimensions. Finally, Sec.~\ref{Sec:Inverse} presents a system over $S^3$ that is an inverse to the system introduced in Sec.~\ref{Sec:1+1d_lattice}; the KS numbers of these two systems are opposite, and when stacked together, they can be deformed to a trivial system.

\subsection{Exactly solvable 1$d$ lattice model over $S^3$}
\label{Sec:1+1d_lattice}

We consider a one-dimensional bosonic system with a single qubit placed at each lattice site, and choose the parameter space $X = S^3$. This choice of parameter space is motivated by the results of 
Kapustin and Spodyneiko, which lead us to expect non-trivial phases over $S^3$ in $d=1$. In more detail, the KS invariant is given by integrating a closed $3$-form with quantized periods over $X$, suggesting a $\Z$ classification over $S^3$. From a different point of view, it is believed that the classifying space of bosonic $d=1$ gapped invertible phases is a $K(\Z,3)$, which also gives a classification by $H^3(S^3, \Z) \cong \Z$.

The Hamiltonian has the form
\begin{equation}
\label{H_1d}
H_{1d}(w)=\sum_{p\in\mathbb Z} H_p^1(w)+\sum_{p\in2\mathbb Z+1} H^{2,+}_{p,p+1}(w)
+\sum_{p\in2\mathbb Z}H^{2,-}_{p,p+1}(w),
\end{equation}
where $w \in S^3$. It is convenient to view $S^3$ as a subspace of $\R^4$ and write $w = (w_1,w_2,w_3,w_4)$ with the constraint $\sum_{i=1}^4 w_i^2=1$. 
The first term in Eq.~\ref{H_1d} is a single-spin term, with 
\begin{equation}
\label{H_1body}
H_p^1(w)=(-1)^p (w_1\sigma_p^1+w_2\sigma_p^2+w_3\sigma_p^3),
\end{equation}
where $p \in \Z$ denotes the lattice site, and $\sigma_p^{1,2,3}$ are Pauli matrices of the qubit at site $p$.
Note that $H_p^1 = 0$ at the two poles $w_4=\pm 1$ of $S^3$, where $w_1=w_2=w_3=0$.
The second and third terms are two-spin terms given by:
\begin{equation}
\label{H_2body}
H_{p,p+1}^{2,\pm}(w)=g^{\pm}(w)\sum_{\mu=1,2,3}\sigma_p^{\mu}\sigma_{p+1}^{\mu}.
\end{equation}
The real functions $g^{\pm}(w)$ are chosen as follows:
\begin{equation}
\label{g+}
g^+(w)=
\begin{cases}
w_4, \quad& 0\le w_4\le 1,\\
0, \quad &\text{otherwise, }
\end{cases}
\end{equation}
and
\begin{equation}
\label{g-}
g^-(w)=
\begin{cases}
-w_4, \quad&-1\le w_4\le 0,\\
0, \quad &\text{otherwise}.
\end{cases}
\end{equation}

We remark that $H(w)$ is continuous but not smooth at 
$w_4=0$. As we will see later, with this choice, 
the higher Berry curvature will 
not be continuous at $w_4=0$. 
If desired, one can of course smooth out the function $g^{\pm}(w)$
at $w_4=0$ so that the higher Berry curvature becomes continuous
at $w_4=0$. However, smoothness of $g^{\pm}(w)$ will not be essential for our purposes. In particular, the KS invariant is an integral of the higher Berry curvature over $S^3$, so the discontinuity at $w_4 = 0$ will not cause problems.

Pictorially, the Hamiltonian in \eqref{H_1d}
for different values of $w_4 \in [-1,1]$ can be visualized as follows: 
\begin{equation}
\label{H_config}
\small
\begin{tikzpicture}

\node at (-60pt,0pt){$0<w_4<1$:};

\notsotiny
\draw (-20pt,0pt) circle (4.5pt);
\node at (-20pt,0pt){$+$};

\draw (0pt,0pt) circle (4.5pt);
\node at (0pt,0pt){$-$};
\draw [thick](4.5pt,0pt)--(15.5pt,0pt);
\draw (20pt,0pt) circle (4.5pt);
\node at (20pt,0pt){$+$};

\draw (40pt,0pt) circle (4.5pt);
\node at (40pt,0pt){$-$};
\draw [thick](44.5pt,0pt)--(55.5pt,0pt);
\draw (60pt,0pt) circle (4.5pt);
\node at (60pt,0pt){$+$};

\draw (80pt,0pt) circle (4.5pt);
\node at (80pt,0pt){$-$};
\draw [thick](84.5pt,0pt)--(95.5pt,0pt);
\draw (100pt,0pt) circle (4.5pt);
\node at (100pt,0pt){$+$};

\draw (120pt,0pt) circle (4.5pt);
\node at (120pt,0pt){$-$};
\draw [thick](124.5pt,0pt)--(135.5pt,0pt);
\draw (140pt,0pt) circle (4.5pt);
\node at (140pt,0pt){$+$};

\begin{scope}[yshift=-22pt]
\small
\node at (-50pt,0pt){$w_4=0$:};

\notsotiny
\draw (-20pt,0pt) circle (4.5pt);
\node at (-20pt,0pt){$+$};

\draw (0pt,0pt) circle (4.5pt);
\node at (0pt,0pt){$-$};

\draw (20pt,0pt) circle (4.5pt);
\node at (20pt,0pt){$+$};

\draw (40pt,0pt) circle (4.5pt);
\node at (40pt,0pt){$-$};

\draw (60pt,0pt) circle (4.5pt);
\node at (60pt,0pt){$+$};

\draw (80pt,0pt) circle (4.5pt);
\node at (80pt,0pt){$-$};

\draw (100pt,0pt) circle (4.5pt);
\node at (100pt,0pt){$+$};

\draw (120pt,0pt) circle (4.5pt);
\node at (120pt,0pt){$-$};

\draw (140pt,0pt) circle (4.5pt);
\node at (140pt,0pt){$+$};

\end{scope}

  \begin{scope}[yshift=-44pt]
  
  \small
\node at (-63pt,0pt){$-1<w_4<0$:};

\notsotiny

\draw (-20pt,0pt) circle (4.5pt);
\node at (-20pt,0pt){$+$};

\draw (0pt,0pt) circle (4.5pt);
\node at (0pt,0pt){$-$};
\draw [thick](4.5-20pt,0pt)--(15.5-20pt,0pt);
\draw (20pt,0pt) circle (4.5pt);
\node at (20pt,0pt){$+$};

\draw (40pt,0pt) circle (4.5pt);
\node at (40pt,0pt){$-$};
\draw [thick](24.5pt,0pt)--(35.5pt,0pt);
\draw (60pt,0pt) circle (4.5pt);
\node at (60pt,0pt){$+$};

\draw (80pt,0pt) circle (4.5pt);
\node at (80pt,0pt){$-$};
\draw [thick](64.5pt,0pt)--(75.5pt,0pt);
\draw (100pt,0pt) circle (4.5pt);
\node at (100pt,0pt){$+$};

\draw (120pt,0pt) circle (4.5pt);
\node at (120pt,0pt){$-$};
\draw [thick](104.5pt,0pt)--(115.5pt,0pt);
\draw (140pt,0pt) circle (4.5pt);
\node at (140pt,0pt){$+$};
   \end{scope}
   
     \begin{scope}[yshift=-66pt]

\draw [thick](-20pt,0pt)--(0pt,0pt);
\draw [thick](20pt,0pt)--(40pt,0pt);
\draw [thick](60pt,0pt)--(80pt,0pt);
\draw [thick](100pt,0pt)--(120pt,0pt);

\node at (140pt,0pt){$\bullet$};
\node at (120pt,0pt){$\bullet$};
\node at (100pt,0pt){$\bullet$};
\node at (80pt,0pt){$\bullet$};
\node at (60pt,0pt){$\bullet$};
\node at (40pt,0pt){$\bullet$};
\node at (20pt,0pt){$\bullet$};
\node at (0pt,0pt){$\bullet$};
\node at (-20pt,0pt){$\bullet$};

\small
\node at (-54pt,0pt){$w_4=-1$:};
\end{scope}

     \begin{scope}[yshift=22pt]

\draw [thick](0pt,0pt)--(20pt,0pt);
\draw [thick](40pt,0pt)--(60pt,0pt);
\draw [thick](80pt,0pt)--(100pt,0pt);
\draw [thick](120pt,0pt)--(140pt,0pt);

\node at (140pt,0pt){$\bullet$};
\node at (120pt,0pt){$\bullet$};
\node at (100pt,0pt){$\bullet$};
\node at (80pt,0pt){$\bullet$};
\node at (60pt,0pt){$\bullet$};
\node at (40pt,0pt){$\bullet$};
\node at (20pt,0pt){$\bullet$};
\node at (0pt,0pt){$\bullet$};
\node at (-20pt,0pt){$\bullet$};

\small
\node at (-52pt,0pt){$w_4=1$:};

\end{scope}

\end{tikzpicture}
\end{equation}
We use ``$+$'' (resp. ``$-$'') to represent a lattice site with non-zero single-spin term $H^1_{p \in 2\mathbb Z}$ 
(resp. $H^1_{p \in 2\mathbb Z+1}$), while ``$\bullet$'' represents a lattice site with vanishing single-spin term, as occurs at $w_4 = \pm 1$. Two-spin terms are represented by solid lines joining pairs of lattice sites. At  $w_4=0$, the system
is composed of decoupled 0$d$ quantum systems over the $S^2$ subspace of $S^3$ obtained by setting $w_4 = 0$. These systems are characterized by a Chern number of $2\pi$ (resp. $-2\pi$) on even (resp. odd) lattice sites.
Then for $0<|w_4|<1$, pairs of nearest-neighbor lattice sites are coupled as represented by solid line segments.
Finally, at the two poles $w_4=\pm 1$, the single-spin terms in \eqref{H_1body}
vanish, and there are only two-spin terms in \eqref{H_2body}.

We emphasize that the 0$d$ example of a spin-1/2 in a Zeeman magnetic field appears as an ingredient in the construction of this 1$d$ system. Indeed, this system is a special case of the suspension construction introduced below in Sec.~\ref{Sec:GeneralConstruct}, which starts with a $d$-dimensional system over $X$ and produces a $(d+1)$-dimensional system over $SX$. In the present case the parameter space of the 0$d$ system is $S^2$, the suspension of which is homeomorphic to the 3-sphere, \emph{i.e.} $S(S^2) \cong S^3$.

The Hamiltonian $H_{1d}(w)$ is exactly solvable because the lattice always decomposes into decoupled dimers. Therefore it is straightforward to verify that the bulk energy spectrum is gapped for all $w \in S^3$ (see Appendix~\ref{Appendix:Berry}).

The properties of the model at a spatial boundary indicate that the system is in a non-trivial phase over $S^3$. We truncate the 1$d$ lattice at the site $p = N$, keeping all lattice sites with $p \leq N$ but removing those with $p > N$. All Hamiltonian terms coupling to sites with $p > N$ are dropped, and all other terms are retained unmodified. This choice of boundary termination does not enlarge the parameter space at the boundary, 
\emph{i.e.} we have $X_{{\rm bulk}} = X_{{\rm bdy}} = S^3$
in the notation introduced in Sec.~\ref{sec:parametrized}.

Now taking $N$ even, the $p = N$ boundary site is decoupled from the bulk for $w_4 \leq 0$, and it is easily seen that the system is gapped for all $w \in S^3$ except at the pole $w_4 = -1$, where $H^1_N(w)$ for the boundary site has a gapless Weyl point. If on the other hand we take $N$ odd, then the boundary site is decoupled from the bulk for $w_4 \geq 0$, and there is a single gapless Weyl point at the opposite pole ($w_4 = 1$).

If we view the boundary as an effective $d=0$ system, the presence of a single Weyl point over $S^3$ is anomalous, and clearly cannot occur for a strictly zero-dimensional system over $S^3$. This can be seen by surrounding the Weyl point with a small 2-sphere $\Sigma \subset S^3$. In a strictly $d=0$ system, the Berry curvature $2$-form $\Omega^{(2)}$ is well-defined on $S^3$ away from the gapless point, and $\int_\Sigma \Omega^{(2)} = \pm 2\pi$ because $\Sigma$ surrounds a single Weyl point. On the other hand, if the $S^3$ has only a single gapless Weyl point, $\Sigma$ can be shrunk to a point (\emph{e.g.} the $w_4 = 1$ pole, opposite the Weyl point), so we must have  $\int_\Sigma \Omega^{(2)} = 0$, a contradiction. This anomalous nature of the boundary indicates that the bulk 1$d$ system is in a non-trivial phase over $S^3$. Moreover, this also indicates that the model generates (under stacking) a factor of $\Z$ in the classification of phases, where the $\Z$ is interpreted as the total integer-valued index of all isolated gapless points at the $d=0$ boundary.

\subsection{Higher Berry curvature and KS invariant}
\label{sec:hbc-1d}

Given the anomalous boundary physics of our model, it is natural to ask whether its non-trivial topological properties can also be understood from a bulk perspective. The desired bulk characterization is provided by the KS invariant obtained from the recently proposed higher Berry curvature.\cite{Kapustin_2020} After reviewing these concepts in Sec.~\ref{Sec:HighBerryIntro}, we compute the KS number in our model and show that it is $2\pi$ (Sec.~\ref{Sec:Berry1d}).

\subsubsection{Review of Kapustin-Spodyneiko results}
\label{Sec:HighBerryIntro}

Here we briefly review Kapustin and Spodyneiko's 
construction of the higher Berry curvature and the associated 
KS invariant.\cite{Kapustin_2020}
For $d$-dimensional lattice systems over $X$, those authors constructed a closed $(d+2)$-form on $X$ with quantized spherical cycles, the higher Berry curvature. The $d=0$ case is simply the usual Berry curvature. While the $(d+2)$-form itself depends on the details of its construction, and moreover is sensitive to deformations of the system within a phase, its cohomology class valued in $H^{d+2}(X,\R)$ is a topological invariant. In this section we focus on the $d=1$ case, discussing $d \geq 2$ in Sec.~\ref{sec:ndKSnumber}.

Consider a $d=0$ gapped system over $X$, where $X$ is a differentiable manifold. It is well-known that one can define the $2$-form Berry curvature as
\begin{equation}
\label{F2}
\Omega^{(2)}=\frac{i}{2}\oint\frac{dz}{2\pi i}\text{Tr}(G \,dH\, G^2 \,dH),
\end{equation}
where $H$ is the Hamiltonian and $G=\frac{1}{z-H}$ the Green's function. Without loss of generality, the ground-state energy is assumed to be constant over $X$, and $\oint$ is a counterclockwise contour integral taken around the ground-state energy. In \eqref{F2}, $d$ denotes the exterior derivative on $X$, \emph{i.e.} in local coordinates $\lambda^a$ we have $d H=\sum_a d\lambda^a\frac{\partial H}{\partial \lambda^a}$.

For an infinite 1$d$ system, the quantity in \eqref{F2} 
may be divergent. For example, if the 1$d$ lattice is composed of
an infinite number of decoupled 0$d$ quantum systems, 
then \eqref{F2} reduces to an ill-defined infinite sum of Berry curvatures over the decoupled systems. A well-defined finite quantity is 
\begin{equation}
\label{F2_pq}
F_{pq}^{(2)}=\frac{i}{2}\oint\frac{dz}{2\pi i}\text{Tr}(G \,dH_p\, G^2 \,dH_q) \text{.}
\end{equation}
Here $p,q \in \Z$, and we have written the Hamiltonian as a sum of local terms, $H = \sum_{p \in \Z} H_p$, where the support of $H_p$  lies within some fixed distance $\xi$ of the site $p$ (\emph{i.e.} $\xi$ is independent of $p$, so the Hamiltonian is finite-ranged).
For gapped Hamiltonians $H$, $F_{pq}^{(2)}$
decays exponentially as a function of $|p-q|$.\cite{Watanabe_2018}

Based on $F_{pq}^{(2)}$, we can construct the well defined $2$-form
\begin{equation}
\label{F2_p}
F_{q}^{(2)}=\sum_{p\in \mathbb Z}F_{pq}^{(2)} = \frac{i}{2}\oint\frac{dz}{2\pi i}\text{Tr}(G \,dH \, G^2 \,dH_q) \text{.}
\end{equation}
The form $F_q^{(2)}$ is not closed, and its exterior derivative can be written as
\begin{equation}
\label{dF2}
dF_{q}^{(2)}=\sum_{p\in \mathbb Z} F_{pq}^{(3)}.
\end{equation}
Here, the $3$-form $F_{pq}^{(3)}$ is anti-symmetric under exchange of site labels $p \leftrightarrow q$, and is given by
\begin{equation}
\label{F3}
\small
\begin{split}
F_{pq}^{(3)}=&\frac{i}{6} \oint\frac{dz}{2\pi i}\text{Tr}
\Big(
G^2dHGdH_p GdH_q-GdHG^2dH_pGdH_q
\Big)\\
&
-(p\leftrightarrow q).
\end{split}
\end{equation}
Similar to $F_{pq}^{(2)}$, the $3$-form $F_{pq}^{(3)}$ is a well defined quantity which
decays exponentially in $|p-q|$ for a gapped Hamiltonian $H$.\cite{Watanabe_2018}
Furthermore, one can check that 
\begin{equation}
\label{dF3}
dF_{pq}^{(3)}=\sum_{r\in\mathbb Z} F_{rpq}^{(4)},
\end{equation}
where $F^{(4)}_{pqr}$ is a $4$-form that is completely antisymmetric in $p$, $q$, $r$.

Using $F_{pq}^{(3)}$ in Eq.\eqref{F3}, one can construct the higher Berry curvature as a closed $3$-form $\Omega^{(3)}$, defined by
\begin{equation}
\label{Omega3}
\Omega^{(3)}(f)=\frac{1}{2}\sum_{p,q\in\mathbb Z}F_{pq}^{(3)} \cdot (f(q)-f(p)),
\end{equation}
where $f$: $\mathbb Z\to\mathbb R$ is a function which is $0$ for $p\ll 0$ and $1$ for $p\gg 0$. The sum converges because $F^{(3)}_{pq}$ decays exponentially in $|p-q|$, and it can be shown using Eq.~\ref{dF3} and the antisymmetry of $F^{(4)}_{pqr}$ that $d \Omega^{(3)}(f) = 0$. A simple choice of $f$ is a step function $f(p)=\Theta(p-a)$ with $a \in \Z + 1/2$, such that $f(p)=0$ for $p<a$ and 
$f(p)=1$ for $p>a$. This results in the simple expression
\begin{equation}
\label{Omega3f}
\Omega^{(3)}(f)=\sum_{p<a,q>a}F_{pq}^{(3)}.
\end{equation}

Kapustin and Spodyneiko argued that the cohomology class $[\Omega^{(3)}(f)/ 2 \pi] \in H^3(X, \R)$ is an invariant of a phase over $X$.\cite{Kapustin_2020} This cohomology class is the KS invariant, and it can be considered as an obstruction to continuously deforming the family to a constant
family of gapped systems over the parameter space.

Moreover, in invertible systems, Kapustin and Spodyneiko argued that the KS invariant is quantized;\cite{Kapustin_2020} in particular, if $X = S^3$, then $\int_{S^3} \Omega^{(3)}(f) \in 2\pi \Z$. In more detail, they showed that $\Omega^{(3)}(f)$ has quantized spherical cycles (in units of $2\pi$). This means that, for any smooth map $\phi : S^3 \to X$, $\int_{S^3} \phi^* \Omega^3(f) \in 2 \pi \Z$, where $\phi^*$ denotes the pullback. At least for $X = S^3$, this allows us to view the cohomology class $[\Omega^{(3)}(f) / 2\pi]$ as an element of $H^3(S^3, \Z) = \Z$, as we now explain. The inclusion $i : \Z \hookrightarrow \R$ induces a map on cohomology $i_* : H^3(X, \Z) \to H^3(X, \R)$. When $X = S^3$, $i_*$ is injective, and the property of quantized spherical cycles implies that $[\Omega^{(3)}(f) / 2\pi]$ lies in the image of $i_*$, so we can view the KS invariant as an element of $H^3(S^3, \Z) = \Z$. For general closed differentiable manifolds $X$ it is expected -- though we are not aware of a proof -- that $[\Omega^{(3)}(f) / 2 \pi]$ again lies in the image of $i_*$, and is quantized in this sense. In the general case $i_*$ is not injective, but its image is instead isomorphic to $H^3(X,\Z)$ modulo its torsion subgroup, so $[\Omega^{(3)}(f) / 2 \pi]$ can be viewed as an element of this quotient group.

In addition, if $X$ is an oriented, closed 3-manifold, we can define the KS number as $KS = \int_{X} \Omega^{(3)}(f) \in 2\pi\Z$, with the sign depending on the orientation. 
More generally, if $\Sigma \subset X$ is an oriented, closed 3-dimensional submanifold, we can define $KS_{\Sigma} = \int_{\Sigma} \Omega^{(3)}(f) \in 2\pi\Z$.

While $\Omega^{(3)}$ depends on the choice of the function $f$, its cohomology class and thus the KS invariant does not. Any two functions $f$ and $f'$ differ by a function $g$ with bounded support on $\Z$, and it is straightforward to check using Eq.~\ref{dF2} that $\Omega^{(3)}(g) = \Omega^{(3)}(f) - \Omega^{(3)}(f')$ is exact.

Another arbitrary choice involved in the construction of $\Omega^{(3)}(f)$ is the choice of local terms $H_p$ needed to write the Hamiltonian as $H = \sum_p H_p$. For a given Hamiltonian, any two such choices can be joined by a continuous path. Because the KS invariant depends continuously on $H_p$, and because the cohomology class $[\Omega^{(3)}(f) / 2 \pi]$ is quantized as described above, we expect that the KS invariant is again unaffected by this arbitrary choice.

\subsubsection{Computation of the KS invariant}
\label{Sec:Berry1d}

Now we compute the higher Berry curvature and KS number of our 1$d$ lattice model as introduced in Sec.~\ref{Sec:1+1d_lattice}.
We show that the KS number of our lattice model is nonzero and takes the value $KS = 2\pi$. 
Later, in 
Sec.~\ref{Sec:Semi-infinite} and \ref{Sec:Clutching}, we discuss how the KS number is 
related to pumping the Chern number of 0$d$ quantum systems
by imposing a spatial boundary.

To compute the $3$-form Berry curvature in \eqref{F3}, we need to choose the 
local Hamiltonian terms $H_p$ for each lattice site $p \in \Z$. 
As the Hamiltonian  \eqref{H_1d} is dimerized, it is convenient to choose $H_p$ with support only on $p$ and its two nearest-neighbors, as follows
\begin{equation}
\label{LocalH}
H_p(w)=H_p^{1}(w)+xH^{2,\pm}_{p,p+1}(w)+(1-x)H^{2,\mp}_{p-1,p}(w), 
\end{equation}
where we take the upper (resp. lower) sign for $p$ odd (resp. even), and where $0\le x\le1$ is a real $w$-independent parameter introduced to illustrate
the ambiguity in choosing local Hamiltonians discussed at the end of Sec.~\ref{Sec:HighBerryIntro}. As discussed there, this ambiguity should not affect the cohomology class of $\Omega^{(3)}(f)$. In fact we find that $\Omega^{(3)}(f)$ does not depend on $x$ at all; however, the higher Berry curvature is expected to depend non-trivially on other variables parametrizing different choices of local Hamiltonians.

We choose the function $f$ needed to construct $\Omega^{(3)}(f)$ to be a step function $f(p)=\Theta(p-a)$, taking either $a\in 2\mathbb Z-1/2$ or $a \in 2\mathbb Z+1/2$. We denote functions for two such choices of $a$ as $f(p)$ and $f'(p)$ respectively. These step functions are illustrated below in relation to the non-zero terms in the Hamiltonian for $w_4 > 0$:
\begin{equation}
\label{H2case}
\begin{tikzpicture}[baseline={(current bounding box.center)}]
\node at (-50pt,0pt){$w_4>0$:};

\node at (38pt,18pt){$f(p)$};
\node at (84pt,-16pt){$f'(p)$};

\notsotiny
\draw (-20pt,0pt) circle (4.5pt);
\node at (-20pt,0pt){$+$};

\draw (0pt,0pt) circle (4.5pt);
\node at (0pt,0pt){$-$};
\draw [thick](4.5pt,0pt)--(15.5pt,0pt);
\draw (20pt,0pt) circle (4.5pt);
\node at (20pt,0pt){$+$};

\draw (40pt,0pt) circle (4.5pt);
\node at (40pt,0pt){$-$};
\draw [thick](44.5pt,0pt)--(55.5pt,0pt);
\draw (60pt,0pt) circle (4.5pt);
\node at (60pt,0pt){$+$};

\draw (80pt,0pt) circle (4.5pt);
\node at (80pt,0pt){$-$};
\draw [thick](84.5pt,0pt)--(95.5pt,0pt);
\draw (100pt,0pt) circle (4.5pt);
\node at (100pt,0pt){$+$};

\draw (120pt,0pt) circle (4.5pt);
\node at (120pt,0pt){$-$};
\draw [thick](124.5pt,0pt)--(135.5pt,0pt);
\draw (140pt,0pt) circle (4.5pt);
\node at (140pt,0pt){$+$};

\draw [thick][black][dashed](-20pt,-13pt)--(50pt,-13pt);
\draw [thick][black][dashed](50pt,-13pt)--(50pt,15pt);
\draw [thick][black][dashed](50pt,15pt)--(140pt,15pt);

\draw [thick][gray][dashed](-20pt,-17pt)--(70pt,-17pt);
\draw [thick][gray][dashed](70pt,-17pt)--(70pt,11pt);
\draw [thick][gray][dashed](70pt,11pt)--(140pt,11pt);

\end{tikzpicture}
\end{equation}
We see that the step in $f(p)$ ``cuts through'' a dimer, while the step in $f'(p)$ lies between dimers. This will have the consequence that $\Omega^{(3)}(f') = 0$ for $w_4 > 0$, while in general $\Omega^{(3)}(f) \neq 0$ for $w_4 > 0$. 

We first compute the higher Berry curvature making the choice $f(p)$.
Given the local Hamiltonian in \eqref{LocalH}, the higher Berry curvature \eqref{Omega3f} can be simplified to
\begin{equation}
\label{F3j}
\Omega^{(3)}(f)=F_{a-1/2,a+1/2}^{(3)}. 
\end{equation}
This follows from the fact that $F_{pq}^{(3)}=0$ for $|p-q|\ge 2$
in our dimerized Hamiltonian with the local Hamiltonian chosen in \eqref{LocalH}.
It is also straightforward to check that
$\Omega^{(3)}(f)$ is only nonzero for $0\le w_4 \le 1$ because for 
$w_4<0$ there is no overlap in the support of local Hamiltonians $H_{a-1/2}$ and $H_{a+1/2}$.

To proceed, it is convenient to change to  
hyperspherical coordinates. We write
\begin{equation}
\label{HyperSphericalCoordinate}
\begin{split}
&w_1=\sin\alpha\cos\theta,\quad
w_2=\sin\alpha\sin\theta\cos\phi,\\
&w_3=\sin\alpha\sin\theta\sin\phi,
\quad w_4=\cos\alpha,
\end{split}
\end{equation} 
where $\alpha, \,\theta\in[0,\pi]$,
and $\phi\in[0,\, 2\pi]$. 
The higher Berry curvature can be explicitly calculated
based on Eqs.\eqref{F3}, \eqref{Omega3f}, and \eqref{F_pq3_sum} in Appendix~\ref{Appendix:Berry}. 
We obtain
$\Omega^{(3)}(f)=\Omega^{(3)}_{\alpha\theta\phi}(f) \, d\alpha\wedge d\theta \wedge d\phi$, with
\begin{equation}
\label{Omega3s}
\Omega^{(3)}_{\alpha\theta\phi}(f)=
\frac{1}{2}\left(
2+\cos\alpha
\right)\cdot \tan^2\frac{\alpha}{2}\cdot \sin\theta,
\end{equation}
where $0\le \alpha \le \frac{\pi}{2}$.
For $\frac{\pi}{2}< \alpha\le \pi$, which corresponds to $-1\le w_4<0$,
one always has 
$\Omega^{(3)}_{\alpha\theta\phi}(f)=0$. 
We note that the $3$-form $\Omega^{(3)}(f)$ is independent of the 
parameter $x$ in \eqref{LocalH}.

To discuss the KS number (denoted $KS$), we fix the orientation on $S^3$ specified by the (locally defined) $3$-form $d\alpha \wedge d\theta \wedge d\phi$. In contrast to the orientation chosen on $S^2$ in Sec.~\ref{sec:review}, this corresponds to the ``inward-normal'' orientation on $S^3$, which is given by viewing $S^3$ as a subspace of $\R^4$ in the usual way, and defining a basis $(b_1,b_2,b_3)$ of the tangent space 
$T_x S^3$ at a unit vector $x\in \R^4$ to be positively oriented if  $(-x,b_1,b_2,b_3)$ has the same orientation as the standard basis of $\R^4$. The reason we choose the inward-normal orientation is for the aesthetic purpose of making the KS number positive; if we instead chose the outward-normal orientation, the sign of the KS number would be reversed.

We then have (using $\int_0^{\pi/2} d\alpha \, (2+\cos \alpha)\tan ^2\frac\alpha 2 = 1$):
\begin{equation}
\label{S3_2pi}
KS = \int_{S^3}\Omega^{(3)}(f)=2\pi .
\end{equation}
 This takes the quantized value $2 \pi$, indicating our lattice model 
is in a non-trivial phase over $S^3$. 

To illustrate the dependence of the higher Berry curvature on the choice of function $f$, we now consider $f'(p) = \Theta(p-a)$ for $a\in 2\mathbb Z+1/2$.   
For $0\le w_4\le 1$ or equivalently $0 \le \alpha \le \frac{\pi}{2}$, we find
$\Omega_{\alpha\theta\phi}^{(3)}(f')=0$.
For $-1\le w_4\le 0$ or equivalently $\frac{\pi}{2}\le \alpha \le \pi$, we obtain
\begin{equation}
\label{Omega3_f2}
\Omega^{(3)}_{\alpha\theta\phi}(f')=\frac{1}{2}\left(
2-\cos\alpha
\right)\cot^2\frac{\alpha}{2}\cdot \sin\theta.
\end{equation}
The higher Berry curvature is now zero in the upper hemisphere and non-zero in the lower hemisphere of $S^3$. 
It can be checked that $\int_{S^3}\Omega^{(3)}(f')=2\pi$, illustrating that the KS number does not depend on the choice of $f$, as explained in general in Sec.~\ref{Sec:HighBerryIntro}.

\subsection{Bulk-boundary correspondence}
\label{Sec:Semi-infinite}

We now return to the properties of a spatial boundary, and make a connection between the bulk KS number and the flow of Chern number to/from the boundary. We formulate a bulk-boundary correspondence, first for our exactly solvable model, and then for a generic gapped 1$d$ system over $S^3$. We restrict attention to the case where the boundary and bulk parameter spaces are identical, \emph{i.e.} $X_{{\rm bdy}} = X_{{\rm bulk}} = S^3$.

\subsubsection{Exactly solvable model}
\label{sec:solvable-bbc}

To formulate a bulk-boundary correspondence for our exactly solvable model, we will need to compare two different systems. One of these will be the infinite system we have been focused on so far, which of course does not have a spatial boundary. The other will be a semi-infinite system with boundary. We consider the same boundary termination at the site $p = N$ introduced in Sec.~\ref{Sec:1+1d_lattice}, and recall that the boundary is gapless at $w_4 = -1$ for $N$ even, and at $w_4 = 1$ for $N$ odd. For concreteness, we now take $N$ even and choose $f(p) = \Theta(p-a)$ with $a \in 2\Z - 1/2$, where $a < N$.

In the semi-infinite system, the higher Berry curvature is only defined on $X_\Delta$, the subspace of $X = S^3$ where the spectrum is gapped. In this case, $X_\Delta = S^3 \backslash \{(0,0,0,-1)\}$, \emph{i.e.} $S^3$ with the $w_4 = -1$ pole removed. Restricting to $X_\Delta$, due to the dimerized nature of our model, it is straightforward to see that the higher Berry curvature is exactly equal to its value in the infinite system. That is,
\begin{equation}
\Omega_{\infty/2}^{(3)}(f) = \Omega_{\infty}^{(3)}(f) \text{,} \label{eqn:omega3-bb-solvable}
\end{equation}
where $\Omega_{\infty/2}^{(3)}(f)$ and $\Omega_{\infty}^{(3)}(f)$ are the higher Berry curvatures of the semi-infinite (``$\infty/2$'') and infinite (``$\infty$'') systems, respectively.
This expression does not hold as an exact equality in a generic system, a point that we return to below.

Now we observe that, based on \eqref{Omega3f}, the higher Berry curvature of the semi-infinite system on $X_\Delta$ can be written
\begin{equation}
\label{Omega3_domega2}
\Omega_{\infty/2}^{(3)}(f)=d\omega^{(2)},
\end{equation}
where 
\begin{equation}
\label{2form_omega}
\omega^{(2)}=\sum_{p>a} F_{p}^{(2)}.
\end{equation}
The sum in \eqref{2form_omega} is finite, and thus well-defined, because the system is semi-infinite.

The $2$-form $\omega^{(2)}$ can be interpreted as a boundary Berry curvature, and its integral over a closed 2-manifold as a boundary Chern number, where the ``boundary'' is considered to be all lattice sites to the right of $a$. This interpretation makes sense because, for those parameter values where the boundary is decoupled from the bulk, $\omega^{(2)}$ reduces to the Berry curvature of the boundary, which is a $0d$ system. For such parameters $\omega^{(2)}$ is clearly closed, and hence $\Omega_{\infty/2}^{(3)}(f) = 0$. Moreover, if $\Sigma \subset X_\Delta$ is a closed oriented 2-manifold such that the bulk and boundary are decoupled for all $x \in \Sigma$, then $\int_{\Sigma} \omega^{(2)}$ is quantized. These statements all hold in a generic system. Of course, the boundary Chern number is not quantized in general, because for some parameter values the boundary and bulk are coupled. 

In the exactly solvable model, the bulk and boundary are decoupled for $w_4 \leq 0$, including the gapless point. 
For $w_4 \leq 0$ we can view the (spatial) boundary as a decoupled 0$d$ system with a gapless Weyl point at $w_4 = -1$, so for $-1 < y \leq 0$ we expect
\begin{equation}
\int_{S^2(y)} \omega^{(2)} = 2\pi \text{,} \label{eqn:int_omega2}
\end{equation}
where $S^2(y) \subset S^3$ is the subspace with $w_4 = y$, and is clearly homeomorphic to the 2-sphere.

In order to make sense of \eqref{eqn:int_omega2}, we need to specify an orientation on $S^2(y)$. We do this by introducing the $3$-ball $D^3(y) \subset S^3$ as the subspace of $S^3$ with $w_4 \geq y$, so then $S^2(y) = \partial D^3(y)$. The subspace $D^3(y)$ inherits an orientation from the given orientation on $S^3$. Its boundary $S^2(y)$ can be oriented by the following general procedure, which we will employ throughout the paper whenever it is necessary to orient the boundary of a manifold with boundary given an orientation on its interior.

We rely on Brown's collaring theorem \cite{brown1962}, which says that any manifold with boundary $X$ possesses a \emph{collar} that is a diffeomorphism $\varphi: \partial X \times [0,1) \to X$ onto an open neighborhood of the boundary. Assuming that $X$ is embedded in an oriented manifold $M$ of equal dimension $m$, the boundary $\partial X$ is oriented in such a way that a basis $(b_1,\ldots,b_{m-1})$ 
of the tangent space $T_x X$ at a boundary point $x\in \partial X$ is positively oriented if and only if $(n(x),b_1,\ldots,b_{m-1})$ is a positively oriented basis of $T_x M$, where $n(x) = - D_{(x,0)} \varphi (0,1)$ is the outward pointing vector at $x$ defined by the collar. Note that this definition of an orientation does not depend on the particular choice of a collar. In the following we will refer to this procedure of assigning an orientation as the \emph{collar method}.
Applying it to $S^2(y) = \partial D^3(y)$ gives the outward-normal orientation described in Sec.~\ref{sec:review}, \emph{i.e.} the same orientation as given by the locally defined $2$-form $d\theta \wedge d\phi$.

Equation~\eqref{eqn:int_omega2} can be verified from an explicit computation of $\omega^{(2)}$ on $X_\Delta$ (see Appendix~\ref{Appendix:Berry}). We find $\omega^{(2)}=\omega_{\theta\phi}^{(2)}d\theta\wedge d\phi$, where 
\begin{equation}
\label{omega_2form}
\omega^{(2)}_{\theta\phi}=\frac{\sin^3\alpha \cdot \sin\theta}{2(1+\cos\alpha)^2},
\end{equation}
for $0\le \alpha \le \frac{\pi}{2}$ (corresponding to $w_4 \geq 0$).
For $\frac{\pi}{2}\leq \alpha <\pi$ (corresponding to $0 \geq w_4 > -1$) we have
\begin{equation}
\label{omega2_Omega2}
\omega^{(2)}_{\theta\phi}=\frac{\sin\theta}{2},
\end{equation}
which is nothing but the Berry curvature of the decoupled $p=N$ qubit at the boundary. This result can be anticipated by observing that each decoupled dimer gives a vanishing contribution to $\omega^{(2)}$, which can be understood by noting that the Hamiltonian for each dimer is invariant under an anti-unitary symmetry, namely the combination of time-reversal and the reflection symmetry exchanging the two sites of the dimer. Therefore, for $w_4 \leq 0$, $\omega^{(2)}=F_{p=N}^{(2)}$.

From \eqref{omega2_Omega2} we can directly check that \eqref{eqn:int_omega2} indeed holds. Moreover, by Stokes theorem, we have
\begin{equation}
\int_{S^2(y)} \omega^{(2)} = \int_{D^3(y)} \Omega_{\infty/2}^{(3)} = \int_{D^3(y)} \Omega_{\infty}^{(3)} \text{.} \label{eqn:bbc-solvable}
\end{equation}
This relationship between the (non-quantized) boundary Chern number over $S^2(y)$ and the integral of the infinite system's higher Berry curvature over $D^3(y)$ can be viewed as a bulk-boundary correspondence, in the special situation of our exactly solvable model. In particular we have
\begin{equation}
\lim_{y \to -1} \int_{S^2(y)} \omega^{(2)} = \int_{S^3} \Omega^{(3)}_{\infty} = 2\pi \text{,}
\end{equation}
where the limit already attains its value for $y \leq 0$. In words, this equation says that the boundary Chern number over a small 2-sphere surrounding the gapless point is equal to the bulk quantized KS invariant.

From the above relations we see that as $y$ decreases continuously from $1$ to $-1$, the boundary Chern number increases from zero to $2\pi$. We can think of this change as a flow of Chern number from the semi-infinite bulk to the zero-dimensional boundary. The boundary Chern number only changes, and is non-quantized, in the interval $y \in [0,1]$, where the bulk and boundary are coupled and a non-zero flow of Chern number is possible. Indeed, combining \eqref{Omega3_domega2} and \eqref{eqn:omega3-bb-solvable}, we have
\begin{equation}
\Omega^{(3)}_{\infty} = d \omega^{(2)} \text{,}
\end{equation}
which says that the higher Berry curvature measures the flow of ordinary Berry curvature to the boundary.

\subsubsection{Generic system over $S^3$}
\label{sec:generic-bbc}

Now we generalize the above discussion to a generic gapped 1$d$ bosonic system over $S^3$. We start with an infinite 1$d$ system, and again consider a semi-infinite system obtained by throwing away lattice sites for $p > N$. For this generic system, we do not specify any particular choice of Hamiltonian terms near the boundary, but we do require $X_{{\rm bdy}} = X_{{\rm bulk}} = S^3$. To discuss the higher Berry curvature in both infinite and semi-infinite systems, we take $f(p) = \Theta(p-a)$, with $a \in \Z - 1/2$ and $a < N$. To simplify the discussion we first assume the infinite system has $KS = 2\pi$ and that the semi-infinite system is gapless only at the pole $w_4 = -1$, later discussing more general situations.

The key difference from the exactly solvable model is that we do not expect \eqref{eqn:omega3-bb-solvable} to hold in a generic system. It is certainly the case that $\Omega_{\infty/2}^{(3)}(f)$ can be influenced by boundary effects when $a$ is close to the boundary. As an extreme example of this, suppose that the infinite system (and $a$) is chosen so that for some values of $w \in S^3$, there is non-zero coupling between  lattice sites lying on different sides of $a$. Then generically $\Omega^{(3)}_{\infty} \neq 0$.  However we can always introduce a boundary termination where the lattice sites with $p > a$ are decoupled from the bulk for all $w \in S^3$, guaranteeing that $\Omega_{\infty/2}^{(3)}(f) = 0$.

However, we can relate $\Omega^{(3)}_{\infty}$ and $\Omega^{(3)}_{\infty/2}(f)$ in a generic system based on the observation that $F^{(3)}_{pq}$ is a local quantity. More precisely, in Appendix~\ref{app:locality} we express $F^{(3)}_{p q}$ as a three-point imaginary-time correlation function of local operators, with the dominant contributions coming from regions of space near $p$ and $q$. In the limit that $p$ and $q$ are both taken far from the boundary, we thus expect $F^{(3)}_{p q}$ computed for the infinite and semi-infinite systems to approach the same value. Then, since \eqref{Omega3f} tells us that the higher Berry curvature is dominated by $F^{(3)}_{p q}$ for $p$ and $q$ near $a$, we conjecture that the infinite and semi-infinite systems have the same higher Berry curvature if $a$ is taken sufficiently far from the boundary. That is,
\begin{equation}
\lim_{a \to -\infty} \Big[ \Omega^{(3)}_{\infty/2}(f) - \Omega^{(3)}_{\infty}(f) \Big] = 0 \text{.} \label{eqn:generic-bb-omega3}
\end{equation}
We expect that the limit -- and other similar limits in $a$ below --  approaches its value exponentially fast, with the scale being set by the system's correlation length. Therefore, practically speaking, we can expect that the bulk-boundary correspondence below will hold to a good approximation if $a$ is taken to lie a few correlation lengths away from the boundary. The expression \eqref{eqn:generic-bb-omega3} should be viewed as replacing \eqref{eqn:omega3-bb-solvable} when considering a generic system.

We introduce the $2$-form $\omega^{(2)}$ as before, and again $d \omega^{(2)} = \Omega_{\infty/2}^{(3)}(f)$. But now we need to take a limit to establish a relation with the bulk higher Berry curvature, namely
\begin{equation} 
\lim_{a \to -\infty} \Omega^{(3)}_{\infty}(f) = 
\lim_{a \to -\infty} d \omega^{(2)} \text{.}
\label{eqn:generic-diff-bbc}
\end{equation}
The integral form of the bulk-boundary correspondence, which generalizes \eqref{eqn:bbc-solvable}, thus takes the form
\begin{equation}
\lim_{a \to -\infty} \int_{S^2(y)} \omega^{(2)} 
= \lim_{a \to -\infty} \int_{D^3(y)} \Omega^{(3)}_{\infty} 
\end{equation}
Taking the limit $y \to -1$ we obtain
\begin{equation}
\lim_{y \to -1} \lim_{a \to -\infty} \int_{S^2(y)} \omega^{(2)} 
= \lim_{a \to -\infty} \int_{S^3} \Omega^{(3)}_{\infty} = 2\pi \text{.} \label{eqn:generic-int-bbc}
\end{equation}

More generally, a boundary termination may be gapless at points other than $w_4 = -1$, and the KS number may take some value other than $2 \pi$, denoted by $KS \in 2\pi\Z$. Supposing we have a boundary termination that is gapless at a finite set of points $C \subset S^3$, we have $X_\Delta = S^3 \backslash C$, and $\Omega_{\infty/2}^{(3)}(f)$ and $\omega^{(2)}$ can be defined on $X_\Delta$. The differential form of the bulk-boundary correspondence \eqref{eqn:generic-diff-bbc} holds without modification. To relate $\omega^{(2)}$ to the bulk KS-invariant, we let $A_{\epsilon} \subset S^3$ be the subspace constructed by removing a small open ball of radius $\epsilon$ surrounding each point in $C$. Then the analog of \eqref{eqn:generic-int-bbc} is
\begin{equation}
\lim_{\epsilon \to 0} \lim_{a \to -\infty} \int_{\partial A_{\epsilon}} \omega^{(2)} = \lim_{a \to -\infty} \int_{S^3} \Omega^{(3)}_{\infty} = KS \text{.}
\end{equation}
Note that the integral over $S^3$ is independent of $a$, so the second limit in $a$ can be dropped. It is straightforward to generalize these statements to the case where $X$ is a general closed oriented 3-dimensional differential manifold.

An important application of the generic form of the bulk boundary correspondence is to show that the boundary must be gapless for some points in $S^3$ when the bulk KS number is nonzero. We start with a gapped 1$d$ system over $S^3$ with KS number $KS \neq 0$ and introduce a boundary at $p = N$ as above. We can then re-introduce lattice sites with $p > N$, and on these sites we place a trivial gapped system over $S^3$. Each individual lattice site with $p > N$ is thus decoupled from the rest of the system. Clearly this does not change the spectrum of the system with boundary, which we are now viewing as an interface between the original bulk system and a trivial system.

Now we consider this interface as a system over $S^3$, with higher Berry curvature $\Omega^{(3)}_{{\rm int}}(f)$, and examine two different choices of the function $f$, namely $f(p) = \Theta(p - a)$ and $f'(p) = \Theta(p - a')$, where $a' > N$ and $a \ll N$. If the interface is a gapped system over $S^3$, then the KS number does not depend on the choice of $f$, and in particular
\begin{equation}
\int_{S^3} \Omega_{{\rm int}}^{(3)}(f) = \int_{S^3} \Omega_{{\rm int}}^{(3)}(f') \text{.}
\end{equation}
But clearly $\Omega_{{\rm int}}^{(3)}(f') = 0$, while $\int_{S^3} \Omega_{{\rm int}}^{(3)}(f) \approx KS \neq 0$, 
where the equality becomes exact in the limit $a \to -\infty$. This is a contradiction, and so evidently the interface cannot be gapped everywhere over $S^3$.

It is straightforward to generalize this argument to more general interfaces between two different gapped 1$d$ systems over some closed oriented 3-manifold $X$. One concludes that if the two KS numbers are different, the interface must be gapless for some $x \in  X$.

\subsection{Clutching construction}
\label{Sec:Clutching}

Here we present a ``clutching construction'' that further illuminates the relationship between the quantized KS number of a 1$d$ system over $S^3$ and the Chern number of a 0$d$ system. Consider a generic gapped 1$d$ system $H$ over $S^3$. If the KS number is non-zero, then the cohomology class of $\Omega^{(3)}$ is non-trivial and we cannot write $\Omega^{(3)} = d\omega^{(2)}$ for a globally defined $2$-form $\omega^{(2)}$ on $S^3$. However, we can cover $S^3$ with charts, on each of which there does exist a well-defined $2$-form $\omega^{(2)}$. The cohomology class of $\Omega^{(3)}$, and the KS number, is determined by these $2$-forms on the overlaps between the charts.

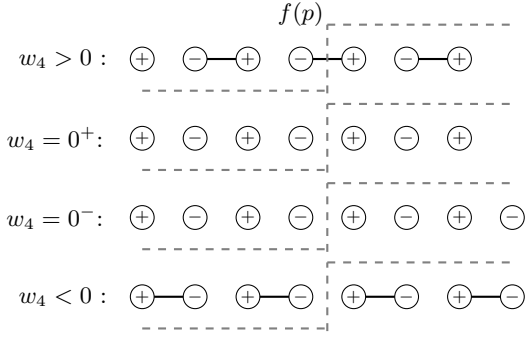
\begin{figure}
\begin{tikzpicture}[baseline={(current bounding box.center)}]
\small
\node at (40pt,17pt){$f(p)$};

\node at (-50pt,0pt){$w_4>0$ :};

\notsotiny
\draw (-20pt,0pt) circle (4.5pt);
\node at (-20pt,0pt){$+$};

\draw (0pt,0pt) circle (4.5pt);
\node at (0pt,0pt){$-$};
\draw [thick](4.5pt,0pt)--(15.5pt,0pt);
\draw (20pt,0pt) circle (4.5pt);
\node at (20pt,0pt){$+$};

\draw (40pt,0pt) circle (4.5pt);
\node at (40pt,0pt){$-$};
\draw [thick](44.5pt,0pt)--(55.5pt,0pt);
\draw (60pt,0pt) circle (4.5pt);
\node at (60pt,0pt){$+$};

\draw (80pt,0pt) circle (4.5pt);
\node at (80pt,0pt){$-$};
\draw [thick](84.5pt,0pt)--(95.5pt,0pt);
\draw (100pt,0pt) circle (4.5pt);
\node at (100pt,0pt){$+$};

\draw [thick][gray][dashed](-20pt,-12pt)--(50pt,-12pt);
\draw [thick][gray][dashed](50pt,-13pt)--(50pt,13pt);
\draw [thick][gray][dashed](50pt,13pt)--(120pt,13pt);


 \begin{scope}[yshift=-30pt]
 \small
 \node at (-53pt,0pt){$w_4=0^+$:};

\notsotiny
\draw [thick][gray][dashed](-20pt,-12pt)--(50pt,-12pt);
\draw [thick][gray][dashed](50pt,-13pt)--(50pt,13pt);
\draw [thick][gray][dashed](50pt,13pt)--(120pt,13pt);

\draw (-20pt,0pt) circle (4.5pt);
\node at (-20pt,0pt){$+$};

\draw (0pt,0pt) circle (4.5pt);
\node at (0pt,0pt){$-$};

\draw (20pt,0pt) circle (4.5pt);
\node at (20pt,0pt){$+$};

\draw (40pt,0pt) circle (4.5pt);
\node at (40pt,0pt){$-$};

\draw (60pt,0pt) circle (4.5pt);
\node at (60pt,0pt){$+$};

\draw (80pt,0pt) circle (4.5pt);
\node at (80pt,0pt){$-$};

\draw (100pt,0pt) circle (4.5pt);
\node at (100pt,0pt){$+$};
 \end{scope}

 \begin{scope}[yshift=-60pt]
 \small
 \node at (-53pt,0pt){$w_4=0^-$:};

\notsotiny
\draw [thick][gray][dashed](-20pt,-12pt)--(50pt,-12pt);
\draw [thick][gray][dashed](50pt,-13pt)--(50pt,13pt);
\draw [thick][gray][dashed](50pt,13pt)--(120pt,13pt);

\draw (-20pt,0pt) circle (4.5pt);
\node at (-20pt,0pt){$+$};

\draw (0pt,0pt) circle (4.5pt);
\node at (0pt,0pt){$-$};

\draw (20pt,0pt) circle (4.5pt);
\node at (20pt,0pt){$+$};

\draw (40pt,0pt) circle (4.5pt);
\node at (40pt,0pt){$-$};

\draw (60pt,0pt) circle (4.5pt);
\node at (60pt,0pt){$+$};

\draw (80pt,0pt) circle (4.5pt);
\node at (80pt,0pt){$-$};

\draw (100pt,0pt) circle (4.5pt);
\node at (100pt,0pt){$+$};

\draw (120pt,0pt) circle (4.5pt);
\node at (120pt,0pt){$-$};

 \end{scope}

  \begin{scope}[yshift=-90pt]
  \small
  \node at (-50pt,0pt){$w_4<0$ :};

\notsotiny
\draw [thick][gray][dashed](-20pt,-12pt)--(50pt,-12pt);
\draw [thick][gray][dashed](50pt,-13pt)--(50pt,13pt);
\draw [thick][gray][dashed](50pt,13pt)--(120pt,13pt);

\draw (-20pt,0pt) circle (4.5pt);
\node at (-20pt,0pt){$+$};

\draw (0pt,0pt) circle (4.5pt);
\node at (0pt,0pt){$-$};
\draw [thick](4.5-20pt,0pt)--(15.5-20pt,0pt);
\draw (20pt,0pt) circle (4.5pt);
\node at (20pt,0pt){$+$};

\draw (40pt,0pt) circle (4.5pt);
\node at (40pt,0pt){$-$};
\draw [thick](24.5pt,0pt)--(35.5pt,0pt);
\draw (60pt,0pt) circle (4.5pt);
\node at (60pt,0pt){$+$};

\draw (80pt,0pt) circle (4.5pt);
\node at (80pt,0pt){$-$};
\draw [thick](64.5pt,0pt)--(75.5pt,0pt);
\draw (100pt,0pt) circle (4.5pt);
\node at (100pt,0pt){$+$};

\draw (120pt,0pt) circle (4.5pt);
\node at (120pt,0pt){$-$};
\draw [thick](104.5pt,0pt)--(115.5pt,0pt);

   \end{scope}
\end{tikzpicture}
\caption{
Depiction of the two boundary terminations used to construct the $2$-forms $\omega_{N,S}^{(2)}$ in the 
clutching construction of the $3$-form Berry curvature $\Omega^{(3)}$ in the exactly solvable 1$d$ model. An extra lattice site is added at the boundary for $w_4 < 0$ (southern hemisphere), as compared to $w_4 > 0$ (northern hemisphere). The form
$\omega_{N,S}^{(2)}$ is globally well defined on the hemisphere $S^3_{N,S}$.
Along the equator $w_4=0$, the difference between the boundary Chern numbers $\int_{S^2} \omega_N^{(2)}$ and $\int_{S^2} \omega_S^{(2)}$ is quantized.
}
\label{FigClutchConstruction}
\end{figure}

We choose $f(p) = \Theta(p-a)$, and we write the KS number as \begin{equation}
\label{ClutchOmega3_a}
\int_{S^3}\Omega^{(3)}=\int_{S_N^3}\Omega^{(3)}+\int_{S_S^3}\Omega^{(3)},
\end{equation}
where $S^3_N$ ($S^3_S$) are northern (southern) hemispheres of $S^3$, \emph{i.e.} the subspaces with $w_4 \geq 0$ ($w_4 \leq 0$). Each hemisphere is contractible, and because there is a trivial classification of 1$d$ gapped bosonic phases without symmetry over a point,\cite{chen2011classification} the restriction of $H$ to either hemisphere is a system in the trivial phase. 
Therefore we can introduce two different boundary terminations, one gapped over $S^3_N$, and the other gapped over $S^3_S$. We denote the corresponding higher Berry curvatures by $\Omega^{(3)}_{\mathrm{bdy},N}(f)$ and $\Omega^{(3)}_{\mathrm{bdy},S}(f)$, respectively. By the discussion of Sec.~\ref{sec:generic-bbc} we have
\begin{equation}
\lim_{a \to -\infty} \Omega^{(3)}(f) = \lim_{a \to -\infty} \Omega^{(3)}_{\mathrm{bdy},N}(f)
= \lim_{a \to -\infty} \Omega^{(3)}_{\mathrm{bdy},S}(f) \text{.}
\end{equation}

We introduce $2$-forms $\omega^{(2)}_{N,S}$ defined on the respective hemispheres, by \eqref{2form_omega}. These $2$-forms can be interpreted as boundary Berry curvatures and satisfy $\Omega^{(3)}_{\mathrm{bdy},N}(f) = d \omega^{(2)}_N$ and $\Omega^{(3)}_{\mathrm{bdy},S}(f) = d \omega^{(2)}_S$. Using Stokes' theorem and taking the limit $a \to -\infty$ we obtain
\begin{equation}
\int_{S^3}\Omega^{(3)} = \lim_{a \to -\infty} \big[ \int_{S^2}\omega_N^{(2)}-\int_{S^2}\omega_S^{(2)} \big] \text{,}
\label{eqn:clutching}
\end{equation}
where $S^2$ is the $w_4 = 0$ ``equator'' of $S^3$, \emph{i.e.} $S^2 = S^3_N \cap S^3_S$. 
The minus sign in \eqref{eqn:clutching} arises because, using the collar method to assign orientations to boundary manifolds, the orientation on $S^2$ induced by $S^2 = \partial S_S^{(3)}$ is opposite to that induced by $S^2 = \partial S_N^{(3)}$.

The clutching construction thus expresses the KS number as a difference of boundary Chern numbers for the two different gapped boundary terminations. Even though the individual boundary Chern numbers are not quantized, evidently their difference is quantized. This is made plausible by writing
\begin{equation}
\omega^{(2)}_N - \omega^{(2)}_S = \sum_{q > a} \big( F^{(2)}_{q,N} - F^{(2)}_{q,S} \big) \text{,}
\end{equation}
and noting that the quantity in parentheses vanishes exponentially as a function of the distance between $q$ and the boundary. Therefore even in the limit $a \to -\infty$ one has a convergent sum dominated by contributions near the boundary, which is suggestive of the Berry curvature of a 0$d$ system.

We now further illustrate the clutching construction by describing it for our solvable lattice model. We consider boundary terminations for the north and south hemispheres as described in Sec.~\ref{Sec:1+1d_lattice}, choosing $N \in 2 \Z$ in the northern hemisphere, and $N \in 2\Z + 1$ in the southern hemisphere, resulting in a gapped boundary in each hemisphere. Evaluating $\omega^{(2)}_N - \omega^{(2)}_S$ is straightforward, because on the $w_4 = 0$ equator all lattice sites are decoupled from one another, with Hamiltonian $H^1_p(w)$. We can thus talk about the Berry curvature of the lattice site $p$, which is simply $\omega^{(2)}_p = (-1)^p \frac{\sin\theta}{2}\,d\theta\wedge d\phi$.

Now, choosing $a \in 2\Z - 1/2$, we see that for the southern hemisphere boundary termination, there are an equal number of sites $p > a$ with $p$ odd and $p$ even, so $\omega^{(2)}_S = 0$ (see Fig.\ref{FigClutchConstruction}). On the other hand, in the northern hemisphere there is one more even than odd site, and $\omega^{(2)}_N = \frac{\sin\theta}{2}\,d\theta\wedge d\phi$. Therefore we find
$\int_{S^2} (\omega^{(2)}_N - \omega^{(2)}_S ) = 2\pi$. Choosing instead $a \in 2\Z + 1/2$, we have $\omega^{(2)}_N = 0$ and $\omega^{(2)}_S = -\frac{\sin\theta}{2}\,d\theta\wedge d\phi$, giving the same result for the integral of $\omega^{(2)}_N - \omega^{(2)}_S$ over $S^2$, which has no dependence on $a$ in the solvable model.

\subsection{Inverse system over $S^3$}
\label{Sec:Inverse}

Here we introduce an inverse $\overbar{H}_{1d}$ of the system $H_{1d}$ over $S^3$ described in Sec.~\ref{Sec:1+1d_lattice}. These two systems are inverses in the sense that upon stacking $H_{1d}$ with $\overbar{H}_{1d}$, the resulting system can be continuously deformed to a trivial system. Moreover, the KS number of $\overbar{H}_{1d}$ is opposite that of $H_{1d}$. The inverse model will also be used in the construction of 
a nontrivial 2$d$ system over $S^4$ in Sec.~\ref{sec:2dmodel}.

We again consider a 1$d$ lattice with a single qubit at each lattice site.  The Hamiltonian is given by
\begin{equation}
\label{H_1d_inverse}
\small
\overbar{H}_{1d}(w)=-\sum_{i\in\mathbb Z} H_i^1(w)+\sum_{i\in2\mathbb Z+1} H^{2,+}_{i,i+1}(w)
+\sum_{i\in2\mathbb Z}H^{2,-}_{i,i+1}(w).
\end{equation}
Comparing with Eq.\eqref{H_1d}, the only difference between $H_{1d}$ and $\overbar{H}_{1d}$ is the sign of the single-spin term, which has been reversed here, \emph{i.e.} $H_i^1(w)\to(-1)H_i^1(w)$.

We first compute the KS number of $\overbar{H}_{1d}$. We choose $f(p) = \Theta(p-a)$ with $a \in 2 \Z - 1/2$.  Following the procedure of Sec.~\ref{Sec:Berry1d},
we find $\Omega^{(3)}(f) = \Omega^{(3)}_{\alpha\theta\phi}(f) d\alpha \wedge d\theta \wedge d\phi$, with $\Omega^{(3)}_{\alpha\theta\phi}(f) = 0$
for $-1\le w_4< 0$, while for  $0\le w_4\le 1$ we have
\begin{equation}
\label{Omega3s_inverse}
\Omega^{(3)}_{\alpha\theta\phi}(f)=
-\frac{1}{2}\left(
2+\cos\alpha
\right)\cdot \tan^2\frac{\alpha}{2}\cdot \sin\theta,
\end{equation}
which differs from \eqref{Omega3s} by a minus sign.
We immediately obtain the KS number
\begin{equation}
\label{S3_2pi_inverse}
\int_{S^3}\Omega^{(3)}(f)=-2\pi \text{,}
\end{equation}
which is opposite that of $H_{1d}$ as claimed.

Next we consider the system obtained by stacking $H_{1d}$ and $\overbar{H}_{1d}$,
with Hamiltonian
\begin{equation}
H_{\text{stack}}(w)=H_{1d}(w) \otimes \mathbbm{1} + \mathbbm{1} \otimes \overbar{H}_{1d}(w) \text{.}
\end{equation}
We deform the Hamiltonian of the stacked system continuously as a function of $t\in[0,1]$ as follows
\begin{eqnarray}
\label{Hwt}
H(w,t)&=&\cos\Big(\frac{\pi t}{2}\Big)\cdot\Big(H_{1d}(w) \otimes \mathbbm{1} + \mathbbm{1} \otimes \overbar{H}_{1d}(w)\Big) \nonumber \\ &+& \sin\Big(\frac{\pi t}{2}\Big)\cdot H_{\text{int}} \text{,}
\end{eqnarray}
for
\begin{equation}
H_{\text{int}}=\sum_{i \in \Z} \sum_{\mu = 1,2,3} \sigma^{\mu}_{i,1} \sigma^{\mu}_{i,2}\,,
\end{equation}
where $\sigma^{\mu}_{i,1(2)}$ are the Pauli operators of layer $1$ (layer $2$) of the stacked system.
Below we illustrate a snapshot of the configuration of $H(w,t)$ in \eqref{Hwt} for $0<w_4 < 1$:
\begin{equation}
\label{1d_stack}
\begin{tikzpicture}

\small
\begin{scope}[yshift=30pt]
\node at (-30pt,0pt){$\overbar{H}_{1d}$:};

\notsotiny
\draw (0pt,0pt) circle (4.5pt);
\node at (0pt,0pt){$+$};
\draw [thick](4.5pt,0pt)--(15.5pt,0pt);
\draw (20pt,0pt) circle (4.5pt);
\node at (20pt,0pt){$-$};

\draw (40pt,0pt) circle (4.5pt);
\node at (40pt,0pt){$+$};
\draw [thick](44.5pt,0pt)--(55.5pt,0pt);
\draw (60pt,0pt) circle (4.5pt);
\node at (60pt,0pt){$-$};

\draw (80pt,0pt) circle (4.5pt);
\node at (80pt,0pt){$+$};
\draw [thick](84.5pt,0pt)--(95.5pt,0pt);
\draw (100pt,0pt) circle (4.5pt);
\node at (100pt,0pt){$-$};

\draw (120pt,0pt) circle (4.5pt);
\node at (120pt,0pt){$+$};
\draw [thick](124.5pt,0pt)--(135.5pt,0pt);
\draw (140pt,0pt) circle (4.5pt);
\node at (140pt,0pt){$-$};

\draw (160pt,0pt) circle (4.5pt);
\node at (160pt,0pt){$+$};

\draw [thick][gray][dashed](0pt,4.5pt)--(0pt,15.5pt);
\draw [thick][gray][dashed](20pt,4.5pt)--(20pt,15.5pt);
\draw [thick][gray][dashed](40pt,4.5pt)--(40pt,15.5pt);
\draw [thick][gray][dashed](60pt,4.5pt)--(60pt,15.5pt);
\draw [thick][gray][dashed](80pt,4.5pt)--(80pt,15.5pt);
\draw [thick][gray][dashed](100pt,4.5pt)--(100pt,15.5pt);
\draw [thick][gray][dashed](120pt,4.5pt)--(120pt,15.5pt);
\draw [thick][gray][dashed](140pt,4.5pt)--(140pt,15.5pt);
\draw [thick][gray][dashed](160pt,4.5pt)--(160pt,15.5pt);
\end{scope}

\begin{scope}[yshift=50pt]

\small
\node at (-30pt,0pt){$H_{1d}$:};

\notsotiny

\draw (0pt,0pt) circle (4.5pt);
\node at (0pt,0pt){$-$};
\draw [thick](4.5pt,0pt)--(15.5pt,0pt);
\draw (20pt,0pt) circle (4.5pt);
\node at (20pt,0pt){$+$};

\draw (40pt,0pt) circle (4.5pt);
\node at (40pt,0pt){$-$};
\draw [thick](44.5pt,0pt)--(55.5pt,0pt);
\draw (60pt,0pt) circle (4.5pt);
\node at (60pt,0pt){$+$};

\draw (80pt,0pt) circle (4.5pt);
\node at (80pt,0pt){$-$};
\draw [thick](84.5pt,0pt)--(95.5pt,0pt);
\draw (100pt,0pt) circle (4.5pt);
\node at (100pt,0pt){$+$};

\draw (120pt,0pt) circle (4.5pt);
\node at (120pt,0pt){$-$};
\draw [thick](124.5pt,0pt)--(135.5pt,0pt);
\draw (140pt,0pt) circle (4.5pt);
\node at (140pt,0pt){$+$};

\draw (160pt,0pt) circle (4.5pt);
\node at (160pt,0pt){$-$};

\end{scope}
\end{tikzpicture}
\end{equation}
Here, the vertical dashed lines correspond to the interaction $H_{\text{int}}$. We note that $H(w,0) = H_{{\rm stack}}(w)$, and so at $t=0$ the KS number of the stacked system is clearly zero, as it is a sum of the opposite KS numbers of $H_{1d}$ and $\overbar{H}_{1d}$.

For $t\in(0,1)$, the system in \eqref{1d_stack} is composed of decoupled clusters, 
each of which contains four qubits, two from each layer. As studied in Appendix~\ref{app:fourspin}, the energy spectrum of each 
cluster is always gapped. Moreover, $H(w,1)$ is independent of $w$ and the ground state is a product state, so for $t=1$ we have a trivial system over $S^3$. This shows that $\overbar{H}_{1d}$ is indeed an inverse of $H_{1d}$.

\section{Chern number pump}
\label{sec:chernpump}

Above in Sec.~\ref{sec:1dmodel-bbc}, we showed that the 1$d$ higher Berry curvature can be understood as a flow of ordinary Berry curvature to/from a spatial boundary. This suggests that it should be possible to construct a system that pumps quantized Chern number to a spatial boundary, in analogy with pumping of charge in the Thouless charge pump. Indeed, in this section we introduce the Chern number pump as a 1$d$ system over $S^2 \times S^1$. It is useful to think of the $S^1$ as a ``time'' parameter, and we will see that upon cycling around the $S^1$ the system pumps one unit of Chern number away from a spatial boundary.

The Chern number pump is a relatively mild modification of the system $H_{1d}$ over $S^3$ introduced in Sec.~\ref{Sec:1+1d_lattice}. As in that system, we place a single qubit at each site of the 1$d$ lattice.  We denote points in parameter space by pairs $(v,t) \in S^2 \times S^1$, with $v = (v_1,v_2,v_3) \in S^2$. It will be convenient to consider $S^1$ as the quotient $S^1 = [-1,1]/\{-1,1\}$, and we think of the time parameter as running from $-1$ to $+1$. Choosing a constant $0 < t_0 < 1$, for $t \in [-t_0, t_0]$ the Hamiltonian $H_{cp}$ of the Chern number pump is given in terms of $H_{1d}$. We define a map $S^2 \times [-t_0, t_0] \to S^3$ with values $w(v,t)$ given by $w_4(v,t) = t/t_0$ and $w_i(v,t) = \sqrt{1 - (t/t_0)^2} v_i$ ($i=1,2,3$).
Therefore, for $t \in [-t_0, t_0]$, we can simply write $H_{cp}(v,t) = H_{1d}(w(v,t))$.

For $t \notin [-t_0, t_0]$, we will choose $H_{cp}(v,t)$ to be independent of $v$. This will immediately imply that the higher Berry curvature is only nonzero on the subspace $S^2 \times [-t_0, t_0] \subset S^2 \times S^1$. At $t = \pm 1$, we choose
\begin{equation}
H_{cp}(v, \pm 1) = \sum_{p \in \Z}(-1)^i\sigma_{p}^{z}
\end{equation}
For other values of $t \notin [-t_0, t_0]$, the Hamiltonian is simply a linear interpolation in $t$. For instance, for $t \in [t_0,1]$ we choose
\begin{equation}
\begin{split}
H_{cp}(v,t) &= \frac{1-t}{1-t_0} \sum_{p \in 2\Z + 1} \sum_{\mu = 1,2,3} \sigma^{\mu}_p \sigma^{\mu}_{p+1} \\
&+  \frac{t-t_0}{1-t_0} \sum_{p \in \Z}(-1)^p \sigma_{p}^{z} \text{.}
\end{split}
\end{equation}
Finally, for $t \in [-1,-t_0]$ we choose
\begin{equation}
\begin{split}
H_{cp}(v,t) &= \frac{1+t}{1-t_0} \sum_{p \in 2\Z} \sum_{\mu = 1,2,3} \sigma^{\mu}_p \sigma^{\mu}_{p+1} \\
&+  \frac{-t - t_0}{1-t_0} \sum_{p \in \Z}(-1)^p \sigma_{p}^{z} \text{.}
\end{split}
\end{equation}
It is straightforward to check that the system $H_{cp}$ thus constructed is gapped for all $(v,t) \in S^2 \times S^1$.

The next step in the analysis of the Chern number pump is to establish that its KS number takes the value $2\pi$. We will need to fix an orientation on $S^2 \times S^1$, and we choose the orientation induced by the locally-defined $3$-form $- dt \wedge d\theta \wedge d\phi$. The reason for the perhaps unnatural-looking minus sign is that with this choice of orientation, the map from $S^2 \times [-t_0,t_0] \to S^3$ used in the construction of $H_{cp}$ becomes an orientation-preserving local diffeomorphism, for the choice of orientation on $S^3$ used in Sec.~\ref{Sec:Berry1d}.
With this choice we thus expect, and indeed find, that $H_{1d}$ over $S^3$ and $H_{cp}$ over $S^2 \times S^1$ have the same value of the KS number. Ultimately the minus sign originates from the fact that $w_4 = \cos \alpha$ decreases as $\alpha$ increases.

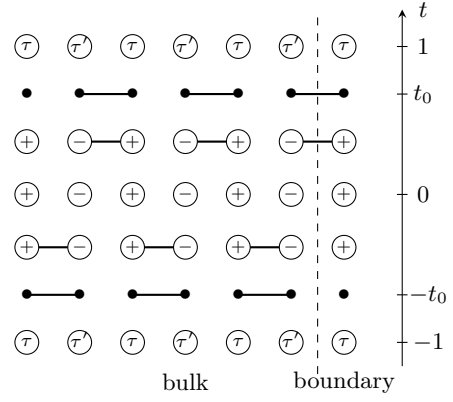
\begin{figure}
\label{H_theta_config}
\centering
    \begin{tikzpicture}
    \small

\begin{scope}[xshift=162pt]
\draw [>=stealth,->](-40pt,-85pt)--(-40pt,50pt);
 \node at (-32pt,50pt){$t$};
 
 \draw (-42pt,-76pt)--(-38pt,-76pt);
 \draw (-42pt,-58pt)--(-38pt,-58pt);
   \draw (-42pt,-20pt)--(-38pt,-20pt);
   \draw (-42pt,18pt)--(-38pt,18pt);
    \draw (-42pt,36pt)--(-38pt,36pt);
    
\node at (-30pt,-76pt){$-1$};
\node at (-30pt,-58pt){$-t_0$};
\node at (-32pt,-20pt){$0$};
\node at (-32pt,18pt){$t_0$};
\node at (-32pt,36pt){$1$};
 \end{scope}

 \begin{scope}[yshift=36pt]

\scriptsize

\draw (100pt,0pt) circle (4.5pt);
\node at (100pt,0pt){$\tau$};
\draw (80pt,0pt) circle (4.5pt);
\node at (80pt,1pt){$\tau'$};
\draw (60pt,0pt) circle (4.5pt);
\node at (60pt,0pt){$\tau$};
\draw (40pt,0pt) circle (4.5pt);
\node at (40pt,1pt){$\tau'$};
\draw (20pt,0pt) circle (4.5pt);
\node at (20pt,0pt){$\tau$};
\draw (0pt,0pt) circle (4.5pt);
\node at (0pt,1pt){$\tau'$};
\draw (-20pt,0pt) circle (4.5pt);
\node at (-20pt,0pt){$\tau$};
 \end{scope}

\begin{scope}[yshift=18pt]
\draw [thick](0pt,0pt)--(20pt,0pt);
\draw [thick](40pt,0pt)--(60pt,0pt);
\draw [thick](80pt,0pt)--(100pt,0pt);

\node at (100pt,0pt){$\bullet$};
\node at (80pt,0pt){$\bullet$};
\node at (60pt,0pt){$\bullet$};
\node at (40pt,0pt){$\bullet$};
\node at (20pt,0pt){$\bullet$};
\node at (0pt,0pt){$\bullet$};
\node at (-20pt,0pt){$\bullet$};
\end{scope}

 \begin{scope}[yshift=0pt]

 \draw [thick](44.5+40pt,0pt)--(55.5+40pt,0pt); 
 \draw [thick](44.5-0pt,0pt)--(55.5-0pt,0pt);
 \draw [thick](44.5-40pt,0pt)--(55.5-40pt,0pt);
 
 \notsotiny
 
\draw (-20pt,0pt) circle (4.5pt);
\node at (-20pt,0pt){$+$};

\draw (0pt,0pt) circle (4.5pt);
\node at (0pt,0pt){$-$};

\draw (20pt,0pt) circle (4.5pt);
\node at (20pt,0pt){$+$};

\draw (40pt,0pt) circle (4.5pt);
\node at (40pt,0pt){$-$};

\draw (60pt,0pt) circle (4.5pt);
\node at (60pt,0pt){$+$};

\draw (80pt,0pt) circle (4.5pt);
\node at (80pt,0pt){$-$};

\draw (100pt,0pt) circle (4.5pt);
\node at (100pt,0pt){$+$};

 \end{scope}

 \begin{scope}[yshift=-20pt]
 
\notsotiny
\draw (-20pt,0pt) circle (4.5pt);
\node at (-20pt,0pt){$+$};

\draw (0pt,0pt) circle (4.5pt);
\node at (0pt,0pt){$-$};

\draw (20pt,0pt) circle (4.5pt);
\node at (20pt,0pt){$+$};

\draw (40pt,0pt) circle (4.5pt);
\node at (40pt,0pt){$-$};

\draw (60pt,0pt) circle (4.5pt);
\node at (60pt,0pt){$+$};

\draw (80pt,0pt) circle (4.5pt);
\node at (80pt,0pt){$-$};

\draw (100pt,0pt) circle (4.5pt);
\node at (100pt,0pt){$+$};

 \end{scope}

  \begin{scope}[yshift=-40pt]
 
 \notsotiny

 \draw [thick](44.5+20pt,0pt)--(55.5+20pt,0pt); 
 \draw [thick](44.5-20pt,0pt)--(55.5-20pt,0pt);
 \draw [thick](44.5-60pt,0pt)--(55.5-60pt,0pt);
 
\draw (-20pt,0pt) circle (4.5pt);
\node at (-20pt,0pt){$+$};

\draw (0pt,0pt) circle (4.5pt);
\node at (0pt,0pt){$-$};

\draw (20pt,0pt) circle (4.5pt);
\node at (20pt,0pt){$+$};

\draw (40pt,0pt) circle (4.5pt);
\node at (40pt,0pt){$-$};

\draw (60pt,0pt) circle (4.5pt);
\node at (60pt,0pt){$+$};

\draw (80pt,0pt) circle (4.5pt);
\node at (80pt,0pt){$-$};

\draw (100pt,0pt) circle (4.5pt);
\node at (100pt,0pt){$+$};

 \end{scope}

  \begin{scope}[yshift=-58pt]

\draw [thick](-20pt,0pt)--(0pt,0pt);
\draw [thick](20pt,0pt)--(40pt,0pt);
\draw [thick](60pt,0pt)--(80pt,0pt);

\node at (100pt,0pt){$\bullet$};
\node at (80pt,0pt){$\bullet$};
\node at (60pt,0pt){$\bullet$};
\node at (40pt,0pt){$\bullet$};
\node at (20pt,0pt){$\bullet$};
\node at (0pt,0pt){$\bullet$};
\node at (-20pt,0pt){$\bullet$};

\draw [dashed](90pt,-30pt)--(90pt,106pt);

   \end{scope}
   
    \begin{scope}[yshift=-76pt]
\scriptsize
\draw (100pt,0pt) circle (4.5pt);
\node at (100pt,0pt){$\tau$};
\draw (80pt,0pt) circle (4.5pt);
\node at (80pt,1pt){$\tau'$};
\draw (60pt,0pt) circle (4.5pt);
\node at (60pt,0pt){$\tau$};
\draw (40pt,0pt) circle (4.5pt);
\node at (40pt,1pt){$\tau'$};
\draw (20pt,0pt) circle (4.5pt);
\node at (20pt,0pt){$\tau$};
\draw (0pt,0pt) circle (4.5pt);
\node at (0pt,1pt){$\tau'$};
\draw (-20pt,0pt) circle (4.5pt);
\node at (-20pt,0pt){$\tau$};

\small

\node at (100pt,-15pt){boundary};
\node at (40pt,-15pt){bulk};
 \end{scope}
\end{tikzpicture}
\caption{Configurations of the Chern number pump Hamiltonian $H_{cp}(v,t)$
 in the case $X_{\text{bdy}}=X_{\text{bulk}}=S^2\times S^1$.
The boundary terminates at site $p=N\in 2\mathbb Z$.
 There is a boundary phase transition at $t=-t_0$, where 
 the boundary becomes gapless. The vertical dashed line represents an arbitrary division of the system into bulk and boundary as in the definition of boundary Berry curvature, and corresponds to $a = N - 1/2$. $\tau$ (resp. $\tau'$) denotes a lattice site $p$ with Hamiltonian $\sigma^z_p$ (resp. $-\sigma^z_p$).
 }
\label{ChernPumpFig}
\end{figure}

Rather than proceeding by a direct calculation of the higher Berry curvature, we will employ the bulk-boundary correspondence as developed in Sec.~\ref{Sec:Semi-infinite}, which will also serve to illuminate the physics of Chern number pumping. As in Sec.~\ref{Sec:1+1d_lattice}, we introduce a boundary by throwing away lattice sites with $p > N$, and dropping all Hamiltonian terms coupling to qubits with $p > N$. We take $N \in 2\Z$, and observe that the boundary is gapless on $S^2 \times \{-t_0 \}$ but is otherwise gapped. To discuss the KS invariant via the bulk-boundary correspondence, we choose $f(p) = \Theta(p-a)$ with $a \in 2\Z - 1/2$ and $a < N$, and introduce the boundary Berry curvature $\omega^{(2)}$ as in \eqref{2form_omega}, which is defined on $X_\Delta = S^2 \times S^1 \backslash S^2 \times \{-t_0 \}$. The Chern number pump with the chosen boundary termination is illustrated in Fig.~\ref{ChernPumpFig}.

Given $t \in S^1$ we have the natural $S^2$ subspace defined by $S^2(t) = S^2 \times \{t \}$. Choosing the orientation on $S^2(t)$ specified by $d\theta \wedge d\phi$, for $t \neq -t_0$ the boundary Chern number is given by $C(t) = \int_{S^2(t)} \omega^{(2)}$. Generically, the boundary Chern number is not quantized. However, it is quantized when there is no coupling between lattice sites on different sides of the position $a$. We thus see that in the model $H_{cp}$ with the given choice of $a$, $C(t)$ is quantized \emph{except} for $t \in [0,t_0]$.

The KS number is given in terms of the boundary Chern number by
\begin{eqnarray}
\label{KS_1d_Chern}
KS &=& \lim_{\epsilon \to 0} \Big[ \int_{S^2(-t_0 + \epsilon)} \omega^{(2)} - \int_{S^2(-t_0 - \epsilon)} \omega^{(2)}\Big] \nonumber \\ &=& \lim_{\epsilon \to 0} \Big[ C(-t_0 + \epsilon) - C(-t_0 - \epsilon) \Big] \label{eqn:ks-bcn} \text{,}
\end{eqnarray}
where the overall sign comes from the chosen orientations on $S^2 \times S^1$ and $S^2(t)$. In principle we should also take the limit $a \to -\infty$, but we will see the result has no dependence on $a$. For $t = -t_0 - \epsilon$, the system consists of decoupled lattice sites whose single-spin Hamiltonians have no dependence on $v \in S^2$, so clearly $C(-t_0 - \epsilon) = 0$. On the other hand, for $t = -t_0 + \epsilon$, to the right of $a$ there are some number of dimerized pairs of lattice sites, as well as the $p = N$ boundary site with Hamiltonian $H^1_{p = N}(w)$. The dimerized pairs do not contribute to $\omega^{(2)}$, so the boundary Chern number is simply the Chern number of the decoupled $p = N$ boundary spin, \emph{i.e.} $C(-t_0 + \epsilon) = 2\pi$. Therefore we find $KS = 2\pi$, as expected. The same result is obtained by a similar analysis if we choose $a \in 2\Z + 1/2$.

More interesting than simply giving the expected result $KS = 2\pi$, \eqref{eqn:ks-bcn} allows us to interpret the KS number in terms of Chern number pumping. As $t$ is increased, starting just above $-t_0$ and wrapping around $S^1$, the boundary Chern number $C(t)$ decreases from $2\pi$ to zero at $t = -t_0 - \epsilon$. Therefore a $2\pi$-quantized amount of Chern number flows from the boundary into the bulk over one cycle of the time parameter. Moreover, because the Hamiltonian depends periodically on $t$, so that the bulk returns to its initial state after a cycle, we can think of the system as a Chern number pump. With this in mind, \eqref{eqn:ks-bcn} tells us that KS number of this system (and indeed of any 1$d$ system over $S^2 \times S^1$) is nothing but a measure of the quantized amount of Chern number pumped over each cycle.

A different but closely related perspective on Chern number pumping is provided by enlarging the boundary parameter space to $X_{{\rm bdy}} = S^2 \times [-1,1]$, where $\pi_{{\rm bdy}}: X_{{\rm bdy}} \to X_{{\rm bulk}} = S^2 \times S^1$ is the identity on $S^2$ and on $[-1,1]$ is the quotient map that identifies the two endpoints to a single point in $S^1$. In words, we have enlarged the boundary parameter space so that $t$ is not required to be a periodic parameter on the boundary (but the system still depends periodically on $t$ in the bulk). Enlarging the parameter space in this way allows us to introduce a fully gapped boundary termination, with the boundary Berry curvature and Chern number thus globally defined over $X_{{\rm bdy}}$.

\begin{figure}
\centering
    \begin{tikzpicture}
    \small
    
\begin{scope}[xshift=162pt]
\draw [>=stealth,->](-40pt,-85pt)--(-40pt,50pt);
 \node at (-32pt,50pt){$t$};
 
 \draw (-42pt,-76pt)--(-38pt,-76pt);
 \draw (-42pt,-58pt)--(-38pt,-58pt);
   \draw (-42pt,-20pt)--(-38pt,-20pt);
   \draw (-42pt,18pt)--(-38pt,18pt);
    \draw (-42pt,36pt)--(-38pt,36pt);
    
\node at (-30pt,-76pt){$-1$};
\node at (-30pt,-58pt){$-t_0$};
\node at (-32pt,-20pt){$0$};
\node at (-32pt,18pt){$t_0$};
\node at (-32pt,36pt){$1$};
 \end{scope}

 \begin{scope}[yshift=36pt]

\scriptsize
\draw (100pt,0pt) circle (4.5pt);
\node at (100pt,0pt){$\tau$};
\draw (80pt,0pt) circle (4.5pt);
\node at (80pt,1pt){$\tau'$};
\draw (60pt,0pt) circle (4.5pt);
\node at (60pt,0pt){$\tau$};
\draw (40pt,0pt) circle (4.5pt);
\node at (40pt,1pt){$\tau'$};
\draw (20pt,0pt) circle (4.5pt);
\node at (20pt,0pt){$\tau$};
\draw (0pt,0pt) circle (4.5pt);
\node at (0pt,1pt){$\tau'$};
\draw (-20pt,0pt) circle (4.5pt);
\node at (-20pt,0pt){$\tau$};
 \end{scope}

\begin{scope}[yshift=18pt]
\draw [thick](0pt,0pt)--(20pt,0pt);
\draw [thick](40pt,0pt)--(60pt,0pt);
\draw [thick](80pt,0pt)--(100pt,0pt);

\node at (100pt,0pt){$\bullet$};
\node at (80pt,0pt){$\bullet$};
\node at (60pt,0pt){$\bullet$};
\node at (40pt,0pt){$\bullet$};
\node at (20pt,0pt){$\bullet$};
\node at (0pt,0pt){$\bullet$};
\node at (-20pt,0pt){$\bullet$};
\end{scope}

 \begin{scope}[yshift=0pt]

 \draw [thick](44.5+40pt,0pt)--(55.5+40pt,0pt); 
 \draw [thick](44.5-0pt,0pt)--(55.5-0pt,0pt);
 \draw [thick](44.5-40pt,0pt)--(55.5-40pt,0pt);
 
 \notsotiny
 
\draw (-20pt,0pt) circle (4.5pt);
\node at (-20pt,0pt){$+$};

\draw (0pt,0pt) circle (4.5pt);
\node at (0pt,0pt){$-$};

\draw (20pt,0pt) circle (4.5pt);
\node at (20pt,0pt){$+$};

\draw (40pt,0pt) circle (4.5pt);
\node at (40pt,0pt){$-$};

\draw (60pt,0pt) circle (4.5pt);
\node at (60pt,0pt){$+$};

\draw (80pt,0pt) circle (4.5pt);
\node at (80pt,0pt){$-$};

\draw (100pt,0pt) circle (4.5pt);
\node at (100pt,0pt){$+$};

 \end{scope}

 \begin{scope}[yshift=-20pt]
 
 \notsotiny

\draw (-20pt,0pt) circle (4.5pt);
\node at (-20pt,0pt){$+$};

\draw (0pt,0pt) circle (4.5pt);
\node at (0pt,0pt){$-$};

\draw (20pt,0pt) circle (4.5pt);
\node at (20pt,0pt){$+$};

\draw (40pt,0pt) circle (4.5pt);
\node at (40pt,0pt){$-$};

\draw (60pt,0pt) circle (4.5pt);
\node at (60pt,0pt){$+$};

\draw (80pt,0pt) circle (4.5pt);
\node at (80pt,0pt){$-$};

\draw (100pt,0pt) circle (4.5pt);
\node at (100pt,0pt){$+$};

 \end{scope}

  \begin{scope}[yshift=-40pt]
 
 \notsotiny

 \draw [thick](44.5+20pt,0pt)--(55.5+20pt,0pt); 
 \draw [thick](44.5-20pt,0pt)--(55.5-20pt,0pt);
 \draw [thick](44.5-60pt,0pt)--(55.5-60pt,0pt);
 
\draw (-20pt,0pt) circle (4.5pt);
\node at (-20pt,0pt){$+$};

\draw (0pt,0pt) circle (4.5pt);
\node at (0pt,0pt){$-$};

\draw (20pt,0pt) circle (4.5pt);
\node at (20pt,0pt){$+$};

\draw (40pt,0pt) circle (4.5pt);
\node at (40pt,0pt){$-$};

\draw (60pt,0pt) circle (4.5pt);
\node at (60pt,0pt){$+$};

\draw (80pt,0pt) circle (4.5pt);
\node at (80pt,0pt){$-$};

\draw (100pt,0pt) circle (4.5pt);
\node at (100pt,0pt){$+$};

 \end{scope}

  \begin{scope}[yshift=-58pt]

\draw [thick](-20pt,0pt)--(0pt,0pt);
\draw [thick](20pt,0pt)--(40pt,0pt);
\draw [thick](60pt,0pt)--(80pt,0pt);

\notsotiny

\draw (100pt,0pt) circle (4.5pt);
\node at (100pt,0pt){$+$};
\node at (80pt,0pt){$\bullet$};
\node at (60pt,0pt){$\bullet$};
\node at (40pt,0pt){$\bullet$};
\node at (20pt,0pt){$\bullet$};
\node at (0pt,0pt){$\bullet$};
\node at (-20pt,0pt){$\bullet$};

\draw [dashed](90pt,-30pt)--(90pt,106pt);

   \end{scope}

    \begin{scope}[yshift=-76pt]

\notsotiny

\draw (100pt,0pt) circle (4.5pt);
\node at (100pt,0pt){$+$};

   \scriptsize

\draw (80pt,0pt) circle (4.5pt);
\node at (80pt,1pt){$\tau'$};
\draw (60pt,0pt) circle (4.5pt);
\node at (60pt,0pt){$\tau$};
\draw (40pt,0pt) circle (4.5pt);
\node at (40pt,1pt){$\tau'$};
\draw (20pt,0pt) circle (4.5pt);
\node at (20pt,0pt){$\tau$};
\draw (0pt,0pt) circle (4.5pt);
\node at (0pt,1pt){$\tau'$};
\draw (-20pt,0pt) circle (4.5pt);
\node at (-20pt,0pt){$\tau$};

\small
\node at (100pt,-15pt){boundary};
\node at (40pt,-15pt){bulk};
 \end{scope}
\end{tikzpicture}
\caption{Configurations of the Chern number pump Hamiltonian $H_{cp}(v,t)$ in the case of enlarged boundary parameter space $X_{{\rm bdy}} = S^2 \times [-1,1]$, for which the boundary is always fully gapped.
}
\label{ChernPumpFig2}
\end{figure}
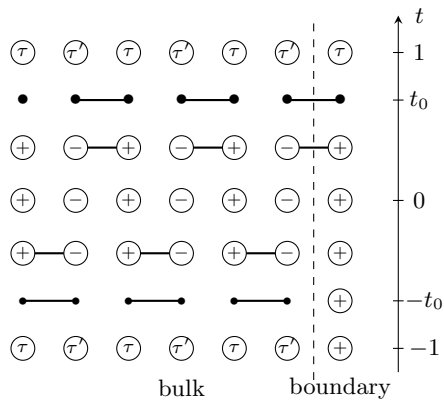

We again terminate the lattice at $p = N$ with $N \in 2 \Z$. The Hamiltonian is chosen as for the previous boundary termination, with the only change being the single-spin Hamiltonian of the $p = N$ boundary spin when $t \in [-1,0]$. We choose that spin's Hamiltonian to be
\begin{equation}
H^1_{p=N}(v,t) = v_1 \sigma^1_N + v_2 \sigma^2_N + v_3 \sigma^3_N \text{,}
\end{equation}
which is independent of $t$. 
Note that we do not make any changes for $t \in [0,1]$, where the $p=N$ spin is coupled to its neighbor at $p=N-1$ in a $t$-dependent manner. This boundary termination is illustrated in Fig.~\ref{ChernPumpFig2}.

We see that initially at $t = -1$, the $p = N$ boundary spin is decoupled from the bulk, and is a non-trivial 0$d$ system over $S^2 \times \{-1\}$ with Chern number $2\pi$. As $t$ is increased, the boundary spin is coupled to the bulk, and its Chern number flows away into the bulk. At $t=1$, the boundary spin is again decoupled from the bulk, but is now a trivial 0$d$ system over $S^2 \times \{ 1 \}$ with Chern number zero. Even though the boundary spin is not periodic in $t$ -- this is the price we pay for a fully gapped boundary -- the bulk is still periodic in $t$. Over one cycle, the Chern number of the boundary spin has thus disappeared into the bulk ``at infinity.''

\section{Classification of invertible quantum systems over $X$}
\label{sec:classification}

Here we discuss the expectation that invertible phases over a space $X$ are classified by a generalized cohomology theory. In Sec.~\ref{sec:gc1}, we take a practical viewpoint and explain the consequences of this expectation for guiding the construction of new examples. Then in Sec.~\ref{sec:gc2}, we explain where this expectation comes from, put it in a broader context, and discuss its status. We also sketch a potential strategy to construct a generalized cohomology theory of parametrized invertible phases (Sec.~\ref{sec:gc3}). We note that Sections~\ref{sec:gc2} and~\ref{sec:gc3} are not needed to follow the remainder of the paper, and can be skipped by readers who are mainly interested in concrete examples.

\subsection{Consequences of generalized cohomology classification}
\label{sec:gc1}

To discuss the classification of parametrized invertible phases, we first fix from the beginning whether we are interested in bosonic or fermionic systems, and the type of symmetry imposed (if any). Given these fixed attributes, the set of $d$-dimensional invertible phases over $X$ is denoted $\GPinv^d(X)$; in fact, $\GPinv^d(X)$ is an abelian group under stacking as noted in Sec.~\ref{sec:parametrized}. The group of ordinary $d$-dimensional phases is $\GPinv^d(\mathrm{pt})$, where $\mathrm{pt}$ is the single-point topological space. As discussed in Sec.~\ref{sec:parametrized}, $\GPinv^d(X)$ is homotopy-invariant; that is, if $X$ and $Y$ are homotopy equivalent spaces, then $\GPinv^d(X) \cong \GPinv^d(Y)$.

The expectation that invertible phases are classified by a generalized cohomology theory means that the groups $\GPinv^d(X)$ obey certain properties (the Eilenberg-Steenrod axioms) that relate groups for different spaces $X$ and in different dimensions $d$. We do not enumerate the Eilenberg-Steenrod axioms here, but instead focus on some particularly useful facts.

Any generalized cohomology theory also comes with a \emph{reduced} version of the same theory. Given a space $X$ and a chosen basepoint $x_0 \in X$, the reduced theory associates an abelian group $\GPinv^d(X, x_0)$. The reduced and unreduced theories are related by
\begin{equation}
\GPinv^d(X) \cong \GPinv^d(\{ x_0 \}) \oplus \GPinv^d(X, x_0) \text{.} \label{eqn:reduced-unreduced}
\end{equation}
If $X$ is path-connected, then $\GPinv^d(X, x_0) \cong \GPinv^d(X, x'_0)$ for two different basepoints $x_0, x'_0 \in X$. Sometimes writing out basepoints is cumbersome, and the notation
\begin{equation}
\tGPinv^d(X) = \GPinv^d(X, x_0)
\end{equation} is convenient, where on the left-hand side $X$ is understood to be a based space, \emph{i.e.} a topological space together with a chosen (but not explicitly written) basepoint. If we only care about the isomorphism class of $\tGPinv^d(X)$, for path-connected spaces we do not need to specify the choice of basepoint.

Both $\GPinv^d(X, x_0)$ and \eqref{eqn:reduced-unreduced} have nice physical interpretations; indeed, as should be apparent from the following discussion, $\GPinv^d(X, x_0)$ can be defined and \eqref{eqn:reduced-unreduced} argued to hold on physical grounds. We consider an invertible system $H$ over $X$ with the additional property that $H(x_0)$ is in the trivial phase over the point $\{ x_0 \}$. (Note that this property is a phase invariant -- if it is satisfied for one system in a given phase over $X$, it holds for any system in the same phase.) We say that $H$ is in a \emph{reduced phase} over $X$ with basepoint $x_0$, and $\GPinv^d(X, x_0)$ is the abelian group of such reduced phases. Equation \eqref{eqn:reduced-unreduced} then has the interpretation that the group of invertible phases over $X$ [$\GPinv^d(X)$] can be decomposed into reduced phases [$\GPinv^d(X, x_0)$] and phases that can be represented by a system with $x$-independent Hamiltonian [$\GPinv^d(\{ x_0 \} )$]. Therefore, a general invertible system over $X$ can always be brought into a reduced phase by stacking with a suitable constant system. From this we see that reduced phases are the most interesting from the point of view of finding physics beyond that of ordinary phases (\emph{i.e.} phases over a point).

Every reduced generalized cohomology theory comes with a \emph{suspension isomorphism}, which is an isomorphism of abelian groups for each space $X$ with basepoint $x_0$:
\begin{equation}
s_X : \GPinv^d(X, x_0) \to \GPinv^{d+1}(\Sigma X, \Sigma x_0) \text{.} \label{eqn:suspension-iso}
\end{equation}
Here $\Sigma X$ is the \emph{reduced suspension} of $X$, and is defined as a quotient space formed from $X \times [-1,1]$ by identifying the subspace $X \times \{ 1 \} \cup X \times \{ -1 \} \cup \{ x_0 \} \times [-1,1]$ to a single point. This point becomes the natural basepoint $\Sigma x_0$.

The reader may wonder about the relationship between $\Sigma X$ and the (unreduced) suspension $SX$ mentioned in Sec.~\ref{sec:intro}. Recall that $S X$ is also formed from $X \times [-1,1]$ by identifying each of the subspaces $X \times \{ 1 \}$ and $X \times \{ -1 \}$ to (distinct) single points. In fact, under some reasonable assumptions about the base point of $x_0 \in X$, $S X$ and $\Sigma X$ are homotopy equivalent. (For example, this holds if $x_0 \in X$ has a neighborhood that deformation retracts onto $x_0$. In particular, this holds for any point $x_0$ of a manifold $X$, or for any vertex in a CW complex.) Therefore, for most of our purposes in this paper, it is not important to distinguish $S X$ and $\Sigma X$; we introduced $\Sigma X$ here to allow for a precise statement of \eqref{eqn:suspension-iso}. In particular, noting that both $\Sigma X$ and $S X$ are always path-connected, we have $\tGPinv^d(SX) \cong \tGPinv^d(\Sigma X)$.

In Sec.~\ref{Sec:GeneralConstruct} we give a proposed physical construction of the suspension isomorphism. This provides a powerful tool to construct new examples of parametrized invertible systems from existing ones, as we now explain. First we note that the suspension of a sphere is homeomorphic to a sphere in one higher dimension: $\Sigma S^n \cong S^{n+1}$. Therefore we can apply the suspension isomorphism $n$ times to obtain $\tGPinv^{d+n}(S^n) \cong \tGPinv^d(S^0)$. But $S^0$ is just the two-point discrete space, so $\tGPinv^d(S^0) = \GPinv^d(\mathrm{pt})$, \emph{i.e.} $\tGPinv^d(S^0)$ is isomorphic to the classification of ordinary non-parametrized $d$-dimensional phases. As a result we have
\begin{equation}
\tGPinv^{d+n}(S^n) \cong \GPinv^d(\mathrm{pt}) \text{,} \label{eqn:s-app-1}
\end{equation}
telling us that the classification of $d+n$-dimensional reduced phases over $S^n$ is the same as that of ordinary $d$-dimensional phases. With an explicit realization of the suspension isomorphism and a sufficient understanding of ordinary $d$-dimensional phases, this allows us to construct an example system for any $d+n$-dimensional reduced phase over $S^n$.

Equation~\eqref{eqn:s-app-1} can be applied to obtain $\tGPinv^d(S^k)$ whenever $d \geq k$. On the other hand, if $d < k$, we instead have
\begin{equation}
\tGPinv^d(S^k) \cong \tGPinv^0(S^{k-d}) \text{,} \label{eqn:s-app-2}
\end{equation}
where the right-hand side is the classification of zero-dimensional reduced phases over $S^{k-d}$. For example, for bosonic systems with no symmetry, zero-dimensional systems over $S^2$ have a $\Z$ classification indexed by the Chern number, and $\tGPinv^0(S^2) \cong \Z$. Then \eqref{eqn:s-app-2} implies that $\tGPinv^d(S^{d+2}) \cong \Z$, where the $d=1$ case corresponds to the KS invariant as discussed above, and for $d > 1$ there are higher KS invariants discussed below in Sec.~\ref{Sec:HigherChernPump}.

To summarize, \eqref{eqn:reduced-unreduced}, \eqref{eqn:s-app-1} and \eqref{eqn:s-app-2} can be used to obtain $\GPinv^d(S^n)$ quite generally, as long as one understands the classifications of ordinary phases and zero-dimensional systems that appear. With the construction of Sec.~\ref{Sec:GeneralConstruct}, beyond merely obtaining the groups $\GPinv^d(S^n)$, one can construct example systems and explore their physical properties. While this paper focuses on bosonic systems with no symmetry, forthcoming work will consider other settings, including fermionic systems and systems with symmetry. As we explain in Sec.~\ref{Sec:GeneralConstruct}, the physics of the reduced phases in $\tGPinv^d(S^n)$ can be viewed in terms of flow of a lower-dimensional invertible phase invariant to/from a spatial boundary, just as 1$d$ systems over $S^3$ with non-zero KS number can be understood in terms of flow of Chern number.

In this paper, we also study systems over $X \times S^1$, which can be interpreted as pumps of a system over $X$ in one dimension lower. From the properties of generalized cohomology theories, it is straightforward to show that 
\begin{align*}
\tGPinv^{d+1}(X \times S^1) 
 &\cong \tGPinv^{d+1}(X) \oplus \GPinv^{d}(\mathrm{pt}) \oplus  \tGPinv^{d}(X) .
\end{align*}
So, $(d+1)$-dimensional phases over $X \times S^1$ come in three flavors, corresponding to each summand:
\begin{enumerate}
\item  A $(d+1)$-dimensional phase over $X$ which is propagated in a constant way over $S^1$, \emph{i.e.} it is possible to choose a representative system whose Hamiltonian does not depend on the $S^1$ parameter. This comes from the map
\[ \tGPinv^{d+1}(X) \to  \tGPinv^{d+1}(X \times S^1), \]
induced by the map of spaces $p_X \colon X \times S^1 \to X$ which sends $(w,t) \mapsto w$.
\item A $d$-dimensional phase over a point, which is suspended to a $(d+1)$-dimensional system over $S^1$ using the suspension construction of Sec.~\ref{Sec:GeneralConstruct}:
\[\GPinv^{d}(\mathrm{pt}) \cong \tGPinv^{d}(S^0) \cong \tGPinv^{d+1}(S^1)  \] and propagated in a constant manner over $X$ (\emph{i.e.}, the Hamiltonian is chosen to have no $X$-dependence).
\item A $d$-dimensional phase over $X$, which is put on $X \times S^1$ by first using the suspension construction to get a $(d+1)$-dimensional phase over $\Sigma X$, and then put on $X \times S^1$ by precomposing the resulting system $H \colon \Sigma X \to \GH$ with the quotient map $q\colon X \times S^1 \to \Sigma X$. In cohomology, this corresponds to the composite
\[\tGPinv^{d}( X) \xrightarrow[\cong]{ \ s_X \ } \tGPinv^{d+1}(\Sigma X) \xrightarrow{\tGPinv^{d+1}(q)} \tGPinv^{d+1}( X \times S^1).\]
\end{enumerate}

Physically the last summand is the most interesting. Evidently, for any reduced phase in $\tGPinv^{d+1}(\Sigma X)$, there is a corresponding reduced phase over $X \times S^1$. That is, there is an injective map from $\tGPinv^{d+1}(\Sigma X)$ to $\tGPinv^{d+1}(X \times S^1)$. If a phase over $\Sigma X$ is characterized in terms of flow of a given $d$-dimensional invertible phase over $X$ to/from a spatial boundary, then the corresponding phase over $X \times S^1$ is a pump of the same $d$-dimensional invertible phase, which we discuss in Sec.~\ref{Sec:GeneralConstruct}. The Chern number pump over $S^2 \times S^1$ (Sec.~\ref{sec:chernpump}), and its close relation to the solvable 1$d$ model over $S^3$ of Sec.~\ref{Sec:1+1d_lattice}, is an illustration of this more general correspondence.

Finally, we briefly note the comparison with cohomological classifications of phases, focusing on bosonic phases without symmetry. Already, general techniques in stable homotopy theory combined with what we know about bosonic phases with no symmetry imply that for any parameter space, $X$, there is a map $H^{d+2}(X;\mathbb{Z}) \to \GPinv^{d}(X)$. The existence of phases ``beyond cohomology'' tells us this map is not an isomorphism. We note however that, when $X$ is a manifold, the differential form underlying the KS invariant gives a map $\GPinv^d(X)\to H^{d+2}(X;\mathbb{R})$. It has been argued \cite{Kapustin_2020} that integrals of this form over spheres $S^{d+2}\to X$ are integer valued. Extending this argument to integrals over arbitrary smooth simplicies $\Delta^{d+2}\to X$ would give comparison in the other direction, namely, a map from $\GPinv^{d}(X)$ to the group  $H^{d+2}(X;\mathbb{Z})$ modulo its torsion.

\subsection{Generalized cohomology: context and status}
\label{sec:gc2}

We now briefly discuss where the generalized cohomology proposal comes from, put it in a larger context, and comment on its current status. In a series of talks, Kitaev proposed that gapped invertible systems form what is known as an $\Omega$-spectrum in homotopy theory (not to be confused with the energy spectrum of a Hamiltonian in quantum mechanics).\cite{kitaevSimonsCenter1,kitaevSimonsCenter2,kitaevIPAM} These ideas were further developed by others.\cite{gaiotto19symmetry,Xiong2018,shiozaki2018generalized} 
We first summarize some of the ideas that have been circulating on this topic, and then emphasize a perspective offered by the study of parametrized systems. 

For simplicity, we focus on bosonic systems without symmetry; the discussion for fermionic no-symmetry systems is identical. We note that the arguments for an $\Omega$-spectrum of no-symmetry invertible systems apply equally well to invertible systems with a fixed symmetry. Moreover, it was argued that a simplification occurs, where the classification of phases with symmetry group $G$ is identical to that of parametrized no-symmetry systems over the classifying space $BG$.\cite{kitaevSimonsCenter1,kitaevSimonsCenter2,kitaevIPAM,gaiotto19symmetry}

One supposes the existence of a \emph{classifying space}  of invertible gapped phases in $d$ spatial dimensions, denoted $\GPinv_d$. 
This means in particular that $\GPinv_d$ is a topological space whose path components correspond to invertible phases of systems. Here, by ``system'' we mean a system over a point, \emph{i.e.} a single gapped Hamiltonian (in contrast to a parametrized system). But the space $\GPinv_d$ will also record other non-trivial topological properties of phases. An initial observation about $\GPinv_d$ is that only the \emph{homotopy type} of $\GPinv_d$ matters, so that there may be different constructions or models for $\GPinv_d$ that give the same classification.

We describe $\GPinv_0$ as an example. A 0$d$ gapped bosonic system without any symmetry over $X$ is a continuous family of self-adjoint $n \times n$ matrices with unique ground states.
The ground states assemble into a line bundle over $X$, whose first Chern class in $H^2(X,\mathbb{Z})$ is a complete phase invariant. (When $X$ is a closed oriented $2$-manifold, this invariant is computed using the Berry curvature $2$-form $\Omega^{(2)}$ by $\int_X \Omega^{(2)}$, as discussed in Sec.~\ref{sec:review}.) So the space
$\GPinv_0$ can be chosen to be the infinite complex projective space (\emph{i.e.} a $K(\Z,2)$),
the classifying space for complex topological line bundles. (More generally, it is expected that all the information about what phase a system is in should be contained in its ground states.  This perspective has been emphasized in this context by Kitaev,\cite{kitaevSimonsCenter2} and embraced in recent related work by Kapustin, Sopenko and Yang.\cite{Kapustin_2021})

Now, we turn to what it means for $\GPinv_{d}$ to form an $\Omega$-spectrum. By definition, this means that we have maps
\[ \rho_d : \GPinv_{d} \xrightarrow{\simeq} \Omega \GPinv_{d+1} \]
 which are \emph{homotopy equivalences}.
Here $\Omega \GPinv_{d+1}$ is the space of loops in $\GPinv_{d+1}$ based at a point representing the phase of a trivial system. This map $\rho_d$ can be loosely interpreted as taking a $d$-dimensional system to a loop in the space  $(d+1)$-dimensional systems. In Refs.~\onlinecite{kitaevSimonsCenter1,kitaevSimonsCenter2,kitaevIPAM,gaiotto19symmetry}, physical pictures were given to motivate these maps and their homotopy inverses, and these pictures form the basis for the constructions of Sec.~\ref{Sec:GeneralConstruct}.

Given the structure of the $\Omega$-spectrum, there exists a \emph{generalized cohomology theory} of gapped invertible phases, as briefly (and incompletely) described above in Sec.~\ref{sec:gc1}. The groups $\GPinv^d(X)$ arise as homotopy classes of maps, in particular
\begin{equation}
\GPinv^d(X) = [X, \GPinv_d ] \text{,}
\end{equation}
where the notation $[X, Y]$ means the set of homotopy classes of continuous maps $f : X \to Y$.

Clearly the generalized cohomology proposal has practical value, for example to understand the classification of invertible phases over $S^n$ as described in Sec.~\ref{sec:gc1}. From a broader perspective, the proposal is important given questions about the relationship between gapped phases of matter and topological quantum field theories (TQFTs). Such questions have become newly pressing given recent progress on fracton phases,\cite{nandkishore19fractons,pretko20fracton} which are gapped but non-invertible phases that lack a TQFT description. Fracton phases are thus counterexamples  to prior conventional wisdom that every gapped phase should be well-approximated by a suitable TQFT at low energies. These examples make it clear that the relationship between TQFT and gapped phases is poorly understood from both physical and mathematical perspectives.

Invertible phases appear to be simpler, and it \emph{is} believed that the universal properties of invertible phases of matter are captured by suitably defined invertible TQFTs. Building on work of Kapustin,\cite{kapustin2014symmetry,kapustin2014bosonic,kapustin2015fermionic} the work of Freed and Hopkins shows that certain families of invertible TQFTs are classified by an $\Omega$-spectrum.\cite{freed2021reflection} This reduces the classification of phases of matter to computations of generalized cohomology theories familiar to stable homotopy theorists.\footnote{With our indexing conventions, the classifying spectrum for bosonic (fermionic) phases with no symmetries is expected to be $\Sigma^2I\mathbb{Z}MSO$ ($\Sigma^2I\mathbb{Z}MSpin$), where here $I\mathbb{Z}$ is the Anderson dual and $MSO$ ($MSpin$) is the oriented bordism (spin) spectrum.} The computed low-dimensional classifications agree with results from the physics literature, but it is not known whether the correspondence between invertible phases and invertible TQFTs holds in arbitrarily high dimension. Determining whether gapped invertible systems indeed form an $\Omega$-spectrum, and if so whether the resulting cohomology theory is the same as that obtained for invertible TQFTs, would represent important progress in better connecting gapped phases and TQFTs.

This brings us to the status of the generalized cohomology proposal, which has not been rigorously established so far. One way forward would be construct the $\Omega$-spectrum and the maps $\rho_d$. For example, one could think of constructing $\GPinv_{d}$ as a space whose points are quantum systems, or ground states, and try to construct the maps $\rho_d$ based on the physical pictures of Refs.~\onlinecite{kitaevSimonsCenter1,kitaevSimonsCenter2,kitaevIPAM,gaiotto19symmetry}. 
Some difficulties do arise in interpreting these physical pictures literally: as noted in Sec.~\ref{subsec:SP-constructions}, this construction depends on more data than the input $d$-dimensional system alone, and thus does not define a function with domain $\GPinv_d$. Work thus remains to prove that the additional data can be chosen consistently and continuously for each point of $\GPinv_d$. Therefore, at this point, a more complete physical argument for the existence of an $\Omega$-spectrum of gapped invertible phases is still needed.

Here, we do not attempt to construct the $\Omega$-spectrum, but rather study directly the generalized cohomology theory of parametrized invertible phases that would arise from the existence of such an $\Omega$-spectrum. 
Although some of this has already been explained in Sec.~\ref{sec:parametrized},  we summarize a strategy for a construction of this cohomology theory directly below in Sec.~\ref{sec:gc3}. Note that the existence of an $\Omega$-spectrum follows automatically from the existence of the cohomology theory by the Brown Representability theorem.\cite{browncohomologytheories1962} The difficulties of directly constructing the classifying spaces $\GPinv_d$ are temporarily avoided. However, understanding their homotopy type remains the key problem for computing classifications.

\subsection{Establishing the generalized cohomology proposal: sketch of a strategy}
\label{sec:gc3}

We continue to assume the existence of a space ${\mathfrak {GH}}_d$ of gapped Hamiltonians. For example, ${\mathfrak {GH}}_d$ may be modeled as a subset of those bounded, finite range interactions that give rise to gapped Hamiltonians,\cite{NaaijkensQSSIL} equipped with an appropriate topology. Here the term ``interaction'' does not have its usual physics meaning, but instead refers to a mathematical object that packages the parameters (\emph{i.e.} coefficients of local operators) defining the Hamiltonian. 
Giving a precise definition of ${\mathfrak {GH}}_d$ is a concrete mathematical problem which we do not undertake here.

For suitable spaces $X$ (e.g., spaces with the homotopy type of a finite CW-complex), gapped systems over $X$ are then  continuous maps from $X$ to $ {\mathfrak {GH}}_d$, the set of which is denoted $C^0(X, {\mathfrak {GH}}_d)$.
If $X$ is a manifold, differentiability conditions may facilitate the study of the system, as for our analysis of the KS invariant, but we do not impose such conditions here.

In Sec.~\ref{sec:parametrized} we defined a \emph{phase} as an equivalence class on $C^0(X, {\mathfrak {GH}}_d)$ generated by deformations (\emph{i.e.} homotopies of systems), isomorphism and stabilization with respect to stacking with trivial systems. 
The stacking operation descends to an operation on phases, and those phases that have an inverse together form an abelian group $\GPinv^d(X)$, the group of invertible phases over $X$.

At this point, it is already clear that the assignment $X \mapsto \GPinv^d(X)$ satisfies many of the properties required of a generalized cohomology theory, \emph{e.g.} functoriality, homotopy invariance, and additivity.  Suitable restrictions on the kind of Hamiltonians one considers as points of $\mathfrak{GH}_d$ may be needed to dispense with systems that would not be physically relevant.

The missing, non-formal ingredient and the key to the program is the suspension isomorphism, which establishes a relationship between phases in dimension $d$ and those in $d+1$.  The general quantum pumping construction of Sec.~\ref{Sec:GeneralConstruct} is a concrete proposal for this isomorphism. It is a straightforward re-interpretation Kitaev's description of the homotopy equivalences $\GPinv_d \simeq \Omega \GPinv_{d+1}$ for the $\Omega$-spectrum of invertible gapped phases. But casting this problem as the construction of a map from $\tGPinv^d(X)$ to $\tGPinv^{d+1}(\Sigma X)$ makes it particularly concrete.

\section{General quantum pumping constructions}
\label{Sec:GeneralConstruct}

In this section we introduce and study constructions that start with a $d$-dimensional invertible system $H_d$ over $X$, and produce $(d+1)$-dimensional invertible systems $S H_d$ over the suspension $S X$ and $P H_d$ over $X \times S^1$. These constructions generalize the 1$d$ models over $S^3$ and $S^2 \times S^1$ introduced in Sections~\ref{Sec:1+1d_lattice} and~\ref{sec:chernpump}, respectively. In the language of the general constructions, one obtains those models by choosing the zero-dimensional system $H_0$ over $S^2$ to be a single spin-1/2 in a Zeeman magnetic field (see Sec.~\ref{sec:review}). Then one obtains 1$d$ systems $S H_0$ over $S(S^2) \cong S^3$ and $P H_0$ over $S^2 \times S^1$.

We propose that the construction $H_d \mapsto S H_d$ realizes the suspension isomorphism in a generalized cohomology theory of gapped invertible phases, as discussed in Sec.~\ref{sec:classification}. We thus also refer to this construction (distinguished from $H_d \mapsto P H_d$) as the suspension construction. Both constructions -- which are very similar to one another -- are described in Sec.~\ref{subsec:SP-constructions}. Physically, $P H_d$ can be understood as a pump of $H_d$ from a spatial boundary into the bulk, and -- roughly -- $S H_d$ can be understood in the same way. We make this precise in Sec.~\ref{subsec:general-bdy-physics}; some of the discussion there is a generalization of results on Chern number pumping described in Sec.~\ref{sec:chernpump}. In Sec.~\ref{subsec:universality} we discuss some expected properties of the constructions, that follow from assuming that $H_d \mapsto S H_d$ realizes the suspension isomorphism. These properties would imply in particular that the pumping phenomena of $S H_d$ and $P H_d$ are characteristic of non-trivial $(d+1)$-dimensional phases over $S X$ and $X \times S^1$.

$H_d$ may be either a bosonic or fermionic system, with or without some internal symmetry (including time-reversal). Spatial (\emph{i.e.} crystalline) symmetries of $H_d$ may introduce some subtleties, and while we expect much of our discussion will still apply, for simplicity we do not consider spatial symmetries.

The constructions of $S H_d$ and $P H_d$ are based on Kitaev's ideas that gapped invertible phases form an $\Omega$-spectrum in homotopy theory (see Sec.~\ref{sec:gc2}).\cite{kitaevSimonsCenter1,kitaevSimonsCenter2,kitaevIPAM} In particular, the key idea enabling these constructions is Kitaev's physical picture motivating the homotopy equivalences $\rho_d : \GPinv_d \to \Omega \GPinv_{d+1}$.

\subsection{Systems over $S X$ and $X \times S^1$}
\label{subsec:SP-constructions}

Here we start with a $d$-dimensional gapped invertible system $H_d$ over $X$, and construct $(d+1)$-dimensional systems $S H_d$ over $S X$ and $P H_d$ over $X \times S^1$. The strategy to construct both these systems will be to first construct a $(d+1)$-dimensional system $\widetilde{H}$ over $X \times I$ (with $I = [-1,1])$, and then exploit the fact that $SX$ and $X \times S^1$ can be obtained from $X \times I$ as a quotient space.

To explain the strategy in more detail, we discuss the case of $SX$, which we recall is the quotient space obtained from $X \times I$ by collapsing $X \times \{1\}$ and $X \times \{-1\}$ to single points. Let $\pi_S : X \times I \to SX$ be the corresponding quotient map. We construct $\widetilde{H}$ so that $\widetilde{H}(x,t)$ is independent of $x$ for $t = 1$ and also for $t = -1$. That is, $\widetilde{H}(x,t)$ is constant over the subspaces of $X \times I$ that are identified to single points in forming the quotient space $SX$. Therefore, we can think of $\widetilde{H}(x,t)$ as a function with domain $SX$, and we thus have a system over $SX$. More formally, there is a unique map $H_S : SX \to \GH_{d+1}$ that satisfies $\widetilde{H} = H_S \circ \pi_S$. The map $H_S$ defines the Hamiltonian of the desired system $S H_d$ over $SX$.

The case of $P H_d$ proceeds very similarly. We view $X \times S^1$ as the quotient space obtained from $X \times I$ by identifying $(x,1)$ with $(x,-1)$ for all $x \in X$, and let $\pi_P : X \times I \to X \times S^1$ be the corresponding quotient map. In this case, $\widetilde{H}$ is chosen to satisfy $\widetilde{H}(x,1) = \widetilde{H}(x,-1)$ for all $x \in X$, so we can think of $\widetilde{H}$ as a system over $X \times S^1$. Or, again more formally, there is a unique map $H_P : X \times S^1 \to \GH_{d+1}$ satisfying $\widetilde{H} = H_P \circ \pi_P$, and which defines the system $P H_d$.

\begin{figure}
    \centering
    \begin{tikzpicture}[x=0.75pt,y=0.75pt,yscale=-1,xscale=1]

\draw  [draw opacity=0][fill={rgb, 255:red, 155; green, 155; blue, 155 }  ,fill opacity=0.61 ] (80,70) -- (120.33,70) -- (120.33,110) -- (80,110) -- cycle ;
\draw    (80,60.33) -- (80.33,201) ;
\draw    (120.67,60) -- (121,200.67) ;
\draw    (160,59.33) -- (160.33,200) ;
\draw    (200.67,59) -- (201,199.67) ;
\draw  [draw opacity=0][fill={rgb, 255:red, 155; green, 155; blue, 155 }  ,fill opacity=0.61 ] (120,150.33) -- (160.33,150.33) -- (160.33,190.33) -- (120,190.33) -- cycle ;
\draw  [draw opacity=0][fill={rgb, 255:red, 155; green, 155; blue, 155 }  ,fill opacity=0.61 ] (160,69.67) -- (200.33,69.67) -- (200.33,109.67) -- (160,109.67) -- cycle ;

\draw  [color={rgb, 255:red, 208; green, 2; blue, 27 }  ,draw opacity=1 ] (112.67,200.67) .. controls (112.67,196.8) and (125.28,193.67) .. (140.83,193.67) .. controls (156.39,193.67) and (169,196.8) .. (169,200.67) .. controls (169,204.53) and (156.39,207.67) .. (140.83,207.67) .. controls (125.28,207.67) and (112.67,204.53) .. (112.67,200.67) -- cycle ;

\draw  [color={rgb, 255:red, 208; green, 2; blue, 27 }  ,draw opacity=1 ] (72.33,60) .. controls (72.33,56.13) and (84.94,53) .. (100.5,53) .. controls (116.06,53) and (128.67,56.13) .. (128.67,60) .. controls (128.67,63.87) and (116.06,67) .. (100.5,67) .. controls (84.94,67) and (72.33,63.87) .. (72.33,60) -- cycle ;

\draw  [color={rgb, 255:red, 208; green, 2; blue, 27 }  ,draw opacity=1 ] (34,200.67) .. controls (34,196.8) and (46.61,193.67) .. (62.17,193.67) .. controls (77.72,193.67) and (90.33,196.8) .. (90.33,200.67) .. controls (90.33,204.53) and (77.72,207.67) .. (62.17,207.67) .. controls (46.61,207.67) and (34,204.53) .. (34,200.67) -- cycle ;
\draw  [color={rgb, 255:red, 255; green, 255; blue, 255 }  ,draw opacity=1 ][fill={rgb, 255:red, 255; green, 255; blue, 255 }  ,fill opacity=1 ] (30.33,191.67) -- (64.17,191.67) -- (64.17,208.67) -- (30.33,208.67) -- cycle ;

\node at (90.5pt,45pt){$\bullet$};
\node at (60.0pt,45pt){$\bullet$};
\node at (120.0pt,45pt){$\bullet$};
\node at (150.5pt,45pt){$\bullet$};


\node at (77.0pt,45-13pt){$\tau^+_2$};
\node at (77.0+60pt,45-13pt){$\tau^+_2$};
\node at (77.0+30pt,45+117pt){$\tau^-_2$};
\node at (77.0-30pt,45+117pt){$\tau^-_2$};
\node at (77.0+85pt,45+117pt){$\tau^-_2$};

\node at (45.0pt,45pt){$\cdots$};
\node at (45.0pt,45+26pt){$\cdots$};

\begin{scope}[yshift=53pt]
\node at (90.5pt,45pt){$\bullet$};
\node at (60.0pt,45pt){$\bullet$};
\node at (120.0pt,45pt){$\bullet$};
\node at (150.5pt,45pt){$\bullet$};
\node at (45.0pt,45pt){$\cdots$};
\node at (45.0pt,45+26pt){$\cdots$};

\scriptsize
\node at (90.5+7pt,45pt){$H_d$};
\node at (60.0+7pt,45pt){$\overbar{H}_d$};
\node at (120.0+7pt,45pt){$\overbar{H}_d$};
\node at (150.5+7pt,45pt){$H_d$};
\end{scope}

\begin{scope}[yshift=106pt,xshift=0.2pt]
\node at (90.5pt,45pt){$\bullet$};
\node at (60.0pt,45pt){$\bullet$};
\node at (120.0pt,45pt){$\bullet$};
\node at (150.5pt,45pt){$\bullet$};

\node at (45.0pt,45pt){$\cdots$};
\end{scope}

\begin{scope}[xshift=118pt]
\draw  [color={rgb, 255:red, 208; green, 2; blue, 27 }  ,draw opacity=1 ] (34,200.67) .. controls (34,196.8) and (46.61,193.67) .. (62.17,193.67) .. controls (77.72,193.67) and (90.33,196.8) .. (90.33,200.67) .. controls (90.33,204.53) and (77.72,207.67) .. (62.17,207.67) .. controls (46.61,207.67) and (34,204.53) .. (34,200.67) -- cycle ;
\begin{scope}[xshift=22pt]
\draw  [color={rgb, 255:red, 255; green, 255; blue, 255 }  ,draw opacity=1 ][fill={rgb, 255:red, 255; green, 255; blue, 255 }  ,fill opacity=1 ] (30.33,191.67) -- (64.17,191.67) -- (64.17,208.67) -- (30.33,208.67) -- cycle ;
\end{scope}
\end{scope}

\begin{scope}[xshift=89pt,yshift=-106pt]
\draw  [color={rgb, 255:red, 208; green, 2; blue, 27 }  ,draw opacity=1 ] (34,200.67) .. controls (34,196.8) and (46.61,193.67) .. (62.17,193.67) .. controls (77.72,193.67) and (90.33,196.8) .. (90.33,200.67) .. controls (90.33,204.53) and (77.72,207.67) .. (62.17,207.67) .. controls (46.61,207.67) and (34,204.53) .. (34,200.67) -- cycle ;
\end{scope}



\begin{scope}[xshift=128pt,yshift=53pt]
\node at (45.0pt,45pt){$\cdots$};
\node at (45.0pt,45+26pt){$\cdots$};
\node at (45.0pt,45+52pt){$\cdots$};
\node at (45.0pt,45-26pt){$\cdots$};
\node at (45.0pt,45-52pt){$\cdots$};
\end{scope}

\begin{scope}[xshift=230pt,yshift=100pt]
\small
\draw [>=stealth,->](-40pt,55pt)--(-40pt,-65pt);
 \node at (-35pt,-67pt){$t$};
 
 \draw (-42pt,-56pt)--(-38pt,-56pt);
  \node at (-33pt,-56pt){$1$};

\draw (-42pt,-56+53pt)--(-38pt,-56+53pt);
\node at (-33pt,-56+53pt){$0$};


\draw (-42pt,-56+106pt)--(-38pt,-56+106pt);
\node at (-31pt,-56+106pt){$-1$}; 

 
 \end{scope}
\end{tikzpicture}
    \caption{Construction of $\widetilde{H}$ and $S H_d$. The vertical axis shows the parameter $t \in I = [-1,1]$ used to construct $\widetilde{H}$ over $X \times I$. At $t=0$, the system alternates between $d$-dimensional layers of systems $H_d$ and $\overbar{H}_d$ over $X$. For $t \in [0,1]$, these systems are coupled in pairs (shaded regions) and deformed to trivial systems $\tau_2^+$ at $t=1$. Similarly, systems are coupled via the opposite pairing in bilayers for $t \in [-1,0]$, with trivial systems $\tau_2^-$ at $t = -1$.}
    \label{fig:Ssd}
\end{figure}
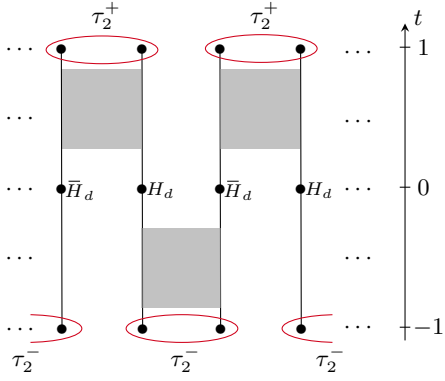

As illustrated in Fig.~\ref{fig:Ssd}, we now construct $\widetilde{H}$ as a one-dimensional lattice of $d$-dimensional layers, with layer index $i \in \Z$.  At $t = 0$, the layers are decoupled and alternate between $H_d$ and $\overbar{H}_d$, where $\overbar{H}_d$ is an inverse for $H_d$.  This gives the Hamiltonian
\begin{equation}
\label{Hd_t_0}
\widetilde{H}(x,0) = \sum_{i \in 2\Z} H_{d,i}(x) + \sum_{i \in 2\Z + 1} \overbar{H}_{d,i}(x) \text{.}
\end{equation}

Now consider a bilayer of adjacent $d$-dimensional layers, with layer indices $(i,i+1)$ and $i$ odd. At $t = 0$, we can think of this bilayer as the stack $\overbar{H}_d \ominus H_d$, so by construction it is in the trivial $d$-dimensional phase over $X$. There thus exists a deformation of $\overbar{H}_d \ominus H_d$ to a trivial system $\tau^+_{2}$. In other words, there exists a homotopy  $H^+_{i,i+1} : X \times [0,1] \to \GH_d$, between the Hamiltonians for $\overbar{H}_d \ominus H_d$ and $\tau^+_{2}$. (In order to construct such a homotopy it may first be necessary to stack $\overbar{H}_d \ominus H_d$ with a trivial system, but we can absorb this into the definition of $H_d$.) 
  We use this homotopy to construct $\widetilde{H}(x,t)$ for $t \in [0,1]$, letting
\begin{equation}
\widetilde{H}(x,t) = \sum_{i \in 2\Z+1} H^+_{i, i+1}(x,t) \text{,}
\end{equation}
where
\begin{equation}
H^+_{i,i+1}(x,0) = \overbar{H}_{d,i}(x) + H_{d,i+1}(x) \label{eqn:hplus1}
\end{equation}
and
\begin{equation}
H^+_{i,i+1}(x,1) = H_{\tau^+_{2}}(x) \text{,} \label{eqn:hplus2}
\end{equation}
the $x$-independent Hamiltonian for the trivial system $\tau^+_{2}$.

The Hamiltonian for $t \in [-1,0]$ is constructed similarly, but in terms of the opposite pairing into bilayers. That is, we consider bilayers $(i,i+1)$ with $i$ even. There exists a homotopy $H^-_{i,i+1}$ along which $H_d \ominus \overbar{H}_d$ deforms into $\tau^-_{2}$, a trivial system on the bilayer. The Hamiltonian for $t \in [-1,0]$ is given by
\begin{equation}
\widetilde{H}(x,t) = \sum_{i \in 2\Z} H^-_{i, i+1}(x,t) \text{,}
\end{equation}
where
\begin{equation}
H^-_{i,i+1}(x,0) = H_{d,i}(x) + \overbar{H}_{d,i+1}(x) \label{eqn:hminus1}
\end{equation}
and
\begin{equation}
H^-_{i,i+1}(x,-1) = H_{\tau^-_{2}}(x) \text{.} \label{eqn:hminus2}
\end{equation}
Note that we do not require any relationship between the homotopies $H^+$ and $H^-$. This completes the construction of $\widetilde{H}$.

We now observe that $\widetilde{H}(x,1)$ and $\widetilde{H}(x,-1)$ are independent of $x$, so as discussed above we obtain from $\widetilde{H}$ the system $S H_d$ over $SX$.

To instead obtain $P H_d$, we need to modify $\widetilde{H}$ so that $\widetilde{H}(x,1) = \widetilde{H}(x,-1)$. The necessary modification is to choose $\tau^{\pm}_{2}$  so that the $(d+1)$-dimensional trivial systems at $t = +1$ and $t = -1$ agree with one another. That is, we choose $\tau^{+}_{2} = \tau' \ominus \tau$ and $\tau^-_2 = \tau \ominus \tau'$, so that at $t = \pm 1$ we have the trivial system $\tau$ (resp. $\tau'$) on all even (resp. odd) layers. 

We note that the construction $H_d \mapsto S H_d$ (resp. $H_d \mapsto P H_d$) depends on more data than just the invertible system $H_d$, and so cannot be thought of as giving a function from $d$-dimensional invertible systems over $X$ to $(d+1)$-dimensional systems over $SX$ (resp. $X \times S^1$). Specifically, beyond the choice of $H_d$, one needs to choose the inverse $\overbar{H}_d$ and the homotopies $H^+$ and $H^-$.

\begin{figure}
    \centering
    \begin{tikzpicture}[x=0.75pt,y=0.75pt,yscale=-1,xscale=1]

\draw  [draw opacity=0][fill={rgb, 255:red, 155; green, 155; blue, 155 }  ,fill opacity=0.61 ] (80,70) -- (120.33,70) -- (120.33,110) -- (80,110) -- cycle ;
\draw    (80,60.33) -- (80.33,201) ;
\draw    (120.67,60) -- (121,200.67) ;
\draw    (160,59.33) -- (160.33,200) ;
\draw    (200.67,59) -- (201,199.67) ;
\draw  [draw opacity=0][fill={rgb, 255:red, 155; green, 155; blue, 155 }  ,fill opacity=0.61 ] (120,150.33) -- (160.33,150.33) -- (160.33,190.33) -- (120,190.33) -- cycle ;
\draw  [draw opacity=0][fill={rgb, 255:red, 155; green, 155; blue, 155 }  ,fill opacity=0.61 ] (160,69.67) -- (200.33,69.67) -- (200.33,109.67) -- (160,109.67) -- cycle ;

\node at (90.5pt,45pt){$\bullet$};
\node at (60.0pt,45pt){$\bullet$};
\node at (120.0pt,45pt){$\bullet$};
\node at (150.5pt,45pt){$\bullet$};


\begin{scope}[yshift=106pt,xshift=0.2pt]

\node at (90.5+7pt,45pt){$\tau$};
\node at (60.0+7pt,45-2pt){$\tau'$};
\node at (120.0+7pt,45-2pt){$\tau'$};
\node at (150.0+7pt,45-2pt){$\tau$};
\end{scope}

\node at (45.0pt,45pt){$\cdots$};
\node at (45.0pt,45+26pt){$\cdots$};

\node at (90.5+7pt,45pt){$\tau$};
\node at (60.0+7pt,45-2pt){$\tau'$};
\node at (120.0+7pt,45-2pt){$\tau'$};
\node at (150.5+7pt,45pt){$\tau$};

\begin{scope}[yshift=53pt]
\node at (90.5pt,45pt){$\bullet$};
\node at (60.0pt,45pt){$\bullet$};
\node at (120.0pt,45pt){$\bullet$};
\node at (150.5pt,45pt){$\bullet$};
\node at (45.0pt,45pt){$\cdots$};
\node at (45.0pt,45+26pt){$\cdots$};

\scriptsize
\node at (90.5+7pt,45pt){$H_d$};
\node at (60.0+7pt,45pt){$\overbar{H}_d$};
\node at (120.0+7pt,45pt){$\overbar{H}_d$};
\node at (150.5+7pt,45pt){$H_d$};
\end{scope}

\begin{scope}[yshift=106pt,xshift=0.2pt]
\node at (90.5pt,45pt){$\bullet$};
\node at (60.0pt,45pt){$\bullet$};
\node at (120.0pt,45pt){$\bullet$};
\node at (150.5pt,45pt){$\bullet$};

\node at (45.0pt,45pt){$\cdots$};
\end{scope}



\begin{scope}[xshift=128pt,yshift=53pt]
\node at (45.0pt,45pt){$\cdots$};
\node at (45.0pt,45+26pt){$\cdots$};
\node at (45.0pt,45+52pt){$\cdots$};
\node at (45.0pt,45-26pt){$\cdots$};
\node at (45.0pt,45-52pt){$\cdots$};
\end{scope}

\begin{scope}[xshift=230pt,yshift=100pt]
\small
\draw [>=stealth,->](-40pt,55pt)--(-40pt,-65pt);
 \node at (-35pt,-67pt){$t$};
 
 \draw (-42pt,-56pt)--(-38pt,-56pt);
  \node at (-33pt,-56pt){$1$};

\draw (-42pt,-56+53pt)--(-38pt,-56+53pt);
\node at (-33pt,-56+53pt){$0$};


\draw (-42pt,-56+106pt)--(-38pt,-56+106pt);
\node at (-31pt,-56+106pt){$-1$}; 

 
 \end{scope}
\end{tikzpicture}
    \caption{Construction of $P H_d$. As compared to the construction of $S H_d$ illustrated in Fig.~\ref{fig:Ssd}, the trivial systems $\tau_2^+$ and $\tau_2^-$ are chosen so that the Hamiltonian is periodic in $t \in [-1,1]$.}
    \label{fig:Psd}
\end{figure}
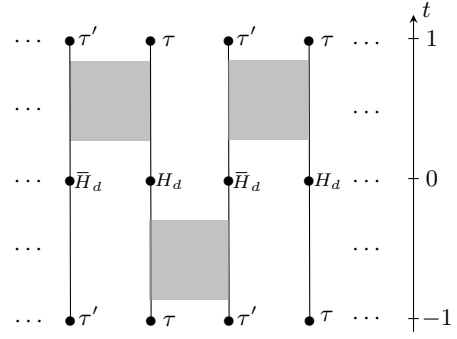

\subsection{Boundary physics and quantum pumping}
\label{subsec:general-bdy-physics}

Here we discuss the boundary physics of the systems $S H_d$ and $P H_d$, and in particular its interpretation in terms of pumping of the invertible phase $H_d$ from the boundary of the system into the bulk. Our use of the term ``pumping'' is somewhat loose in the case of $S H_d$, because the bulk does not depend periodically on the pumping parameter.

\begin{figure}
    \centering
   \begin{tikzpicture}[x=0.75pt,y=0.75pt,yscale=-1,xscale=1]
   \normalsize

\draw  [draw opacity=0][fill={rgb, 255:red, 155; green, 155; blue, 155 }  ,fill opacity=0.61 ] (80,70) -- (120.33,70) -- (120.33,110) -- (80,110) -- cycle ;
\draw    (80,60.33) -- (80.33,201) ;
\draw    (120.67,60) -- (121,200.67) ;
\draw    (160,59.33) -- (160.33,200) ;
\draw    (200.67,59) -- (201,199.67) ;
\draw  [dash pattern={on 4.5pt off 4.5pt}]  (180.67,49.67) -- (180.33,210) ;
\draw  [draw opacity=0][fill={rgb, 255:red, 155; green, 155; blue, 155 }  ,fill opacity=0.61 ] (120,150.33) -- (160.33,150.33) -- (160.33,190.33) -- (120,190.33) -- cycle ;
\draw  [draw opacity=0][fill={rgb, 255:red, 155; green, 155; blue, 155 }  ,fill opacity=0.61 ] (160,69.67) -- (200.33,69.67) -- (200.33,109.67) -- (160,109.67) -- cycle ;

\node at (90.5pt,45pt){$\bullet$};
\node at (60.0pt,45pt){$\bullet$};
\node at (120.0pt,45pt){$\bullet$};
\node at (150.5pt,45pt){$\bullet$};

\node at (90.5+7pt,45pt){$\tau$};
\node at (60.0+7pt,45-2pt){$\tau'$};
\node at (120.0+7pt,45-2pt){$\tau'$};
\node at (150.5+7pt,45pt){$\tau$};

\node at (45.0pt,45pt){$\cdots$};
\node at (45.0pt,45+26pt){$\cdots$};

\begin{scope}[yshift=53pt]
\node at (90.5pt,45pt){$\bullet$};
\node at (60.0pt,45pt){$\bullet$};
\node at (120.0pt,45pt){$\bullet$};
\node at (150.5pt,45pt){$\bullet$};
\node at (45.0pt,45pt){$\cdots$};
\node at (45.0pt,45+26pt){$\cdots$};

\scriptsize

\node at (90.5+7pt,45pt){$H_d$};
\node at (60.0+7pt,45pt){$\overbar{H}_d$};
\node at (120.0+7pt,45pt){$\overbar{H}_d$};
\node at (150.5+7pt,45pt){$H_d$};
\end{scope}

\small

\begin{scope}[yshift=106pt,xshift=0.2pt]
\node at (90.5pt,45pt){$\bullet$};
\node at (60.0pt,45pt){$\bullet$};
\node at (120.0pt,45pt){$\bullet$};
\node at (150.5pt,45pt){$\bullet$};

\node at (90.5+7pt,45pt){$\tau$};
\node at (60.0+7pt,45-2pt){$\tau'$};
\node at (120.0+7pt,45-2pt){$\tau'$};
\node at (150.5+7pt,45pt){$\tau$};

\node at (45.0pt,45pt){$\cdots$};
\end{scope}

\node at (106pt,165pt){$\text{bulk}$};
\node at (106+55pt,165pt){$\text{boundary}$};
\node at (106+52pt,175pt){$\text{layer}$};
\node at (150.8pt,125pt){$\mathbf{\times}$};

\begin{scope}[xshift=220pt,yshift=100pt]
\small
\draw [>=stealth,->](-40pt,55pt)--(-40pt,-65pt);
 \node at (-35pt,-67pt){$t$};
 
 \draw (-42pt,-56pt)--(-38pt,-56pt);
  \node at (-33pt,-56pt){$1$};

\draw (-42pt,-56+53pt)--(-38pt,-56+53pt);
\node at (-33pt,-56+53pt){$0$};

\draw (-42pt,-56+81pt)--(-38pt,-56+81pt);
\node at (-31pt,-56+81pt){$t_0$}; 

\draw (-42pt,-56+106pt)--(-38pt,-56+106pt);
\node at (-31pt,-56+106pt){$-1$}; 
 
 \end{scope}
 \small
 \node at (150.5+2pt,25pt){$i=N$};
  \node at (150.5+2pt,35pt){even};
\end{tikzpicture}
    \caption{Boundary physics of the system $P H_d$, in the case of boundary parameter space $X_{{\rm bdy}} = X \times S^1$. The dashed line shows the division of the system into bulk and a boundary layer, with the $i = N$ boundary layer immediately to the right. The bulk extends infinitely to the left. The boundary phase transition between $\tau$ and $H_d$ occurs at $t = t_0$, denoted by $\times$. 
    Shaded regions indicate coupling between neighboring layers.}
    \label{fig:Psd_pumping_periodic}
\end{figure}
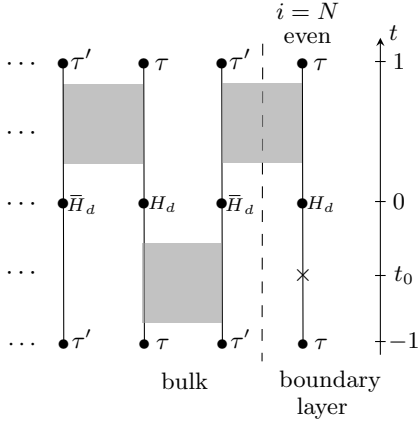

We first examine the boundary physics of $P H_d$, in the case of identical bulk and boundary parameter spaces (\emph{i.e.} $X_{{\rm bdy}} = X_{{\rm bulk}} = X \times S^1$), as illustrated in Fig.~\ref{fig:Psd_pumping_periodic}. We denote the coordinate of $S^1$ by $t \in [-1,1]$, identifying $t=1$ and $t=-1$. To introduce a boundary, we consider a semi-infinite lattice of $d$-dimensional layers, indexed by $i$ with $-\infty < i \leq N$, taking $N$ even.
For $t \in [0,1]$ the Hamiltonian is given by
\begin{equation}
H(x,t) = \sum_{i < N, i \text{ odd}} H^+_{i,i+1}(x,t) \text{.} \label{eqn:Hplus-P-bdy}
\end{equation}
For $t \in [-1,0]$, on the other hand, we modify the Hamiltonian on the boundary, \emph{i.e.} on the $i = N$ layer:
\begin{equation}
H(x,t) = \sum_{i < N, i \text{ even}} H^-_{i,i+1}(x,t) + H^{{\rm bdy}}_N(x,t) \text{.} \label{eqn:Hminus-P-bdy}
\end{equation}
Here $H^{{\rm bdy}}_N(x,t)$ is the Hamiltonian for the $i=N$ layer.

For consistency with the Hamiltonian chosen for $t \in [0,1]$, we must have
\begin{equation}
    H^{{\rm bdy}}_N(x,0) = H_{d,N}(x) \text{,}
\end{equation}
and
\begin{equation}
H^{{\rm bdy}}_N(x,-1) = H_{\tau,N}(x) \text{,}
\end{equation}
where $H_{\tau,N}$ is the Hamiltonian for the trivial system $\tau$.
That is, the $N$th layer goes from the trivial system $\tau$ at $t = -1$, to $H_d$ at $t=0$. Moreover, the $N$th layer is decoupled from the rest of the system for all $t \in [-1,0]$. Therefore, if $H_d$ is in a non-trivial phase, the boundary must become gapless or undergo a first-order phase transition for some $(x,t)$ with $t \in (-1,0)$. For concreteness, we assume the simplest scenario, where this only occurs for a single value $t_0 \in (-1,0)$.

Therefore, for small positive $\epsilon$, as $t$ increases from $t_0 + \epsilon$ to $t_0 -\epsilon$, we can think of the $d$-dimensional phase invariant of $H_d$ as being pumped into the bulk, leaving the boundary layer in the trivial phase over $X$. Note that we can just as well think of the inverse $\overbar{H}_d$ as being pumped \emph{from} the bulk and \emph{to} the boundary. This pumping can occur without closing a gap for $t \notin (t_0-\epsilon, t_0 + \epsilon)$ because, when the boundary layer is coupled to the bulk, $d$-dimensional phase invariants of the boundary layer are not well-defined. At the same time, these invariants are well-defined when the boundary layer is decoupled from the bulk for $t \in [-1,0]$, and the $d$-dimensional boundary layer phases at $t_0 \pm \epsilon$ can be meaningfully compared.

By enlarging the boundary parameter space, we can access essentially the same physics, but avoiding the complication of boundary phase transitions, as illustrated in Fig.~\ref{fig:Psd_pumping_open}. Keeping $X_{{\rm bulk}} = X \times S^1$, we take $X_{{\rm bdy}} = X \times I$, with $\pi_{{\rm bdy}} : X_{{\rm bdy}} \to X_{{\rm bulk}}$ the map that identifies the endpoints of $I = [-1,1]$. We proceed as above, choosing $H(x,t)$ to be of the same form as in Eqs.~(\ref{eqn:Hplus-P-bdy}) and~(\ref{eqn:Hminus-P-bdy}). We modify only the choice of $H^{{\rm bdy}}_N(x,t)$ for $t \in [-1,0]$, which we now take to be
\begin{equation}
H^{{\rm bdy}}_N(x,t) = H_{d, N}(x) \text{,}
\end{equation}
independent of $t$ (for $t \in [-1,0])$.  Clearly the gap remains open for all $(x,t) \in X \times I$.

\begin{figure}
    \centering
    \begin{tikzpicture}[x=0.75pt,y=0.75pt,yscale=-1,xscale=1]

\draw  [draw opacity=0][fill={rgb, 255:red, 155; green, 155; blue, 155 }  ,fill opacity=0.61 ] (80,70) -- (120.33,70) -- (120.33,110) -- (80,110) -- cycle ;
\draw    (80,60.33) -- (80.33,201) ;
\draw    (120.67,60) -- (121,200.67) ;
\draw    (160,59.33) -- (160.33,200) ;
\draw    (200.67,59) -- (201,199.67) ;
\draw  [dash pattern={on 4.5pt off 4.5pt}]  (180.67,49.67) -- (180.33,210) ;
\draw  [draw opacity=0][fill={rgb, 255:red, 155; green, 155; blue, 155 }  ,fill opacity=0.61 ] (120,150.33) -- (160.33,150.33) -- (160.33,190.33) -- (120,190.33) -- cycle ;
\draw  [draw opacity=0][fill={rgb, 255:red, 155; green, 155; blue, 155 }  ,fill opacity=0.61 ] (160,69.67) -- (200.33,69.67) -- (200.33,109.67) -- (160,109.67) -- cycle ;

\node at (90.5pt,45pt){$\bullet$};
\node at (60.0pt,45pt){$\bullet$};
\node at (120.0pt,45pt){$\bullet$};
\node at (150.5pt,45pt){$\bullet$};

\node at (90.5+7pt,45pt){$\tau$};
\node at (60.0+7pt,45-2pt){$\tau'$};
\node at (120.0+7pt,45-2pt){$\tau'$};
\node at (150.5+7pt,45pt){$\tau$};

\node at (45.0pt,45pt){$\cdots$};
\node at (45.0pt,45+26pt){$\cdots$};

\begin{scope}[yshift=53pt]
\node at (90.5pt,45pt){$\bullet$};
\node at (60.0pt,45pt){$\bullet$};
\node at (120.0pt,45pt){$\bullet$};
\node at (150.5pt,45pt){$\bullet$};
\node at (45.0pt,45pt){$\cdots$};
\node at (45.0pt,45+26pt){$\cdots$};

\scriptsize
\node at (90.5+7pt,45pt){$H_d$};
\node at (60.0+7pt,45pt){$\overbar{H}_d$};
\node at (120.0+7pt,45pt){$\overbar{H}_d$};
\node at (150.5+7pt,45pt){$H_d$};
\end{scope}

\small

\begin{scope}[yshift=106pt,xshift=0.2pt]
\node at (90.5pt,45pt){$\bullet$};
\node at (60.0pt,45pt){$\bullet$};
\node at (120.0pt,45pt){$\bullet$};
\node at (150.5pt,45pt){$\bullet$};

\node at (90.5+7pt,45pt){$\tau$};
\node at (60.0+7pt,45-2pt){$\tau'$};
\node at (120.0+7pt,45-2pt){$\tau'$};

\scriptsize
\node at (150.5+7pt,45pt){$H_d$};

\node at (45.0pt,45pt){$\cdots$};
\end{scope}

\node at (106pt,165pt){$\text{bulk}$};
\node at (106+55pt,165pt){$\text{boundary}$};
\node at (106+52pt,175pt){$\text{layer}$};

\begin{scope}[xshift=220pt,yshift=100pt]
\small
\draw [>=stealth,->](-40pt,55pt)--(-40pt,-65pt);
 \node at (-35pt,-67pt){$t$};
 
 \draw (-42pt,-56pt)--(-38pt,-56pt);
  \node at (-33pt,-56pt){$1$};

\draw (-42pt,-56+53pt)--(-38pt,-56+53pt);
\node at (-33pt,-56+53pt){$0$};


\draw (-42pt,-56+106pt)--(-38pt,-56+106pt);
\node at (-31pt,-56+106pt){$-1$}; 
 
 \end{scope}
 \small
 \node at (150.5+2pt,25pt){$i=N$};
  \node at (150.5+2pt,35pt){even};
\end{tikzpicture}
    \caption{Gapped quantum pumping at the boundary of $P H_d$, in the case of boundary parameter space $X_{{\rm bdy}} = X \times I$. The dashed line again shows the division of the system into bulk and a boundary layer, with the $i = N$ boundary layer immediately to the right.  The boundary layer begins at $t=-1$ as the $d$-dimensional system $H_d$, and ends -- after coupling to the bulk -- as the $d$-dimensional trivial system $\tau$ at $t=1$. Note while the bulk is periodic in $t$, the boundary layer is not.}
    \label{fig:Psd_pumping_open}
\end{figure}

This modified system realizes \emph{gapped quantum pumping}, where the $d$-dimensional phase invariant of $H_d$ is pumped into the bulk as $t$ is increased from $-1$ to $1$. 
At $t=-1$, the boundary layer is decoupled from the system and is the system $H_d$ over $X$. As $t$ is increased, bulk and boundary are coupled for $t \in (0,1)$, and at $t=1$ the boundary layer is in the trivial system $\tau$. Meanwhile, the bulk is periodic in $t$; that is, the bulk system over $X$ is the same for $t = -1$ and $t = 1$.

The boundary physics of the system $S H_d$ over $S X$ can also be somewhat loosely understood in terms of gapped quantum pumping. We have $X_{{\rm bulk}} = S X$, and take $X_{{\rm bdy}} = X \times I$, with $\pi_{{\rm bdy}} : X \times I \to SX$ the usual quotient map. Once again, $t \in I$ will play the role of time in a pumping process. Strictly speaking, referring to such a system as a pump is an abuse of terminology, because the bulk is not periodic in $t$. Nonetheless, the physics is very similar to that of $P H_d$, so we freely use the language of pumping. Alternatively, we can speak of a \emph{flow} of a $d$-dimensional phase invariant from the boundary into the bulk (or vice versa).

We again consider a semi-infinite lattice with $-\infty < i \leq N$ with $N$ even. For $t \in [-1,0]$, we take the Hamiltonian to be
\begin{equation}
H(x,t) = \sum_{i < N, i \text{ even}} H^-_{i,i+1}(x,t) + H^{{\rm bdy}}_N(x,t) \text{,}
\end{equation}
with $H^-_{i,i+1}(x,t)$ as in Eqs.~(\ref{eqn:hminus1}) and~(\ref{eqn:hminus2}), and $H^{{\rm bdy}}_N(x,t) = H_{d, N}(x)$, independent of $t$. It is important to note that this choice of $H^{{\rm bdy}}_N(x,t)$ may be $x$-dependent at $t=-1$, which would not have been well-defined if we had $X_{{\rm bdy}} = X_{{\rm bulk}} = SX$.  For $t \in [0,1]$ we have
\begin{equation}
H(x,t) = \sum_{i < N, i \text{ odd}} H^+_{i,i+1}(x,t) \text{,}
\end{equation}
where $H^+_{i,i+1}(x,t)$ is as in Eqs.~(\ref{eqn:hplus1}) and~(\ref{eqn:hplus2}). We write $\tau_2^+$ as a stack of trivial systems on bilayers, $\tau_2^+ = \tau' \ominus \tau$, as illustrated for the $N-1$st and $N$th layers in Fig.~\ref{fig:Ssd_pumping}.
We see that the $i=N$ boundary layer goes from the sytem $H_d$ at $t = -1$ to the trivial system $\tau$ at $t=1$, and we can think of the phase invariant of $H_d$ as flowing into the $(d+1)$-dimensional bulk. 

\begin{figure}
    \centering
    \begin{tikzpicture}[x=0.75pt,y=0.75pt,yscale=-1,xscale=1]

\draw  [draw opacity=0][fill={rgb, 255:red, 155; green, 155; blue, 155 }  ,fill opacity=0.61 ] (80,70) -- (120.33,70) -- (120.33,110) -- (80,110) -- cycle ;
\draw    (80,60.33) -- (80.33,201) ;
\draw    (120.67,60) -- (121,200.67) ;
\draw    (160,59.33) -- (160.33,200) ;
\draw    (200.67,59) -- (201,199.67) ;
\draw  [dash pattern={on 4.5pt off 4.5pt}]  (180.67,49.67) -- (180.33,210) ;
\draw  [draw opacity=0][fill={rgb, 255:red, 155; green, 155; blue, 155 }  ,fill opacity=0.61 ] (120,150.33) -- (160.33,150.33) -- (160.33,190.33) -- (120,190.33) -- cycle ;
\draw  [draw opacity=0][fill={rgb, 255:red, 155; green, 155; blue, 155 }  ,fill opacity=0.61 ] (160,69.67) -- (200.33,69.67) -- (200.33,109.67) -- (160,109.67) -- cycle ;

\draw  [color={rgb, 255:red, 208; green, 2; blue, 27 }  ,draw opacity=1 ] (112.67,200.67) .. controls (112.67,196.8) and (125.28,193.67) .. (140.83,193.67) .. controls (156.39,193.67) and (169,196.8) .. (169,200.67) .. controls (169,204.53) and (156.39,207.67) .. (140.83,207.67) .. controls (125.28,207.67) and (112.67,204.53) .. (112.67,200.67) -- cycle ;

\draw  [color={rgb, 255:red, 208; green, 2; blue, 27 }  ,draw opacity=1 ] (72.33,60) .. controls (72.33,56.13) and (84.94,53) .. (100.5,53) .. controls (116.06,53) and (128.67,56.13) .. (128.67,60) .. controls (128.67,63.87) and (116.06,67) .. (100.5,67) .. controls (84.94,67) and (72.33,63.87) .. (72.33,60) -- cycle ;

\draw  [color={rgb, 255:red, 208; green, 2; blue, 27 }  ,draw opacity=1 ] (34,200.67) .. controls (34,196.8) and (46.61,193.67) .. (62.17,193.67) .. controls (77.72,193.67) and (90.33,196.8) .. (90.33,200.67) .. controls (90.33,204.53) and (77.72,207.67) .. (62.17,207.67) .. controls (46.61,207.67) and (34,204.53) .. (34,200.67) -- cycle ;
\draw  [color={rgb, 255:red, 255; green, 255; blue, 255 }  ,draw opacity=1 ][fill={rgb, 255:red, 255; green, 255; blue, 255 }  ,fill opacity=1 ] (30.33,191.67) -- (64.17,191.67) -- (64.17,208.67) -- (30.33,208.67) -- cycle ;

\node at (90.5pt,45pt){$\bullet$};
\node at (60.0pt,45pt){$\bullet$};
\node at (120.0pt,45pt){$\bullet$};
\node at (150.5pt,45pt){$\bullet$};

\node at (120.0+7pt,45-2pt){$\tau'$};
\node at (150.5+7pt,45pt){$\tau$};

\node at (77.0pt,45-13pt){$\tau^+_2$};
\node at (77.0+30pt,45+117pt){$\tau^-_2$};
\node at (77.0-30pt,45+117pt){$\tau^-_2$};

\node at (45.0pt,45pt){$\cdots$};
\node at (45.0pt,45+26pt){$\cdots$};

\begin{scope}[yshift=53pt]
\node at (90.5pt,45pt){$\bullet$};
\node at (60.0pt,45pt){$\bullet$};
\node at (120.0pt,45pt){$\bullet$};
\node at (150.5pt,45pt){$\bullet$};
\node at (45.0pt,45pt){$\cdots$};
\node at (45.0pt,45+26pt){$\cdots$};

\scriptsize
\node at (90.5+7pt,45pt){$H_d$};
\node at (60.0+7pt,45pt){$\overbar{H}_d$};
\node at (120.0+7pt,45pt){$\overbar{H}_d$};
\node at (150.5+7pt,45pt){$H_d$};
\end{scope}

\begin{scope}[yshift=106pt,xshift=0.2pt]
\node at (90.5pt,45pt){$\bullet$};
\node at (60.0pt,45pt){$\bullet$};
\node at (120.0pt,45pt){$\bullet$};
\node at (150.5pt,45pt){$\bullet$};

\scriptsize
\node at (150.5+10pt,45pt){$H_d$};

\node at (45.0pt,45pt){$\cdots$};
\end{scope}

\node at (90pt,175pt){$\text{bulk}$};
\node at (106+55pt,168pt){$\text{boundary}$};
\node at (106+52pt,178pt){$\text{layer}$};

\begin{scope}[xshift=220pt,yshift=100pt]
\small
\draw [>=stealth,->](-40pt,55pt)--(-40pt,-65pt);
 \node at (-35pt,-67pt){$t$};
 
 \draw (-42pt,-56pt)--(-38pt,-56pt);
  \node at (-33pt,-56pt){$1$};

\draw (-42pt,-56+53pt)--(-38pt,-56+53pt);
\node at (-33pt,-56+53pt){$0$};


\draw (-42pt,-56+106pt)--(-38pt,-56+106pt);
\node at (-31pt,-56+106pt){$-1$}; 

\draw [blue](-69.5pt,51pt) circle (4.5pt);
 
 \end{scope}
 \small
 \node at (150.5+2pt,25pt){$i=N$};
  \node at (150.5+2pt,35pt){even};
\end{tikzpicture}
    \caption{Boundary physics of $S H_d$, in the case of boundary parameter space $X_{{\rm bdy}} = X \times I$. The dashed line again shows the division of the system into bulk and a boundary layer, with the $i = N$ boundary layer immediately to the right.  The boundary layer begins at $t=-1$ as the $d$-dimensional system $H_d$, and ends -- after coupling to the bulk -- as the $d$-dimensional trivial system $\tau$ at $t=1$. 
    A boundary termination in the case $X_{{\rm bdy}} = X_{{\rm bulk}}$ can be obtained from the same picture by choosing the Hamiltonian of the boundary layer to be strictly zero at $t = -1$ (blue circle), as described in the text.   
    }
    \label{fig:Ssd_pumping}
\end{figure}

Finally, we complete our discussion by considering the boundary physics of $S H_d$ with $X_{{\rm bdy}} = X_{{\rm bulk}} = SX$. The only modification needed to the case above is to ensure that $H^{{\rm bdy}}_N(x,t)$ is independent of $x$ at $t = -1$. In fact, if $H_d$ is a constant system over $X$, then no modification is needed. Therefore, if $H_d$ is constant, $S H_d$ admits a trivially gapped boundary to vacuum when $X_{{\rm bdy}} = X_{{\rm bulk}}$, so based on the bulk-boundary correspondence as formulated in Sec.~\ref{sec:parametrized}, we expect $S H_d$ is in the trivial phase in this case. (Indeed, without assuming the bulk-boundary correspondence, this can be straightforwardly shown by exhibiting a homotopy to a trivial system.) However, if $H_d$ is not constant over $X$, we need to modify the Hamiltonian for the boundary layer. One simple choice -- albeit one that results in highly fine-tuned boundary physics -- is to take (for $t \in [-1,0])$
\begin{equation}
H^{{\rm bdy}}_N(x,t) = (1+t) H_{d, N}(x) \text{.}
\end{equation}
This gives $H^{{\rm bdy}}_N(x,-1) = 0$, so that the boundary is gapless and in general highly degenerate at $t=-1$. While, more generically, this degeneracy must be split somehow, as long as $S H_d$ is in a non-trivial phase, the bulk-boundary correspondence tells us it should not be possible to trivially gap the boundary. As discussed below, we expect -- but have not yet established -- that $S H_d$ is in a non-trivial phase if $H_d$ is in a non-trivial reduced phase.

\subsection{Universality and properties of quantum pumping constructions}
\label{subsec:universality}

At this stage, it is not clear to what extent the various pumping phenomena above are universal, or are special non-universal features of the systems $P H_d$ and $S H_d$. This issue is closely related to a number of properties of the quantum pumping constructions that follow if we assume our proposal that $H_d \mapsto S H_d$ realizes the suspension isomorphism in a generalized cohomology theory of invertible phases. These properties are simple consequences of the above assumption, but we list them here explicitly to give a kind of roadmap to what one might want to prove to make progress towards establishing the generalized cohomology proposal. We note that property \#1 is straightforward to show if one assumes the bulk-boundary correspondence as stated in Sec.~\ref{sec:parametrized}.
If properties \#1-\#3 indeed hold, the quantum pumping phenomena of $SH_d$ and $PH_d$ are associated with non-trivial $(d+1)$-dimensional phases over $SX$ and $X \times S^1$, respectively.
These properties will not be used in the remainder of the paper, except as a part of the motivation for the construction of solvable models in $d > 1$.

\begin{enumerate}

     \item \emph{The constructions $H_d \mapsto S H_d$ and $H_d \mapsto P H_d$ give well-defined functions on invertible phases denoted $s_X : \GPinv^d(X) \to \GPinv^{d+1}(S X)$ and $p_X : \GPinv^d(X) \to \GPinv^{d+1}(X \times S^1)$.} This means that the phase of $S H_d$ (and of $P H_d$) is invertible and only depends on the phase of $H_d$. 
 
   \item \emph{The constructions respect the stacking operation, and thus $s_X$ and $p_X$ are homomorphisms of abelian groups.} Assuming \#1, this is easily shown to hold.

     \item \emph{$S H_d$ is in a non-trivial phase whenever $H_d$ is in a non-trivial reduced phase. Moreover, if $H_d$ is in a non-trivial (not necessarily reduced) phase, then $P H_d$ is in a non-trivial phase.} In other words, $p_X$ is injective, and $s_X$ is injective when restricted to the group of reduced phases $\GPinv^d(X, x_0) \subset \GPinv^d(X)$ for any basepoint $x_0 \in X$. We note that it is easily shown that if $H_d$ is a constant system (possibly in a non-trivial phase), then $S H_d$ is in the trivial phase.
    
    \item \emph{$s_X : \tGPinv^d(X) \to \tGPinv^{d+1}(SX) \cong \tGPinv^{d+1}(\Sigma X)$ is an isomorphism.} Given the above properties, this could be established by exhibiting a (two-sided) inverse construction, which may be possible based on  Kitaev's proposal for the homotopy inverse mapping $\Omega \GPinv_{d+1} \to \GPinv_d$.\cite{kitaevSimonsCenter1,kitaevSimonsCenter2,kitaevIPAM}
\end{enumerate}

\section{Higher Berry curvature flow and pumping in higher dimensions}
\label{Sec:HigherChernPump}

In this section, we consider the higher-dimensional descendants of 
the one-dimensional systems introduced in Sec.~\ref{sec:1dmodel-bbc} and Sec.~\ref{sec:chernpump}. That is, we discuss systems with nonzero KS number in $2d$ and higher dimensional lattice systems. These are $d$-dimensional systems over a parameter space which is a $(d+2)$-manifold. One family of such systems can be constructed by repeatedly applying the suspension construction to the $0d$ system over $S^2$ of a spin-$1/2$ in a Zeeman magnetic field.

We first give a review of $(d+2)$-form higher Berry curvature
and KS invariant in families of $d$-dimensional lattice systems  (Sec.~\ref{Sec:KS_higherD}).
In Sec.~\ref{sec:2dmodel}, based on the suspension construction as introduced in Sec.~\ref{Sec:GeneralConstruct}, we give an explicit construction of a $2d$ lattice system over $S^4$ that has non-zero KS number, and argue that its spatial boundaries are anomalous. We establish that the KS number of this system is nonzero in Sec.~\ref{Sec:2d_KS_bbc} via the bulk boundary correspondence, which also gives the higher Berry curvature $\Omega^{(4)}$ the interpretation of a flow of $1d$ higher Berry curvature to/from the spatial boundary. The corresponding interpretation also holds in arbitrary dimensions as discussed in Sec.~\ref{sec:ndKSnumber}, where we show that a gapped invertible $d$-dimensional system $H_d$ over a closed oriented $(d+2)$-manifold $X$ has the same KS number as the $(d+1)$-dimensional system $SH_d$ over $SX$ obtained from the suspension construction. In general $SX$ is not a topological manifold (except when $X$ is a sphere), so this requires us to generalize slightly the definition of KS number. Finally, we discuss the system $PH_d$ over $X \times S^1$ as a pump of KS number in Sec.~\ref{subSec:KSpump}.

\subsection{Review of higher Berry curvature in higher dimensions}
\label{Sec:KS_higherD}

Kapustin and Spodyneiko constructed the $(d+2)$-form higher Berry curvature and associated KS invariant in $d$-dimensional lattice systems.\cite{Kapustin_2020} Here we give a brief review focusing on those results needed below. 

The $2$-form $F^{(2)}_p$ in \eqref{F2_p} is related to a sequence of $n$-forms $F^{(n)}_{p_0 \cdots p_{n-2}}$ by the descent equations
\begin{equation}
\label{descentEq}
dF^{(n)}=\partial F^{(n+1)} \text{.}
\end{equation}
Here $n \geq 2$ is an integer, and $F^{(n)}$ is a short-hand notation for 
the $n$-form $F^{(n)}_{p_0\cdots p_{n-2}}$, which depends in a totally antisymmetric manner on the $n-1$ lattice sites $p_0, \dots, p_{n-2}$. Here $d$ is the usual exterior derivative, and the operator $\partial$ is defined by $(\partial F)_{p_1\cdots p_{n}}=\sum_{p_{0}} F_{p_0 p_1\cdots  p_{n}}$. Similar descent equations were found earlier by Kitaev in the context of Euclidean lattice systems.\cite{Kitaev2019differential} Below, we discuss how the higher Berry curvature $(d+2)$-form $\Omega^{(d+2)}$ is constructed from $F^{(d+2)}$.

In Ref.~\onlinecite{Kapustin_2020}, explicit formulas for $F^{(n)}$ were given that solve the descent equations and generalize the expression for $F^{(2)}$ given in \eqref{F2_p}. Taking $F^{(2)}$ as given by \eqref{F2_p}, there is an ambiguity in using the descent equations to obtain $F^{(n)}$ with $n>2$ that we now discuss. We imagine solving the descent equations iteratively, starting with $F^{(2)}$ and obtaining $F^{(3)}$, then obtaining $F^{(4)}$, and so on. At each stage of this procedure, $F^{(n)}$ is determined in terms of $F^{(n-1)}$ up to the ambiguity
\begin{equation}
F^{(n)}_{p_0 \cdots p_{n-2}} \to F^{(n)}_{p_0 \cdots p_{n-2}} + B^{(n)}_{p_0 \cdots p_{n-2}} \text{,}
\end{equation}
where $\partial B^{(n)} = 0$. Kapustin and Spodyneiko have argued that the equation $\partial B^{(n)} = 0$ has only ``trivial'' solutions of the form $B^{(n)} = \partial C^{(n)}$.\cite{kapustin_private_comm} If we shift $F^{(n)} \to F^{(n)} + \partial C^{(n)}$, the descent equation at level $n$ is modified to
\begin{equation}
d F^{(n)} = \partial( F^{(n+1)} - d C^{(n)} ) \text{.}
\end{equation}
This change can be compensated by shifting $F^{(n+1)} \to F^{(n+1)} + d C^{(n)}$, which does not affect descent equations at higher levels since $d^2 = 0$. It is also natural to allow for an ambiguity in $F^{(2)}$, namely $F^{(2)} \to F^{(2)} + \partial C^{(2)}$, since in a $d=0$ system this does not affect the Berry curvature $\Omega^{(2)} = \sum_q F^{(2)}_q$. Therefore, given a sequence $F^{(n)}$ ($n \geq 2$) of solutions to the descent equations, the full ambiguity is
\begin{equation}
F^{(2)} \to F^{(2)} + \partial C^{(2)} \text{,}
\end{equation}
and, for $n > 2$,
\begin{equation}
F^{(n)} \to F^{(n)} + d C^{(n-1)} + \partial C^{(n)} \text{.} \label{eqn:full-ambiguity}
\end{equation}

Based on the discussion of Appendix~\ref{app:locality}, we expect that  $F^{(n)}_{p_0 \cdots p_{n-2}}$ is a local quantity, in a sense that we now describe.  First of all, for $n = 2$ and $n = 3$, $F^{(n)}_{p_0 \cdots p_{n-2}}$ is expressed as a sum of imaginary-time-ordered correlation functions of $n$ local operators, where $n-1$ of the local operators are supported near the lattice sites $p_0, \dots, p_{n-2}$, and the remaining local operator comes from a sum of local operators over all sites.  It is expected that similar expressions hold for all $n$, but become more complicated as $n$ increases. Moreover, it is also expected that $F^{(n)}_{p_0\cdots p_{n-2}}$ decays exponentially if any two points of $p_0,\cdots, p_{n-2}$ are far away from each other compared to the correlation length of the system.\cite{Kapustin_2020} 
For $n=2$ this is an empty statement, and for $n=3$ it follows from the discussion of Appendix~\ref{app:locality}, and in fact is proved rigorously based on Ref.~\onlinecite{Watanabe_2018}.
For each $n > 3$ this statement could in principle be shown by expressing $F^{(n)}_{p_0 \cdots p_{n-2}}$ in terms of imaginary-time-ordered correlation functions and studying the resulting expression, although this brute-force method will rapidly become impractical with increasing $n$. We leave a careful study of these issues for future work.

To construct the higher Berry curvature, it is convenient to employ the language of homology and cohomology. More specifically, as noted in Ref.~\onlinecite{Kapustin_2020}, the relevant mathematical theory appears to be that of coarse homology/cohomology,\cite{roe2003lectures} although connecting this theory with the treatment of Ref.~\onlinecite{Kapustin_2020} in a fully precise way is an open problem. Here we give only a very brief description of some key aspects of the chain and cochain complexes that arise. 
A more thorough discussion can be found in Ref.~\onlinecite{Kapustin_2020}.  

We consider a chain complex whose $k$-chains are totally antisymmetric functions $A_{p_0 \cdots p_{k}}$ of $k+1$ lattice sites, taking values in some real vector space $V$, which is a fixed property of a given chain complex (\emph{i.e.} the vector space $V$ is the same for all chains in a given complex, independent of the chain degree). It is required that $k$-chains decay exponentially away from the diagonal $p_0 = \cdots = p_{k}$. As above, $\partial$ is given by $(\partial A)_{p_1 \cdots p_{k}} = \sum_{p_0} A_{p_0 \cdots p_{k}}$; this lowers the chain degree by one, and it is easily seen that $\partial^2 = 0$. Now, we view $F^{(n)}$ as a chain of degree $k = n-1$, in a chain complex where the vector space $V$ is the space of differential $n$-forms on the parameter space $X$. It should be noted that $F^{(n)}$ and $F^{(m)}$ for $n \neq m$ belong to different chain complexes with different vector spaces of $n$-forms and $m$-forms, respectively.

The $k$-cochains are bounded, totally antisymmetric real-valued functions of $k+1$ lattice sites. This is similar to the definition of $k$-chains, with $V = \R$. However, an important difference between chains and cochains is that the exponential decay condition on chains is replaced with a condition on the support of cochains described in Ref.~\onlinecite{Kapustin_2020}.
There is a pairing between $k$-chains and $k$-cochains given by
\begin{equation}
\langle A,\alpha\rangle=\frac{1}{(k+1)!}\sum_{p_0,\cdots,p_k}A_{p_0,\cdots,p_k}\alpha(p_0,\cdots,p_k).
\end{equation}
There is a coboundary operator $\delta$ on cochains, so that if $\alpha$ is an $k$-cochain, $\delta \alpha$ is a $(k+1)$-cochain, and $\delta^2 = 0$. The above pairing satisfies 
\begin{equation}
\label{pairing_bdy}
\langle \partial A,\, \alpha\rangle=\langle A,\,\delta\alpha\rangle.
\end{equation}

Using this language, the higher Berry curvature $\Omega^{(d+2)}$ is a $(d+2)$-form on $X$ given by the following formula:\cite{Kapustin_2020}
\begin{equation}
\label{d+2_form}
\Omega^{(d+2)}(f_1,\cdots,f_d)=
\langle
F^{(d+2)}, \delta f_1\cup \cdots \cup \delta f_d
\rangle \text{.}
\end{equation}
Here, $\delta f_1\cup \cdots \cup \delta f_d$ is a $d$-cochain that we now describe. We choose $d$ bounded, real-valued functions of a single lattice site, with values written $f_{\mu}(p)$ ($\mu = 1,\dots,d$). We assume that $f_{\mu}(p)=f(x_{\mu}(p))$ and $x_{\mu}(p)$ is the $\mu$-coordinate of $p$. Moreover we assume that $f_{\mu}(p)=0$ for $x_{\mu}(p)\ll 0$ and $f_{\mu}(p)=1$ for $x_{\mu}(p)\gg 0$. For simplicity, one may choose step functions $f_{\mu}(p)=\Theta(x_{\mu}(p)-a_{\mu})$, where $a_{\mu}\in \mathbb Z+1/2$.
From each of these functions we construct a 1-cocycle $\delta f_{\mu}$, defined as\textbf{}
\begin{equation}
(\delta f_{\mu})(p,q)=f_{\mu}(q)-f_{\mu}(p) \text{.}
\end{equation}
Even though we use the notation of the coboundary operator, $\delta f_{\mu}$ is not exact, because $f_{\mu}$ does not have finite support and thus does not satisfy all the conditions to be a 0-cochain. However $\delta f_{\mu}$ is a cocycle, \emph{i.e.} $\delta (\delta f_{\mu}) = 0$. Finally $\delta f_1\cup \cdots \cup \delta f_d$ is a $d$-cocycle obtained as a cup product of 1-cocycles as defined in Ref.~\onlinecite{Kapustin_2020}.

The higher Berry curvature is a closed form, \emph{i.e.} $d \Omega^{(d+2)} = 0$, which is easily checked using Eqs.~\eqref{descentEq}, \eqref{d+2_form}, and \eqref{pairing_bdy}. We also note that the ambiguity in $F^{(d+2)}$ given in Eq.~\eqref{eqn:full-ambiguity} results in the shift
\begin{equation}
\Omega^{(d+2)} \to \Omega^{(d+2)} + d \langle C^{(d+1)} , \delta f_1\cup \cdots \cup \delta f_d
\rangle \text{.}
\end{equation}
This is a shift of the higher Berry curvature by an exact form, so the cohomology class of $\Omega^{(d+2)}$ is unaffected. In addition, the cohomology class does not depend on the choice of the functions $f_{\mu}$ ($1\le \mu \le d$) in \eqref{d+2_form}, as long as 
$f_{\mu}(p)=0$ for $x_{\mu}(p)\ll 0$ and $f_{\mu}(p)=1$ 
for $x_{\mu}(p)\gg 0$. In more detail, if two choices $f_{\mu}(p)$ and $f'_{\mu}(p)$ differ by a function $g_{\mu}(p)$ with compact support, then one can find $\Omega^{(d+2)}(f_1,\cdots,f_{\mu},\cdots,f_d)$ and
$\Omega^{(d+2)}(f_1,\cdots,f'_{\mu},\cdots,f_d)$ differ by an exact form $d\omega^{(d+1)}$.

The KS invariant is precisely the cohomology class $[\Omega^{(d+2)}/2\pi]$. If $X$ is an oriented $(d+2)$-dimensional manifold, then the KS number can be defined as
\begin{equation}
KS=\int_{X}\Omega^{(d+2)}.
\end{equation}
For invertible systems, it is argued that the KS number over $X=S^{d+2}$
is always quantized in units of $2\pi$.\cite{Kapustin_2020} Therefore, since $F^{(d+2)}$ and hence $\Omega^{(d+2)}$ depends continuously on $x \in X$, the KS number can be considered as 
an obstruction to deforming a system over $S^{d+2}$ to a constant system. For 
invertible systems over general closed oriented differentiable manifolds, it is also expected that the KS number is quantized in units of $2\pi$, though we are not aware of a proof.

\subsection{$2d$ lattice model over $S^4$}
\label{sec:2dmodel}

Now we construct a non-trivial $2d$ lattice system $H_{2d}$ over $S^4$ based on the suspension construction as introduced in Sec.~\ref{Sec:GeneralConstruct}. To make the suspension construction, we need an input 1$d$ system, and we choose $H_{1d}$ over $S^3$ introduced in Sec.~\ref{sec:1dmodel-bbc}. 
We then argue that a spatial boundary of $H_{2d}$ is anomalous, suggesting that $H_{2d}$ is in a non-trivial phase over $S^4$. This is confirmed in Sec.~\ref{Sec:2d_KS_bbc} where the KS number is found to be $2\pi$.

We consider a $2d$ square lattice with sites labeled by $p = (x_1, x_2) \in \Z \times \Z$, and place a single qubit on each site.  Given a lattice site $p$, sometimes it is convenient to denote the two coordinates by $x_1(p)$ and $x_2(p)$.  The parameter space is $S^4$, and we denote points $w \in S^4$ by $w = (w_1,\dots,w_5)$ with $\sum_{i = 1}^5 w_i^2 = 1$.  On the ``equator'' $S^3 \subset S^4$ defined by $w_5 = 0$, the system is chosen to be a lattice of $1d$ layers, with each layer a column extending along the $x_2$-direction.  The layers alternate between the systems $H_{1d}$ (for $x_1$ even) and $\overbar{H}_{1d}$ (for $x_1$ odd) over $S^3$.

The Hamiltonian is
\begin{equation}
\label{Hamiltonian2d}
\small
\begin{split}
&H_{2d}(w)=\sum_{x_1 \in 2\mathbb Z} H_{1d, x_1}(w)+\sum_{x_1 \in 2\mathbb Z+1} \overbar{H}_{1d,x_1}(w)\\
&+\sum_{x_1\in2\mathbb Z+1, x_2\in \mathbb Z} H^{2,+}_{(x_1,x_2); (x_1+1,x_2)}(w)\\
&+\sum_{x_1\in2\mathbb Z, x_2\in\mathbb Z}H^{2,-}_{(x_1,x_2); (x_1+1,x_2)}(w).
\end{split}
\end{equation}
Here the first two terms are Hamiltonians for the $1d$ layers summed over even and odd $x_1$ coordinates, with $H_{1d,x_1}(w)$ and $\overbar{H}_{1d,x_1}(w)$  defined in 
\eqref{H_1d} and \eqref{H_1d_inverse}, respectively. While these Hamiltonians are defined for $w \in S^3$ with $w = (w_1, \dots, w_4)$, they are given explicitly as functions of the coordinates $w_1, \dots, w_4$, which continue to make sense when $w_1^2 + \cdots + w_4^2 < 1$.  Note in particular that $H_{1d,x_1}(w) = \overbar{H}_{1d,x_1}(w) = 0$ when $w_5 = \pm 1$.  

The remaining terms couple neighboring layers and are 
\begin{equation}
\label{V_couple}
H_{(x_1,x_2);(x_1+1,x_2)}^{2,\pm}(w)=h^{\pm}(w)\sum_{\mu=1,2,3}\sigma_{x_1,x_2}^{\mu}\sigma_{x_1+1,x_2}^{\mu},
\end{equation}
where $\sigma_{x_1,x_2}^{1,2,3}$ are Pauli matrices of the qubit at site $(x_1,x_2)$.
The real functions $h^{\pm}(w)$ are chosen as follows:
\begin{equation}
\label{h+}
h^+(w)=
\begin{cases}
w_5, \quad& 0\le w_5\le 1,\\
0, \quad &\text{otherwise, }
\end{cases}
\end{equation}
and
\begin{equation}
\label{h-}
h^-(w)=
\begin{cases}
-w_5, \quad&-1\le w_5\le 0,\\
0, \quad &\text{otherwise}.
\end{cases}
\end{equation}
It can be checked that $H_{2d}(w)$ is gapped everywhere on $S^4$ (see Appendix~\ref{app:fourspin}).

Similar to the $1d$ case as discussed in Sec.~\ref{Sec:1+1d_lattice}, 
$H_{2d}(w)$ is not smooth at $w_4=0$ and $w_5=0$,
which may result in a discontinuous distribution of higher Berry curvature. One can always smooth out the functions $g^{\pm}(w)$ and 
$h^{\pm}(w)$ at $w_4=0$ and $w_5=0$, respectively, and then the 
higher Berry curvature becomes continuous at $w_{4(5)}=0$.
Here we are mainly interested in the cohomology class of the higher Berry curvature as measured by the KS number, and therefore the discontinuity of the higher Berry curvature will not cause any problems.

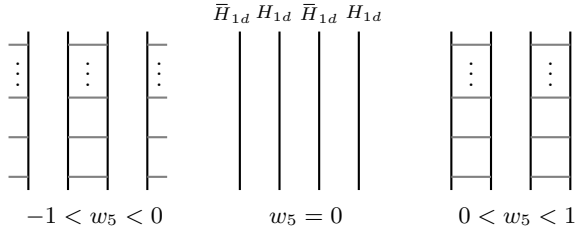
\begin{figure}
\begin{tikzpicture}[baseline={(current bounding box.center)}]

\small

\begin{scope}[yshift=50pt,xshift=160pt]

\draw [thick](0pt,0pt)--(0pt,60pt);
\draw [thick](15pt,0pt)--(15pt,60pt);

\draw [thick](30pt,0pt)--(30pt,60pt);
\draw [thick](45pt,0pt)--(45pt,60pt);

\draw [thick][gray](0pt,5pt)--(15pt,5pt);
\draw [thick][gray](0pt,20pt)--(15pt,20pt);
\draw [thick][gray](0pt,35pt)--(15pt,35pt);
\node at (7.5pt,47pt){$\vdots$};
\draw [thick][gray](0pt,55pt)--(15pt,55pt);

\draw [thick][gray](0+30pt,5pt)--(15+30pt,5pt);
\draw [thick][gray](0+30pt,20pt)--(15+30pt,20pt);
\draw [thick][gray](0+30pt,35pt)--(15+30pt,35pt);
\node at (7.5+30pt,47pt){$\vdots$};
\draw [thick][gray](0+30pt,55pt)--(15+30pt,55pt);

\node at (25pt,-10pt){$0< w_5<1$ };

\scriptsize

\end{scope}

\begin{scope}[yshift=50pt,xshift=80pt]

\draw [thick](0pt,0pt)--(0pt,60pt);
\draw [thick](15pt,0pt)--(15pt,60pt);

\draw [thick](30pt,0pt)--(30pt,60pt);
\draw [thick](45pt,0pt)--(45pt,60pt);

\node at (25pt,-10pt){$ w_5=0$ };

\scriptsize
\node at (47pt,67pt){$H_{1d}$ };
\node at (30pt,67pt){$\overbar{H}_{1d}$ };
\node at (13pt,67pt){$H_{1d}$ };
\node at (-3pt,67pt){$\overbar{H}_{1d}$ };

\end{scope}

\begin{scope}[yshift=50pt]

\draw [thick](0pt,0pt)--(0pt,60pt);
\draw [thick](15pt,0pt)--(15pt,60pt);

\draw [thick](30pt,0pt)--(30pt,60pt);
\draw [thick](45pt,0pt)--(45pt,60pt);

\draw [thick][gray](45-52.5pt,5pt)--(7.5+45-52.5pt,5pt);
\draw [thick][gray](45-52.5pt,20pt)--(7.5+45-52.5pt,20pt);
\draw [thick][gray](45-52.5pt,35pt)--(7.5+45-52.5pt,35pt);
\node at (3.5+45-52.5pt,47pt){$\vdots$};
\draw [thick][gray](0+45-52.5pt,55pt)--(7.5+45-52.5pt,55pt);

\draw [thick][gray](0+15pt,5pt)--(15+15pt,5pt);
\draw [thick][gray](0+15pt,20pt)--(15+15pt,20pt);
\draw [thick][gray](0+15pt,35pt)--(15+15pt,35pt);
\node at (7.5+15pt,47pt){$\vdots$};
\draw [thick][gray](0+15pt,55pt)--(15+15pt,55pt);

\draw [thick][gray](45pt,5pt)--(7.5+45pt,5pt);
\draw [thick][gray](45pt,20pt)--(7.5+45pt,20pt);
\draw [thick][gray](45pt,35pt)--(7.5+45pt,35pt);
\node at (4.5+45pt,47pt){$\vdots$};
\draw [thick][gray](0+45pt,55pt)--(7.5+45pt,55pt);

\node at (25pt,-10pt){$-1< w_5<0$ };

\scriptsize

\end{scope}

\end{tikzpicture}
\caption{Illustration of the dependence on $w_5$ of the $2d$ lattice system $H_{2d}$ of \eqref{Hamiltonian2d}. At $w_5=0$, the $2d$ lattice is composed of decoupled 
$1d$ lattice systems $H_{1d}$ and $\overbar{H}_{1d}$ running along the vertical direction. These $1d$ systems are coupled in neighboring pairs for $w_5 \neq 0$, with two different pairings appearing, depending on the sign of $w_5$.}
\label{Fig_2d_lattice}
\end{figure}

The dependence of the Hamiltonian on $w_5$ is visualized in Fig.\ref{Fig_2d_lattice}.
At $w_5 = 0$, the system is composed of decoupled $1d$ quantum systems over the equatorial $S^3 \subset S^4$. These $1d$ systems are characterized by a KS number of 
$2\pi$  (resp. $-2\pi$) on even (resp. odd) layers.

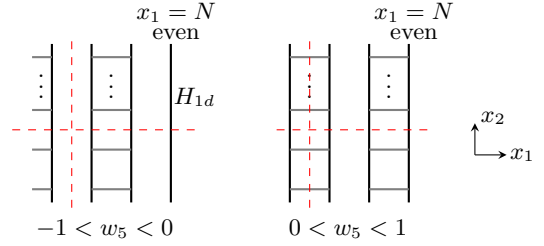
\begin{figure}[htp]
\begin{tikzpicture}[baseline={(current bounding box.center)}]
\small

\begin{scope}[yshift=50pt]

\draw [thick](0pt,0pt)--(0pt,60pt);
\draw [thick](15pt,0pt)--(15pt,60pt);

\draw [thick](30pt,0pt)--(30pt,60pt);
\draw [thick](45pt,0pt)--(45pt,60pt);

\draw [thick][gray](45-52.5pt,5pt)--(7.5+45-52.5pt,5pt);
\draw [thick][gray](45-52.5pt,20pt)--(7.5+45-52.5pt,20pt);
\draw [thick][gray](45-52.5pt,35pt)--(7.5+45-52.5pt,35pt);
\node at (3.5+45-52.5pt,47pt){$\vdots$};
\draw [thick][gray](0+45-52.5pt,55pt)--(7.5+45-52.5pt,55pt);

\draw [thick][gray](0+15pt,5pt)--(15+15pt,5pt);
\draw [thick][gray](0+15pt,20pt)--(15+15pt,20pt);
\draw [thick][gray](0+15pt,35pt)--(15+15pt,35pt);
\node at (7.5+15pt,47pt){$\vdots$};
\draw [thick][gray](0+15pt,55pt)--(15+15pt,55pt);

\node at (20pt,-10pt){$-1< w_5<0$ };

\small
\node at (47pt,72pt){$x_1=N$ };
\node at (47pt,64pt){$\text{even}$ };

\small
\node at (54pt,40pt){$H_{1d}$ };

\draw [red][dashed](7.5pt,-2pt)--(7.5pt,62pt);
\draw [red][dashed](-15pt,27.5pt)--(55pt,27.5pt);

\end{scope}

\begin{scope}[xshift=160pt,yshift=60pt]
\draw [>=stealth,->] (0pt,8pt)--(12pt,8pt);
\draw [>=stealth,->] (0pt,8pt)--(0pt,20pt);

\node at (18pt,8pt){$x_1$ };
\node at (7pt,22pt){$x_2$ };

\end{scope}

\begin{scope}[yshift=50pt,xshift=90pt]

\draw [thick](0pt,0pt)--(0pt,60pt);
\draw [thick](15pt,0pt)--(15pt,60pt);

\draw [thick](30pt,0pt)--(30pt,60pt);
\draw [thick](45pt,0pt)--(45pt,60pt);

\draw [thick][gray](0pt,5pt)--(15pt,5pt);
\draw [thick][gray](0pt,20pt)--(15pt,20pt);
\draw [thick][gray](0pt,35pt)--(15pt,35pt);
\node at (7.5pt,47pt){$\vdots$};
\draw [thick][gray](0pt,55pt)--(15pt,55pt);

\draw [thick][gray](0+30pt,5pt)--(15+30pt,5pt);
\draw [thick][gray](0+30pt,20pt)--(15+30pt,20pt);
\draw [thick][gray](0+30pt,35pt)--(15+30pt,35pt);
\node at (7.5+30pt,47pt){$\vdots$};
\draw [thick][gray](0+30pt,55pt)--(15+30pt,55pt);

\node at (22pt,-10pt){$0< w_5<1$ };

\scriptsize

\draw [red][dashed](7.5pt,-2pt)--(7.5pt,62pt);
\draw [red][dashed](-5pt,27.5pt)--(55pt,27.5pt);

\small
\node at (47pt,72pt){$x_1=N$ };
\node at (47pt,64pt){$\text{even}$ };

\end{scope}
\end{tikzpicture}
\caption{The $2d$ lattice model with a spatial boundary terminated at the layer $x_1=N\in 2\mathbb Z$.
The red dashed lines indicate the steps in the functions 
$f_1(p)$ and $f_2(p)$.
}
\label{f1f2}
\end{figure}

Now we impose a spatial boundary by truncating the lattice to a half-plane, retaining only those sites with $x_1 \leq N$ (see Fig.~\ref{f1f2}).  All Hamiltonian terms coupling to sites with $x_1>N$ are dropped, and all other terms are retained unmodified. This choice of boundary termination does not enlarge the parameter space at the boundary, \emph{i.e.} we have $X_\text{bulk} = X_{\text{bdy}} = S^4$ in the notation of Sec.~\ref{sec:parametrized}.

If $N$ is even, the $x_1 = N$ boundary layer is decoupled from the bulk for $w_5\le 0$ (see Fig.~\ref{f1f2}). The energy spectrum of the $2d$ system with boundary is gapped everywhere except at $w_5 = -1$, where the Hamiltonian for the boundary layer goes to zero and the system is gapless.  Similarly, if $N$ is odd, the boundary layer is decoupled from the bulk for $w_5\ge 0$, and there is a single gapless point at the opposite pole $w_5=1$.

The gapless point in the system with boundary is closely analogous to the gapless Weyl point appearing when $H_{1d}$ over $S^3$ is studied in a semi-infinite system, as discussed in Sec.~\ref{Sec:1+1d_lattice}.  In the $1d$ case, we argued that the $0d$ spatial boundary is anomalous because a single Weyl point cannot occur for a strictly $0d$ system over $S^3$.  This also led us to expect that the bulk $1d$ system is in a non-trivial phase over $S^3$.

We now give a parallel discussion for the $2d$ lattice model, leading to the conclusion that the $1d$ boundary is anomalous. For concreteness we take $N$ even.  We view the boundary as an effective $1d$ system, and note that it is gapless 
\textit{only} at the pole $w_5=-1$, where $w_1=w_2=w_3=w_4=0$.  Now we consider the family of $S^3$ subspaces defined by fixing $w_5$, for any $w_5 \in (-1,0]$.  For such values of $w_5$, the $x_1 = N$ boundary layer is decoupled from the bulk, and has a well-defined higher Berry curvature $\Omega^{(3)}$. As studied in Sec.~\ref{sec:hbc-1d}~one has $\int_{S^3}\Omega^{(3)}=2\pi$. 
However, we can shrink the $S^3$ subspace to a point (for instance by continuously increasing the fixed value of $w_5$ until $w_5 = 1$), upon which we must have $\int_{S^3}\Omega^{(3)}=0$.  In a strictly $1d$ system this is a contradiction, because the value of $\int_{S^3}\Omega^{(3)}$ is quantized.  Therefore the $1d$ boundary is anomalous, as it exhibits behavior impossible in a strictly $1d$ system. As we will see below in Sec.~\ref{Sec:2d_KS_bbc}, this anomaly is related to the topologically nontrivial nature of the $2d$ bulk.

\subsection{Calculation of KS invariant from bulk-boundary correspondence}
\label{Sec:2d_KS_bbc}

Now we study the KS number of the $2d$ lattice system $H_{2d}$ over $S^4$.
For a $2d$ lattice system defined on an infinite plane, 
according to \eqref{d+2_form}, the 4-form higher Berry curvature can be expressed as
\begin{equation}
\label{Omega4}
\Omega^{(4)}
=\big\langle F^{(4)},
\delta f_1\cup \delta f_2
\big\rangle .
\end{equation}
We choose $f_1$ and $f_2$ to be step functions depending on $x_1$ and $x_2$, respectively, namely $f_{1,2}(p) = \Theta(x_{1,2}(p)-a_{1,2})$, with $a_{1,2} \in \Z + \frac{1}{2}$. The cup product of the two 1-cocycles $\delta f_1$ and $\delta f_2$ is given explicitly by
\begin{equation}
    \begin{split}
(\delta f_1 \cup \delta f_2)(p_0, p_1, p_2) &= \frac{1}{6} \sum_{\sigma \in S_3} (\operatorname{sgn} \sigma) [f_1(p_{\sigma(1)}) - f_1(p_{\sigma(0)})] \\
&\times [f_2(p_{\sigma(2)}) - f_2(p_{\sigma(1)})] \text{,}
\end{split} \label{eqn:cup2}
\end{equation}
where the sum is over permutations of the three-element set $\{0,1,2\}$.

To determine $\Omega^{(4)}$, we need to make a choice of local Hamiltonians $H_p$ so that $H = \sum_p H_p$.  We choose $H_p$ to have support only on the site $p=(x_1,x_2)$ and its four nearest-neighbor sites as 
\begin{equation}
\label{Hp_2d}
\begin{split}
    &H_{p}(w)= H_{(x_1,x_2)}^{1}(w)\\
    +&xH^{2,\pm}_{(x_1,x_2);(x_1,x_2+1)}(w)
    +(1-x)H^{2,\mp}_{(x_1,x_2);(x_1,x_2-1)}(w)\\+&yH^{2,\pm}_{(x_1,x_2);(x_1+1,x_2)}(w)
    +(1-y) H^{2,\pm}_{(x_1,x_2);(x_1-1,x_2)}(w),
\end{split}
\end{equation}
where the single-spin and two-spin terms can be found in \eqref{H_1body}, \eqref{H_2body}, and \eqref{V_couple}, and where $0\le x,y\le 1$ are real $w$-independent parameters introduced to illustrate the ambiguity in choosing local Hamiltonians. In the discussion below, we will only need to use the fact that when $w_5 = 0$, the support of $H_p$ lies within the $1d$ layer containing $p$.

It is quite involved to obtain an analytical expression for $\Omega^{(4)}$, so we instead evaluate the KS number $KS = \int_{S^4} \Omega^{(4)}$ using the bulk-boundary correspondence. We first discuss the bulk-boundary correspondence for a general gapped $2d$ parametrized system over $X$, and then specialize to the specific system $H_{2d}$ over $S^4$.

We consider both infinite and semi-infinite systems, \emph{i.e.} systems without and with a boundary. We choose $f_\mu(p)=\Theta(x_\mu(p)-a_\mu)$ with $\mu=1,\,2$. For concreteness, we take the boundary to run along the vertical direction; that is, the system with boundary is defined by retaining only those lattice sites with $x_1 \leq N$. We focus on the case $X = X_{{\rm bdy}} = X_{{\rm bulk}}$, where $X$ is a differentiable manifold.

The higher Berry curvatures in the infinite and semi-infinite systems are denoted $\Omega^{(4)}_{\infty}$ and $\Omega^{(4)}_{\infty/2}$, respectively. These 4-forms are both defined on the subspace $X_\Delta \subset X$ for which the semi-infinite system is gapped. We have
\begin{equation}
\label{4form_semi}
\lim_{a_1 \ll N} \Big[ \Omega^{(4)}_{\infty/2}(f) - \Omega^{(4)}_{\infty}(f) \Big] = 0 \text{,} 
\end{equation}
where the limit means that $a_1$ is taken deep inside the bulk, far away from the spatial boundary.

The formula \eqref{4form_semi} is expected to hold for a general gapped $2d$ parametrized system. To see this, we first observe that the two step functions $f_1$ and $f_2$ divide the $2d$ plane into four quadrants meeting at the point $(a_1, a_2)$. Moreover, by \eqref{eqn:cup2},
$(\delta f_1 \cup \delta f_2) (p,q,r)$ is only non-zero when the lattice sites $p$, $q$ and $r$ all lie in different quadrants. We next consider the expression
\begin{equation}
\Omega^{(4)} = \frac{1}{3!} \sum_{p,q,r} F^{(4)}_{pqr} (\delta f_1 \cup \delta f_2)(p,q,r) \text{,}
\end{equation}
and observe that the expected exponential decay of $F^{(4)}_{pqr}$ as discussed above in Sec.~\ref{Sec:KS_higherD} implies the sum is dominated by contributions where $p$, $q$ and $r$ are all near the point $(a_1,a_2)$. Finally, the expected locality of $F^{(4)}_{pqr}$ (see Sec.~\ref{Sec:KS_higherD}) implies that, in the limit where $(a_1,a_2)$ is far away from the spatial boundary, the dominant local contributions to $\Omega^{(4)}$ in the infinite and semi-infinite systems become the same, and we have $\Omega^{(4)}_{\infty/2}(f) - \Omega^{(4)}_{\infty}(f) \to 0$.

Next, using \eqref{pairing_bdy} and \eqref{descentEq}, we express $\Omega_{\infty/2}^{(4)}$ in terms of a $3$-form defined by $\omega^{(3)} = \langle F^{(3)}, f_1\cup \delta f_2\rangle$ by writing
\begin{equation}
\begin{split}
\label{Omega4_semi_bdy}
    \Omega_{\infty/2}^{(4)}(f)=&\langle F^{(4)}, \delta (f_1\cup \delta f_2)\rangle=\langle \partial F^{(4)}, f_1\cup \delta f_2\rangle\\
    =&\langle dF^{(3)}, f_1\cup \delta f_2\rangle=
    d\langle F^{(3)}, f_1\cup \delta f_2\rangle\\
    = &d\omega^{(3)}(f).
\end{split}    
\end{equation}
These manipulations only make sense in the semi-infinite system, because the sum in the pairing defining $\omega^{(3)}$ would be divergent in an infinite system. This can be seen noting that
\begin{equation}
(f_1 \cup \delta f_2)(p_0, p_1) = \frac{1}{2} [ f_1(p_0) + f_1(p_1)] [ f_2(p_1) - f_2(p_0) ] \text{.}
\end{equation}
This is non-zero whenever $x_2(p_0)$ and $x_2(p_1)$ lie on opposite sides of $a_2$, and if $x_1(p_0) > a_1$ or $x_1(p_1) > a_1$. In a semi-infinite system, the finite width of the strip with $a_1 < x_1 \leq N$ combined with the exponential decay of $F^{(3)}_{p_0 p_1}$ ensures that $\omega^{(3)}(f)$ is well-defined.

We can view $\omega^{(3)}(f)$ as a $3$-form boundary higher Berry curvature, where in this case the effectively one-dimensional ``boundary'' is considered to be all lattice sites with $x_1 > a_1$. This point of view is justified by considering the situation (at some value of parameters) where the boundary is decoupled from the bulk (\emph{i.e.} lattice sites with $x_1 < a_1$), so the boundary is truly a one-dimensional system. In this case, $F^{(3)}_{p_0 p_1} = 0$ if $p_0$ lies in the boundary and $p_1$ lies in the bulk (or vice versa). Then we have
\begin{equation}
\label{2d_omega3}
\omega^{(3)}=\langle F^{(3)},\delta f_2\rangle=\frac{1}{2}\sum_{p_0, p_1} \big[
f_2(p_1)-f_2(p_0)
\big] 
F_{p_0p_1}^{(3)},
\end{equation}
where the sum is over sites lying in the boundary, \emph{i.e.} those with $x_1 > a_1$. This is nothing but the $3$-form higher Berry curvature for the (decoupled) one-dimensional boundary. In particular, $d \omega^{(3)}(f) = 0$ whenever the boundary is decoupled from the bulk. 

To summarize, \eqref{4form_semi} and \eqref{Omega4_semi_bdy} provide us with a bulk-boundary correspondence relating the bulk 4-form higher Berry curvature to the $3$-form boundary higher Berry curvature. In particular, $\Omega^{(4)}$ can be interpreted as a flow of $3$-form higher Berry curvature to/from a spatial boundary. 

We now specialize again to the system $H_{2d}$ over $S^4$ and use these results to compute the KS number. In order to proceed, we need to give $S^4$ an orientation. We do this by viewing $S^4$ as the suspension $S^4 \cong S(S^3)$, taking the orientation on $S^3$ given in Sec.~\ref{Sec:Berry1d}, and using this to specify an orientation on $S(S^3)$ as described in a more general situation below in Sec.~\ref{sec:ndKSnumber}. Here, we specifically note that the coordinate $t$ in the suspension construction can be identified with the coordinate $w_5$ when we view $S^4$ as a subspace $S^4 \subset \R^5$; if we start with $S^4$ and remove the two poles $w_5 = \pm 1$, the resulting space is homeomorphic to the product $S^3 \times (-1,1)$.
When that space is endowed with the product orientation of the given one on $S^3$ and the upward one on $(-1,1)$, the homeomorphism equips $S^4$ minus the poles with an orientation which can be extended to $S^4$ and is the one we choose.

We impose a spatial boundary as described in Sec.~\ref{sec:2dmodel} (see Fig.\ref{f1f2}), taking $N$ even for concreteness. Moreover, we choose $a_1 \in 2 \Z - \frac{1}{2}$ with $a_1 < N$. We note that \eqref{4form_semi} holds without taking the limit $a_1 \ll N$, because neighboring $1d$ layers are only coupled in pairs. At $w_5 = 0$, we have decoupled $1d$ systems $H_{1d}$ and $\overbar{H}_{1d}$ running along the $x_2$-direction, and the boundary higher Berry curvature is simply a sum of higher Berry curvatures of these decoupled systems. As a result we have $\omega^{(3)}(f) = \Omega^{(3)}_{1d}(f_2)$, the higher Berry curvature of $H_{1d}$. This is so because, apart from the $x_2 = N$ layer on the boundary, the systems $H_{1d}$ and $\overbar{H}_{1d}$ come in pairs with equal and opposite contributions to $\omega^{(3)}(f)$.

Now we let $S^3(y) \subset S^4$ be the subspace defined by setting $w_5 = y$, and similarly define $D_4(y) \subset S^4$ as the subspace with $w_5 \geq y$. We have $S^3(y) = \partial D^4(y)$, which gives $S^3(y)$ an orientation inherited from that of $S^4$ by the collar method of Sec.~\ref{sec:solvable-bbc}. The discussion above implies that $\int_{S^3(0)} \omega^{(3)}(f) = 2\pi$; this is nothing but the expression for the KS number of $H_{1d}$. More generally, we have
\begin{equation}
    \int_{S^3(y)}\omega^{(3)}=2\pi \quad \text{for}\quad -1<y\le 0 \text{.}
\end{equation} 
This holds because for $-1 < w_5 \leq 0$ there is no coupling between boundary and bulk, so we can think of $y$ as parametrizing a deformation of the gapped $1d$ boundary system over $S^3$. Under such a deformation, the $1d$ KS number is invariant.

Using \eqref{Omega4_semi_bdy} and applying Stokes theorem gives
\begin{equation}
    \int_{S^3(y)}\omega^{(3)}=\int_{D^4(y)}\Omega^{(4)}_{\infty/2}=\int_{D^4(y)}\Omega^{(4)}_{\infty} \text{.}
\end{equation}
Finally, taking the limit $y \to -1$, we obtain the KS number
\begin{equation}
    \int_{S^4}\Omega^{(4)}_{\infty}=\lim_{y\to -1}\int_{S^3(y)}\omega^{(3)}=2\pi.
\end{equation}
That is, the boundary KS number over a small 3-sphere surrounding the gapless point equals the quantized KS number of the bulk.

We remark that the same analysis applies if we start with a general invertible gapped $1d$ system $H$ over $X$, where $X$ is a closed oriented differentiable 3-manifold, and consider the $2d$ system $SH$ over $SX$ obtained by the suspension construction. That is, the $2d$ KS number of $SH$ is equal to the $1d$ KS number of $H$.

We further remark that the clutching construction of Sec.~\ref{Sec:Clutching} can be generalized to the present setting of a $2d$ gapped invertible system $H$ over $S^4$, with higher Berry curvature $\Omega^{(4)}$. In the 2d case, we need the further assumption that $H$ is in a reduced phase (see Sec.~\ref{sec:gc1}); recall this means that $H(w)$ is in a trivial phase over $\{ w \}$ for any $w \in S^4$. We cover $S^4$ with two subspaces $S^4_N$ and $S^4_S$ defined by $w_5 \geq 0$ and $w_5 \leq 0$, respectively. These subspaces are contractible, and using the assumption that $H$ is in a reduced phase, it follows that the restrictions of $H$ to $S^4_N$ and $S^4_S$ are in the trivial phase. Therefore we can introduce two different gapped semi-infinite systems $H_{N,S}$ over $S^4_{N,S}$, whose Hamiltonians are identical to $H$ in the bulk, and differ only near the spatial boundary, which for concreteness we define by retaining only lattice sites with $x_1 \leq N$. The higher Berry curvatures of the semi-infinite systems satisfy
\begin{equation}
\Omega^{(4)}_{N,S} = d \omega^{(3)}_{N,S} \text{,}
\end{equation}
where the $3$-forms $\omega^{(3)}_{N,S}$ have the interpretation of boundary $1d$ higher Berry curvatures. Then the KS invariant is given by
\begin{equation}
KS = \int_{S^4} \Omega^{(4)} = \lim_{a_1 \ll N} \Big[ \int_{S^3} \omega^{(3)}_N - \int_{S^3} \omega^{(3)}_{S} \Big] \text{,}
\end{equation}
where $S^3 = S^4_N \cap S^4_S$. For the system $H_{2d}$, and more generally for any $2d$ system obtained via the suspension construction, the clutching construction provides an alternative method to compute the KS invariant.

\subsection{KS number in $(d+1)$ dimensional lattice systems from suspension construction}
\label{sec:ndKSnumber}

So far we have given an explicit construction of $1d$ and $2d$ lattice models that have nonzero KS number.
One can go on to use these lower dimensional lattice models and their inverses to construct higher dimensional systems with nonzero KS number, based on the suspension construction as introduced in Sec.~\ref{Sec:GeneralConstruct}. Here, given a $d$-dimensional gapped invertible system $H_d$ over $X$, and the $(d+1)$-dimensional system $S H_d$ over $S X$ obtained from $H_d$ by suspension construction,  we will show that 
\begin{equation}
\label{KS_bbc}
KS_{d+1}(SH_d) = \int_{SX}\Omega^{(d+3)} =  \int_{X} \Omega^{(d+2)} = KS_d(H_d) \text{,}
\end{equation}
where $\Omega^{(d+2)}$ is the higher Berry curvature of $H_d$ and $\Omega^{(d+3)}$ is that of $S H_d$. That is, the KS number of the $(d+1)$-dimensional system $SH_d$ over $SX$ is identical to that of the $d$-dimensional system $H_d$ over $X$.

Here we assume that $X$ is a closed oriented differentiable manifold, with orientation specified by a nowhere-vanishing $(d+2)$-form $\tau$. There is a subtlety, because in general $SX$ is not a topological manifold; in fact, $SX$ is a closed topological manifold if and only if $X$ is a sphere. Therefore in order to make sense of the claim \eqref{KS_bbc}, we have to give a definition of the KS number on $SX$ and show it is a phase invariant. Below, we first discuss the simpler case where $X$ is a sphere, employing the clutching construction (see Sec.~\ref{Sec:Clutching} and Sec.~\ref{Sec:2d_KS_bbc}). Next, we treat the general case where $X$ is an arbitrary closed oriented differentiable manifold. 

Now letting $X = S^n$, the result \eqref{KS_bbc} holds trivially if $n \neq d+2$, since both KS numbers are zero.  Therefore we assume $X = S^{d+2}$. We fix the orientation on $SX \cong S^{d+3}$ by choosing the $(d+3)$-form $- dt \wedge \tau$, which is defined away from the $t = \pm 1$ poles. Below, while we do assume $X = S^{d+2}$, we write $X$ and $SX$ instead of $S^{d+2}$ and $S(S^{d+2}) \cong S^{d+3}$, to help make contact with the more general discussion to follow.

As described in Sec.~\ref{Sec:GeneralConstruct}, $S H_d$ can be viewed as a one-dimensional lattice of $d$-dimensional layers. We choose coordinates so that the layers are stacked along the $x_1$-direction. Let us consider a semi-infinite system over $S X$ by terminating the system at layer $x_1=N$. Taking $a_1 < N$ in the step function $f_1(p)=\Theta(x_1(p)-a_1)$, and restricting to the subspace of $SX$ over which the semi-infinite system is gapped, we have
\begin{equation}
\label{Omega_d+3}
    \Omega^{(d+3)}_{\infty}=\Omega^{(d+3)}_{\infty/2},
\end{equation}
where $\Omega_{\infty}^{(d+3)}$ and $\Omega_{\infty/2}^{(d+3)}$ denote the higher Berry curvatures of the infinite and semi-infinite systems, respectively. Here, the equality is expected to be exact without taking the limit $a_1 \ll N$ because layers are only coupled in pairs. The result \eqref{Omega_d+3} relies on the expectation that $F^{(d+3)}$ (in $\Omega^{(d+3)}=\langle F^{(d+3)}, \delta f_1\cup \cdots \cup \delta f_{d+1}\rangle$) is a local quantity in the sense discussed in Appendix \ref{app:locality}.

In the semi-infinite system, the higher Berry curvature can be written
\begin{equation}
\label{d+3_form_bdy}
\begin{split}
    \Omega_{\infty/2}^{(d+3)}=&\langle F^{(d+3)}, \delta (f_1\cup \delta f_2\cup \cdots \cup \delta f_{d+1})\rangle\\
    =&d \langle F^{(d+2)}, f_1\cup \delta f_2\cup \cdots \cup \delta f_{d+1}\rangle\\
    =& \, d\omega^{(d+2)},
    \end{split}
\end{equation}
where the $(d+2)$-form
\begin{equation}
\omega^{(d+2)} = \langle F^{(d+2)}, f_1\cup \delta f_2\cup \cdots \cup \delta f_{d+1}\rangle
\end{equation} is a boundary higher Berry curvature, and is expected, with appropriate choice of local Hamiltonian terms $H_p$, to reduce to the higher Berry curvature of a $d$-dimensional system for parameters where the boundary region ($x_1 > a_1)$ is decoupled from the bulk ($x_1 < a_1$). In the second step of \eqref{d+3_form_bdy}, we have used \eqref{descentEq} and \eqref{pairing_bdy}. The relations \eqref{Omega_d+3} and \eqref{d+3_form_bdy} can be considered as a bulk-boundary correspondence for the higher Berry curvature, and tell us that the $(d+3)$-form higher Berry curvature measures the flow of $(d+2)$-form Berry curvature to the boundary.

If the KS number $KS_{d+1}(SX) = \int_{SX}\Omega_{\infty}^{(d+3)}$ is nonzero, then the cohomology class of $\Omega_{\infty}^{(d+3)}$ is non-trivial, and
$\omega^{(d+2)}$ is not globally well defined over $SX$.
Our strategy is to cover $SX$ by two charts as $S X=S X_+ \cup S X_-$, where $S X_+$ ($S X_-$) denotes the subspace of $S X$ with $t\ge 0$ ($t\le 0$).
On each chart a gapped boundary termination can be chosen (\emph{i.e.} we consider two different gapped semi-infinite systems, one over each chart), and we have well-defined $(d+2)$-forms $\omega^{(d+2)}_{\pm}$ on $SX_{\pm}$.
More concretely, on $S X_{+}$ ($S X_{-}$), we choose the boundary terminated at layer $x_1=N\in 2\mathbb Z$ ($x_1=N-1\in 2\mathbb Z-1$), which results in gapped systems over $SX_{\pm}$ (see Sec.~\ref{Sec:GeneralConstruct}).
Denoting the higher Berry curvature on $SX_{\pm}$ by $\Omega^{(d+3)}_{\pm}$, we have 
\begin{equation}
\label{dOmega_d+3_bdy}
  \Omega^{(d+3)}_{\pm}=
  d\omega^{(d+2)}_{\pm} \text{.}
\end{equation}

The KS invariant of $SH_d$ can then be expressed
\begin{equation}
    \int_{SX} \Omega^{(d+3)}_{\infty}=
    \int_{SX_{+}} \Omega^{(d+3)}_{+}+\int_{SX_{-}} \Omega^{(d+3)}_{-},
\end{equation}
where we have used \eqref{Omega_d+3}.
Then, by \eqref{dOmega_d+3_bdy} and Stokes theorem, we obtain
\begin{equation}
\label{SX_split}
    \int_{SX} \Omega^{(d+3)}_{\infty}=\int_X \omega_{+}^{(d+2)}-\int_X \omega_{-}^{(d+2)},
\end{equation}
where $X= S X_{+}\cap S X_{-}$ is the $t=0$ ``equator" of $SX$. The minus sign in \eqref{SX_split} arises because 
the orientation on $X$ induced by $X = \partial (S X_{-})$ according to the collar method  is opposite to the given orientation of $X$.

At $t=0$, the system $S H_d$ is composed of decoupled layers that alternate between $H_{d}$ and $\overbar{H}_d$. Therefore,
\begin{equation}
\label{d+2_form_t_0}
\begin{split}
   \int_X \omega_{+}^{(d+2)}=\sum_{x_1\in\mathbb Z,a_1<x_1\le N}(-1)^{x_1}  \int_X\Omega^{(d+2)},\\
      \int_X \omega_{-}^{(d+2)}=\sum_{x_1\in \mathbb Z, a_1<x_1\le N-1}(-1)^{x_1}  \int_X\Omega^{(d+2)},
   \end{split}
\end{equation}
where $\Omega^{(d+2)}$ is the higher Berry curvature of $H_{d}$. Here we are assuming the local Hamiltonian terms $H_p$ are chosen so that, at $t=0$, the support of $H_p$ lies within the $d$-dimensional layer containing $p$. We are also assuming for notational convenience that $\overbar{H}_{d}$ has higher Berry curvature (not just KS number) equal and opposite  to that of $H_{d}$; this assumption is easily relaxed. From \eqref{SX_split} it then follows that 
\begin{equation}
    KS_{d+1}(SH_d)=KS_d(H_d) \text{,}
\end{equation} 
which is the desired relation between KS invariants in the case where $X = S^{d+2}$.

We now address the case when $X$ is not a sphere. In this case, $SX$ is not a closed manifold: The north ($t=1$) and south ($t=-1$) poles are singular points. 
We consider a general $d$-dimensional reduced invertible system $H$ over $SX$.
Provided that $H$ is differentiable away from the singularities of $SX$, we can enlarge our definition of the KS invariant so that the results above continue to hold. 

For $0 < \epsilon < 1$, define a function $r: SX \to SX$ by
\begin{equation}
r(x,t)= \begin{cases} (x,(1-\epsilon)^{-1} t) & |t|<1-\epsilon \\
N=\{X \times \{1\}\} & t \geq 1-\epsilon \\
S=\{X \times \{-1\}\} & t \leq -1+\epsilon \\
\end{cases}
\end{equation}
where $N$ and $S$ are the equivalence classes of points representing the north and south poles. 
The system $H \circ r$ is locally constant and trivial outside of the subset $X \times [-1+\epsilon, 1-\epsilon]$.
Furthermore, $H \circ r$ is in the same phase as $H$, as can be seen by tuning $\epsilon$ back to zero in the definition above. 

For $H\circ r$, the higher Berry curvature $\Omega^{(d+2)}(f)$ is defined on $X \times (-1,1)$. It is a closed  compactly supported differential form on the non-compact manifold $X \times (-1,1)$. So, we get an element of $H_{c, dR}^{d+2}(X\times (-1,1))$, where the subscripts denote compactly supported de Rham cohomology. But this group is isomorphic to $H^{d+2}([X \times (-1,1)]^*,\{ * \})$, where the asterisk denotes both the one-point compactification and the single added point (see Lemma 11 of Chapter 6, Section 6 in Ref.~\onlinecite{spanierbook}), and in turn $H^{d+2}([X \times (-1,1)]^*,\{ * \}) \cong H^{d+2}(SX)$ for all $d \geq 0$. The latter group is where the KS invariant should live.

The KS number is then defined by
\begin{equation}
KS = \int_{X \times (-1,1)} \Omega^{(d+2)}(f) \text{,}
\end{equation}
which is well-defined because $\Omega^{(d+2)}(f)$ is compactly supported.
We argue that $KS$ is quantized in integer multiples of $2\pi$, making the assumption that the KS number of an invertible system over a closed oriented manifold is similarly quantized. The idea is to obtain a system over the manifold $X \times S^1$ whose KS number is equal to $KS$ defined above. We introduce a continuous function $h_{NS} : [0,1] \to \GH_d$ so that $h_{NS}(0) = H(N)$ and $h_{NS}(1) = H(S)$; this is possible because both $H(N)$ and $H(S)$ are trivial systems over a point. We then define a new system $\widetilde{H}_\delta$ over $SX$ by
\begin{equation}
\widetilde{H}_\delta (x,t)= \begin{cases} h_{NS}\Big( \delta \frac{t - (1-\epsilon)}{\epsilon} \Big)  & t \geq 1-\epsilon \\
H\circ r(x,t) & t < 1-\epsilon
\end{cases} \text{,}
\end{equation}
where $\delta \in [0,1]$. Since $\widetilde{H}_0 = H$,  we see that $\widetilde{H}_\delta$ is in the same phase as $H\circ r$ by tuning $\delta$ to zero.
Now at $\delta = 1$ we have the property $\widetilde{H}_1(S) = \widetilde{H}_1(N) = H(S)$. Therefore we can view $\widetilde{H}_1$ as a system over $X \times S^1$. Moreover, the higher Berry curvatures of $\widetilde{H}_\delta$ and $H$ are same for all $\delta$; for $t > 1-\epsilon$ the Hamiltonian only depends on $t$ and thus the higher Berry curvature is zero there. The KS number of $H$ is thus the same as that of $\widetilde{H}_1$ viewed as a system over $X \times S^1$.

Next we need to verify that $KS$ is an invariant for phases over $SX$. If $H$ can be deformed to $H'$ via a continuous homotopy $h : SX \times I \to \mathfrak{GH}_d$, then $h\circ r$ deforms $H \circ r$ to $H' \circ r$ via a homotopy of systems which are equal to a constant trivial system outside of $X \times [-1+\epsilon, 1-\epsilon]$. Because the KS number is quantized, it follows that the systems $H\circ r$ and $H'\circ r$ have the same KS number.

With this definition of the higher Berry curvature and KS number for a system over $SX$, the discussion above applies to any closed manifold $X$, giving the relationship between the KS numbers of $H_d$ and $SH_d$ stated in \eqref{KS_bbc}. We note that the boundary higher Berry curvature $\omega^{(d+2)}_{+}$ on $SX_{+}$ is compactly supported on $X \times [0,1-\epsilon] \subset SX_+$, because the semi-infinite system over $SX_+$ is trivial and constant outside $X \times [0,1-\epsilon]$. The analogous result holds for $\omega^{(d+2)}_-$ on $SX_-$. Equation~\eqref{SX_split} thus continues to hold because \emph{e.g.} the integral of $\omega^{(d+2)}_+$ over $X \times \{ 1-\epsilon \}$ vanishes.

Different choices of conventions would result in the slightly different relationship $KS_{d+1}(SH_d)=-KS_d(H_d)$. For instance, the minus sign appears if we reverse the chosen orientation on $SX$. Another way to get a minus sign is to exchange the roles of $H_d$ and $\overbar{H}_d$ in the suspension construction.

\subsection{KS number pump in $(d+1)$ dimensions}
\label{subSec:KSpump}

In Sec.~\ref{sec:chernpump}, we studied a $1d$ system $H_{cp}$ over $S^2 \times S^1$ that we interpreted as a Chern number pump. Upon adiabatically cycling the coordinate of the $S^1$, a quantized $2\pi$ Chern number over $S^2$ was pumped from the boundary of the system into the bulk. This was generalized in Sec.~\ref{subsec:general-bdy-physics}, where given a gapped $d$-dimensional invertible system $H_d$ over $X$, we showed that the $(d+1)$-dimensional system $P H_d$ over $X \times S^1$ can be interpreted as a pump of the $d$-dimensional phase invariant of $H_d$ from the boundary into the bulk. Therefore if $H_d$ is characterized by the $d$-dimensional KS number $KS_d$, we expect that $P H_d$ is a KS number pump. That is, for each cycle around $S^1$, $P H_d$ pumps a $d$-dimensional KS number $KS_d$ from the spatial boundary into the bulk. We now briefly establish this expectation.

We consider a gapped boundary termination with $Y_{{\rm bdy}} = X \times [-1,1]$ and $Y_{{\rm bulk}} = X \times S^1$, as described in Sec.~\ref{subsec:general-bdy-physics} and illustrated in Fig.~\ref{fig:Psd_pumping_open}. We suppose $X$ is an oriented closed differentiable $(d+2)$-manifold, with orientation given by a nowhere-vanishing $(d+2)$-form $\tau$.  We then give $X \times [-1,1]$ the orientation specified by the $(d+3)$-form $-dt \wedge \tau$, where $t \in [-1,1]$. This also gives $X \times S^1$ an orientation by viewing $S^1 = [-1,1]/\{-1,1\}$.

We let $\Omega^{(d+3)}_{\infty/2}$ denote the higher Berry curvature of the semi-infinite system, and observe that the cohomology class of $\Omega^{(d+3)}_{\infty/2}$ is trivial for dimensional reasons. This is so because $Y_{{\rm bdy}} = X \times I$ is homotopy equivalent to the $(d+2)$-manifold $X$. Therefore, $\Omega^{(d+3)}_{\infty/2} = d \omega^{(d+2)}$, where $\omega^{(d+2)}$ is globally well-defined on $Y_{{\rm bdy}}$ and has the interpretation of boundary higher Berry curvature.

Now we let $\Omega^{(d+3)}$ be the higher Berry curvature of the infinite $(d+1)$-dimensional system $P H_d$.  Applying the bulk-boundary correspondence and Stokes theorem, we have
\begin{equation}
\begin{split}
KS_{d+1} &= \int_{X \times S^1} \Omega^{(d+3)} = \int_{X \times [-1,1]} \Omega^{(d+3)}_{\infty/2} \\
&= - \int_{X \times \{1\}} \omega^{(d+2)} + \int_{X \times \{-1\}} \omega^{(d+2)}
= KS_d
\end{split}
\end{equation}
The last equality holds because the system is trivial at $t=1$, while at $t=-1$ the system is trivial apart from the $x_1 = N$ boundary layer, which is in the $d$-dimensional system $H_d$. Not only do we see that $KS_{d+1} = KS_d$, but we also see that $KS_{d+1}$ can be interpreted as the $d$-dimensional KS number pumped from the spatial boundary into the bulk over a single cycle of the $S^1$ parameter.

\section{Discussion}
\label{sec:discussion}

In this paper, we explored the physics of parametrized quantum systems, focusing in particular on invertible phases of parametrized systems. We developed a bulk-boundary correspondence for the Kapustin-Spodyneiko higher Berry curvature, which clarifies its physical interpretation. Moreover, we introduced a pair of quantum pumping constructions that take as input a $d$-dimensional invertible system over $X$ and produce a $(d+1)$-dimensional system either over the suspension $SX$, or over $X \times S^1$. These constructions can be used to generate many examples, and we used them in our discussion of $d$-dimensional systems over $S^{d+2}$ with non-zero KS invariant. Moreover, the construction producing a system over $SX$, referred to as the suspension construction, is proposed to realized the suspension isomorphism in a generalized cohomology theory of parametrized invertible phases.

Many related directions remain to be explored in future work on parametrized quantum systems. For instance, some results on invertible such systems with discrete symmetries have been reported, which are also related (under duality) to parametrized systems with spontaneously broken symmetry, and (upon gauging symmetry) to parametrized non-invertible systems.\cite{Hermele2021,shiozaki2021adiabatic} Much work remains to be done in understanding the bulk and anomalous boundary properties of such systems. More generally, parametrized non-invertible systems are even less explored and understood than their invertible counterparts.

We close with a brief remark on interfaces between systems with different KS numbers. As discussed at the end of Sec.~\ref{sec:generic-bbc}, for two $1d$ systems over the same closed oriented 3-manifold $X$, if their KS numbers are different, then an interface between them must be gapless for some $x \in  X$.
For $d$ dimensional systems over a closed oriented $(d+2)$-dimensional manifold $X$, we argue that this conclusion 
is still true, \emph{i.e.}, the $(d-1)$ dimensional interface must be gapless for some $x \in X$ (or, there must be a first-order interface phase transition for some $x \in X$) if the two 
$d$-dimensional systems have different KS numbers. We note that a similar conclusion was reached in Ref.~\onlinecite{Hsin_2020} for systems over $S^{d+2}$. In particular, this rules out the possibility of having a topologically ordered system on the $(d-1)$ dimensional interface.
This is in contrast to invertible phases over a point, where the interface between two invertible systems 
in different phases can sometimes host a gapped, anomalous topologically ordered system if $d \geq 3$.

The argument is as follows. We consider two gapped (but not necessarily invertible) $d$-dimensional systems over $X$ as introduced in Sec.~\ref{sec:ndKSnumber}, with an interface between them located near the hyperplane $x_1=0$. We suppose the interface is gapped everywhere over $X$, and consider higher Berry curvatures defined for two different choices of $a_1$ in the function $f_1 = \Theta(x_1 - a_1)$. In one case we choose $a_1 \ll 0$, and in the other we choose $a_1 \gg 0$, so that the two higher Berry curvatures (as local quantities) are those of the two $d$-dimensional systems and do not depend on the properties of the interface. These two $(d+2)$-forms differ only by an exact form, and thus must give rise to the same KS number. We emphasize that these arguments continue to hold if the interface is gapped but topologically ordered. Therefore, if the two systems have different KS numbers, their interface must become gapless at some $x\in X$, or a first-order phase transition must occur at the interface for some $x \in X$.

\acknowledgments We are grateful to Anton Kapustin and Lev Spodyneiko for useful discussions. This material is based upon work supported by the National Science Foundation under Grant No.~DMS 2055501 awarded to AB, MH and MJP. The research of MQ is supported by the NDSEG program.  This work was also partly supported by the Simons Collaboration on Ultra-Quantum Matter, which is a grant from the Simons Foundation (651440, MH, XW; 618615, AV, XW).

\appendix

\section{Higher Berry curvature of $1d$ lattice model}
\label{Appendix:Berry}

Here we give some details on the calculation of the
 higher Berry curvature in the 1d lattice model $H_{1d}$ studied in Sec.~\ref{sec:1dmodel-bbc}.  For the Hamiltonian $H_{1d}(w)$ in \eqref{H_1d}, one can find the
energy spectrum is gapped everywhere over $S^3$.
Recalling that for any $w = (w_1,\dots,w_4) \in S^3$, spins are only coupled in dimers, it  is enough to check the energy spectrum for a single dimer. This is easily found to be 
\begin{equation}
E_n=-2-|w_4|, \, |w_4|, \, |w_4|, 2-|w_4| \text{.}
\label{eqn:dimer-spectrum}
\end{equation}
Therefore, the system is gapped for all $w\in S^3$. At the $w_4 = 0$ equator, the spins are decoupled, and the energy spectrum for a each spin is $E_n = \pm 1$, which reproduces \eqref{eqn:dimer-spectrum}  for a pair of spins (at $w_4 = 0$).

To obtain the higher Berry curvature $\Omega^{(3)}$, we consider the $3$-form $F_{pq}^{(3)}$ in \eqref{F3}. By inserting
complete sets of basis and performing the contour integral, one can obtain
\begin{equation}
\label{F_pq3_sum}
\begin{split}
F_{pq}^{(3)}=&\frac{i}{6}\big[
-2\langle  dH \,G_0^2 \, dH_p \, G_0 \, dH_q \rangle
-\langle dH\, G_0\, dH_p\, G_0^2\, dH_q\rangle\\
&+2\langle dH_p\, G_0\, dH_q\, G_0^2\,dH\rangle
+\langle dH_p\, G_0^2\, dH_q\, G_0\, dH\rangle\\
&+\langle dH_q\, G_0^2\, dH\, G_0\, dH_p\rangle
-\langle dH_q \, G_0\, dH\, G_0^2\, dH_p \rangle\\
&+3\langle dH\, G_0^3 \,dH_p \rangle\langle dH_q\rangle
-3\langle dH_q\, G_0^3 \, dH\rangle\langle dH_p\rangle\big]\\
&-(p\leftrightarrow q),
\end{split}
\end{equation}
where $\langle \cdots \rangle$ denotes the ground-state expectation value, and we have defined
\begin{equation}
\label{G0_def}
G_0=\sum_{n\neq 0}\frac{|n\rangle\langle n|}{E_0-E_n} \text{,}
\end{equation}
where $\sum_{n\neq 0}$ denotes the summation  over all excited states,
$E_0$ is the ground state energy, and
$E_{m\neq 0}$ is the energy of the $m$-th excited state. Finally, $H_p$ is the local Hamiltonian defined in \eqref{LocalH}.

Using \eqref{F_pq3_sum}, we evaluated $\Omega^{(3)}$ as expressed in \eqref{F3j}, and 
obtained the explicit expression \eqref{Omega3s}. The same procedure can be used to obtain the higher Berry curvature for other choices of $a$ in the step function $f(p) = \Theta(p-a)$, and for the inverse system $\overbar{H}_{1d}$ discussed in Sec.~\ref{Sec:Inverse}.

Now we consider the boundary Berry curvature $2$-form $\omega^{(2)}$ as studied in Sec.~\ref{Sec:Semi-infinite} and defined by \eqref{2form_omega} in terms of $F^{(2)}_q = \sum_{p \in \Z} F^{(2)}_{p q}$.
By inserting complete sets of energy eigenstates in Eq.\eqref{F2} and performing the contour integral, 
one can rewrite the $2$-forms $F_{pq}^{(2)}$
explicitly as
\begin{equation}
\label{F2Green}
\begin{split}
F_{pq}^{(2)}=&\frac{i}{2}\big[
\langle dH_p\, G_0^2\, dH_q\rangle+\langle dH_q\, G_0^2\, dH_p\rangle
\big] \text{.}
\end{split}
\end{equation}
Since $H_{1d}$ is composed of decoupled dimers, 
the energy eigenstates are tensor products over dimers, and it is
straightforward to check that $F_{pq}^{(2)}=0$ if $|p-q|>1$. 
For $w_4 \geq 0$, the expression \eqref{2form_omega} for $\omega^{(2)}$ can be simplified to $\omega^{(2)}=F^{(2)}_{a+1/2}$, because decoupled dimers that lie fully to the right of $a$ give vanishing contributions. Similarly, for $-1 < w_4 \leq 0$ we have $\omega^{(2)} = F^{(2)}_{N}$.
Using Eq.\eqref{F2Green}, we obtained $\omega^{(2)}$ as given in \eqref{omega_2form}.

\section{Energy spectrum of four-spin clusters}
\label{app:fourspin}

Here we show that the energy spectrum of $H(w,t)$ in 
\eqref{Hwt} is gapped for arbitrary $w\in S^3$ and $t\in[0,1]$.
In fact, the spectrum of $H(w,t)$ is directly related to the energy spectrum of the $2d$ lattice model $H_{2d}$ introduced in Sec.~\ref{sec:2dmodel}, by identifying 
$w_5=\sin\frac{\pi t}{2}$ in \eqref{Hamiltonian2d}.
It is noted that the lattice models in \eqref{Hwt} and \eqref{Hamiltonian2d} are
always composed of decoupled clusters (each of which contains 4 sites). Therefore,  
it is enough to study the energy spectrum of a single cluster.

\begin{figure}[h]
\center
\centering
\includegraphics[width=3.25in]{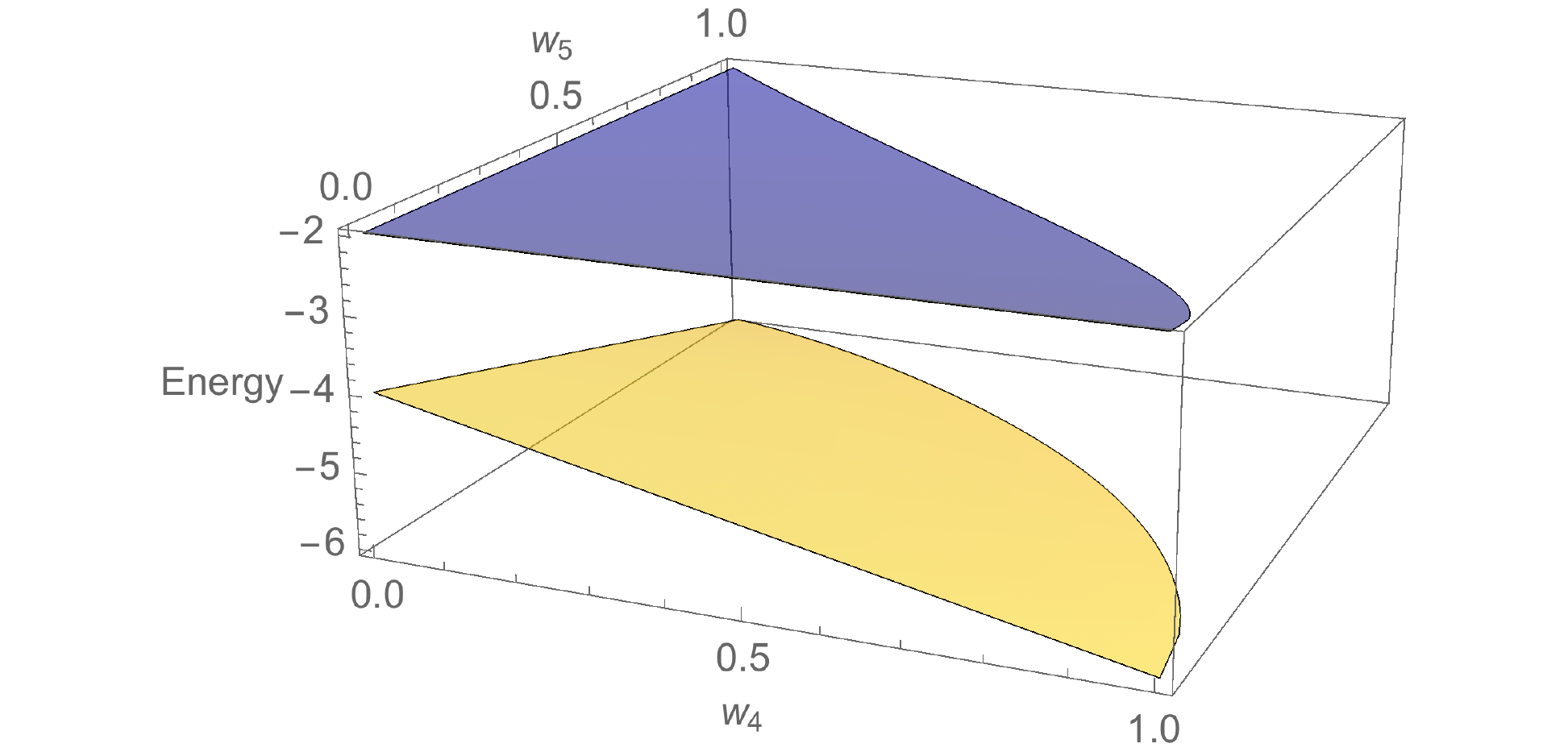}
\caption{
Energies of the ground state and the first excited state of the Hamiltonian in \eqref{RotateH}
 as a function of $w_4$ and $w_5$, with $w_4,\, w_5\in[0,1]$.
}
\label{Energy2d}
\end{figure}

Consider the cluster composed of four sites at $(i,\, j)$ where $i$, $j\in\{1,2\}$ in \eqref{Hamiltonian2d}.
By making a unitary transformation to rotate the on-site magnetic field $\vec{w}=(w_1,w_2,w_3)$
to the positive $z$ direction, the Hamiltonian has the simple form:
\begin{equation}
\label{RotateH}
\small
\begin{split}
H=&\sum_{i,j\in \{1,2\}} (-1)^{i+j} w\, \sigma^3_{i,j}\\
&+w_4 \sum_{\mu=1,2,3}\sigma_{1,1}^{\mu}\sigma_{2,1}^{\mu}+w_4\sum_{\mu=1,2,3}\sigma_{1,2}^{\mu}\sigma^{\mu}_{2,2}\\
&+w_5\sum_{\mu=1,2,3}\sigma_{1,1}^{\mu}\sigma_{1,2}^{\mu}+w_5\sum_{\mu=1,2,3}\sigma_{2,1}^{\mu}\sigma^{\mu}_{2,2},\\
\end{split}
\end{equation}
where $w=\sqrt{1-w_4^2-w_5^2}$, and $w_4,\, w_5\in[0,1]$. We computed the energy spectrum of the 
Hamiltonian in \eqref{RotateH} numerically, as shown in Fig.\ref{Energy2d}, and found that the spectrum is gapped everywhere. 
While the plot is for positive $w_4$ and $w_5$, the same property holds when when $w_4$ or $w_5$ is negative.

\section{Locality of the higher Berry curvature}
\label{app:locality}

Here we express both $F^{(2)}_q$ and $F^{(3)}_{pq}$ in terms of imaginary-time-ordered correlation functions of local operators. These expressions show that these quantities are local, in the sense that they are dominated by contributions from regions of space near $q$, or near $p$ and $q$, respectively. The result for $F^{(3)}$ is used in Sec.~\ref{sec:generic-bbc} to obtain a bulk-boundary correspondence for generic parametrized systems in $d=1$. It is expected from the derivation of these results that similar local expressions exist for $F^{(n)}$ when $n > 3$.

We first consider $F^{(2)}_q = \sum_{p \in \Z} F^{(2)}_{pq}$, which from \eqref{F2Green} can be written
\begin{equation}
F^{(2)}_q = \frac{i}{2} \Big(  \langle dH G_0^2 dH_q \rangle + \langle dH_q G_0^2 dH \rangle \Big) \text{.}
\end{equation}
Moreover, defining $\widetilde{dH} = dH - \langle dH \rangle$ and $\widetilde{dH}_q = dH_q - \langle dH_q \rangle$, we have
\begin{equation}
F^{(2)}_q = \frac{i}{2}\Big( \langle \widetilde{dH} G_0^2 \widetilde{dH}_q \rangle + \langle \widetilde{dH}_q G_0^2 \widetilde{dH} \rangle \Big) \text{.} \label{eqn:F2-G0}
\end{equation}

For operators $A$ and $B$ with vanishing ground-state expectation value, the identity
\begin{eqnarray}
\langle A G_0^2 B \rangle &=& \int_0^\infty d\tau \, \tau  \langle A(\tau) B(0) \rangle \\
&=& \int_0^\infty d\tau \, \tau  \langle A e^{-\tau H} B \rangle \label{eqn:AGB}
\end{eqnarray}
can be established by inserting complete sets of energy eigenstates and evaluating the $\tau$ integral. Here, for instance, $A(\tau) = e^{\tau H} A e^{-\tau H}$.

Using \eqref{eqn:AGB} in \eqref{eqn:F2-G0}, we obtain the result
\begin{equation}
F^{(2)}_q = \frac{i}{2} \int_0^{\infty} d\tau \, \tau \Big[
\langle \widetilde{dH}(\tau) \widetilde{dH}_q \rangle + \langle \widetilde{dH}_q(\tau) \widetilde{dH} \rangle \Big] \text{.}
\end{equation}
Because $\widetilde{dH}$ is a sum of local operators, this expresses $F^{(2)}_q$ in terms of a sum of imaginary-time-ordered correlation functions of local operators whose ground-state expectation values vanish.  Therefore we see that the dominant contributions to $F^{(2)}_q$ come from regions of space near $q$, while far-away contributions are exponentially suppressed in a gapped system.  

Now we proceed to obtain a similar result for $F^{(3)}_{pq}$. Starting from \eqref{F_pq3_sum} we have 
\begin{eqnarray}
F^{(3)}_{pq} &=& \frac{i}{6} \Big[ -2 \langle \widetilde{dH} G_0^2 \widetilde{dH}_p G_0 \widetilde{dH}_q \rangle
- \langle \widetilde{dH} G_0 \widetilde{dH}_p G_0^2 \widetilde{dH}_q \rangle  \nonumber  \\
&+& 2 \langle \widetilde{dH}_p G_0 \widetilde{dH}_q G_0^2 \widetilde{dH} \rangle + \langle \widetilde{dH}_p G_0^2 \widetilde{dH}_q G_0 \widetilde{dH} \rangle \nonumber \\
&+& \langle \widetilde{dH}_q G_0^2 \widetilde{dH} G_0 \widetilde{dH}_p \rangle - \langle \widetilde{dH}_q G_0 \widetilde{dH} G_0^2 \widetilde{dH}_p \rangle \nonumber \\
&-& (p \leftrightarrow q) \Big]  \text{.} \label{eqn:F3-useful}
\end{eqnarray}
Each term in this expression can be written as an imaginary-time-ordered correlation function of local operators with vanishing ground-state expectation value. This is easily seen via the identity
\begin{equation}
\label{ABC_correlation}
\begin{split}
\langle A G^a_0 B G^b_0 C \rangle = \frac{(-1)^a (-1)^b}{(a-1)! (b-1)!} \int_0^\infty d\tau_1 \int_0^{\infty} d\tau_2 \\
\tau_1^a \tau_2^b \langle A(\tau_1 + \tau_2) B(\tau_2) C \rangle \text{,}
\end{split}
\end{equation}
where $A$, $B$ and $C$ are operators with vanishing ground-state expectation value, and where $a$ and $b$ are positive integers. The identity can be established by inserting complete sets of energy eigenstates on the right-hand side, and evaluating the integrals. 
The quantity in \eqref{ABC_correlation} decays exponentially to zero if any two operators of $A$, $B$ and $C$ are far away from each other in space. See, \emph{e.g.} Ref.~\onlinecite{Watanabe_2018} for a more detailed discussion. So we see that $F^{(3)}_{p q}$ is dominated by contributions of space from regions near $p$ and $q$. 

We expect that similar expressions can be obtained for arbitrary $F^{(n)}_{p_0 \cdots p_{n-2}}$. In more detail, starting from the expression in Equation~(21) of Ref.~\onlinecite{Kapustin_2020}, we expect an expression similar to \eqref{eqn:F3-useful} can be obtained. In turn, it should be possible to express each term in the resulting expression as an imaginary-time-ordered correlation function of $\widetilde{dH}$ and the $n-1$ operators $\widetilde{dH}_{p_i}$ for $i=0,\dots,n-2$.

\bibliography{ref}

\end{document}